\documentclass[]{aastex631}
\usepackage{graphicx}

\shorttitle{Age Dependence of Infrared Star Formation Rate Indicators}
\shortauthors{Calzetti et al.}

\begin{document}

\title{Quantification of The Age Dependence of Mid--Infrared Star Formation Rate Indicators}

\author[0000-0002-5189-8004]{Daniela Calzetti}
\affiliation{Department of Astronomy, University of Massachusetts Amherst, 710 North Pleasant Street, Amherst, MA 01003, USA}

\author[0000-0001-5448-1821]{Robert C. Kennicutt}
\affiliation{Department of Physics and Astronomy, Texas A\&M University, 578 University Drive, College Station, TX 77843-4242, USA}
 \affiliation{Steward Observatory, University of Arizona, 933 N Cherry Avenue, Tucson, AZ 85721, USA}

\author[0000-0002-8192-8091]{Angela Adamo}
\affiliation{Department of Astronomy, The Oskar Klein Centre, Stockholm University, AlbaNova, SE-10691 Stockholm, Sweden}

\author[0000-0002-4378-8534]{Karin Sandstrom}
\affiliation{Department of Astronomy \& Astrophysics, University of California, San Diego, 9500 Gilman Drive, La Jolla, CA 92093}

\author[0000-0002-5782-9093]{Daniel~A.~Dale}
\affiliation{Department of Physics and Astronomy, University of Wyoming, Laramie, WY 82071, USA}

\author[0000-0002-1723-6330]{Bruce Elmegreen}
\affiliation{Katonah, NY 10536, USA}

\author[0000-0001-8608-0408]{John S. Gallagher}
\affiliation{Department of Astronomy, University of Wisconsin--Madison, 475 N. Charles Street, Madison, WI 53706--1507 USA}
\affiliation{Department of Physics and Astronomy, Macalester University, 1600 Grand Avenue, Saint Paul, MN 55105-1899 USA}

\author[0000-0003-4910-8939]{Benjamin Gregg}
\affiliation{Department of Astronomy, University of Massachusetts Amherst, 710 North Pleasant Street, Amherst, MA 01003, USA}

\author[0009-0008-4009-3391]{Varun Bajaj}
\affiliation{Space Telescope Science Institute, 3700 San Martin Drive Baltimore, MD 21218, USA}

\author[0000-0002-5666-7782]{Torsten B\"oker}
\affiliation{European Space Agency, c/o STSCI, 3700 San Martin Drive, Baltimore, MD 21218, USA}

\author[0009-0003-6182-8928]{Giacomo Bortolini}
\affiliation{Department of Astronomy, The Oskar Klein Centre, Stockholm University, AlbaNova, SE-10691 Stockholm, Sweden}

\author[0000-0003-4850-9589]{Martha Boyer}
\affiliation{Space Telescope  Science Institute, 3700 San Martin Drive, Baltimore, MD 21218, USA}

\author[0000-0001-6464-3257]{Matteo Correnti}
\affiliation{INAF Osservatorio Astronomico di Roma, Via Frascati 33, 00078, Monteporzio Catone, Rome, Italy}
\affiliation{ASI-Space Science Data Center, Via del Politecnico, I-00133, Rome, Italy}

\author[0000-0001-9419-6355]{Ilse De Looze}
\affiliation{Department of Physics and Astronomy, University of Ghent, Proeftuinstraat 86, 9000 Gent , Belgium}

\author[0000-0002-0846-936X]{Bruce T. Draine}
\affiliation{Department of Astrophysical Sciences, Princeton University, 4 Ivy Lane, Princeton, NJ 08544, USA}

\author[0000-0002-5259-4774]{Ana Duarte-Cabral}
\affiliation{Cardiff Hub for Astrophysics Research and Technology (CHART), School of Physics \& Astronomy, Cardiff University, The Parade, CF24 3AA Cardiff, UK}

\author[0000-0002-2199-0977]{Helena Faustino Vieira}
\affiliation{Department of Astronomy, The Oskar Klein Centre, Stockholm University, AlbaNova, SE-10691 Stockholm, Sweden}

\author[0000-0002-3247-5321]{Kathryn~Grasha}
\altaffiliation{ARC DECRA Fellow}
\affiliation{Research School of Astronomy and Astrophysics, Australian National University, Canberra, ACT 2611, Australia}   
\affiliation{ARC Centre of Excellence for All Sky Astrophysics in 3 Dimensions (ASTRO 3D), Australia}   

\author[0000-0001-9162-2371]{L.~K. Hunt}
\affiliation{INAF -- Osservatorio Astrofisico di Arcetri, Largo E. Fermi 5, 50125 Firenze, Italy}

\author[0000-0001-8348-2671]{Kelsey E. Johnson}
\affiliation{Department of Astronomy, University of Virginia, Charlottesville, VA, USA}

\author[0000-0002-0560-3172]{Ralf S.\ Klessen}
\affiliation{Universit\"{a}t Heidelberg, Zentrum f\"{u}r Astronomie, Institut f\"{u}r Theoretische Astrophysik, Albert-Ueberle-Str.\ 2, 69120 Heidelberg, Germany}
\affiliation{Universit\"{a}t Heidelberg, Interdisziplin\"{a}res Zentrum f\"{u}r Wissenschaftliches Rechnen, Im Neuenheimer Feld 225, 69120 Heidelberg, Germany}
\affiliation{Harvard-Smithsonian Center for Astrophysics, 60 Garden Street, Cambridge, MA 02138, USA}
\affiliation{Elizabeth S. and Richard M. Cashin Fellow at the Radcliffe Institute for Advanced Studies at Harvard University, 10 Garden Street, Cambridge, MA 02138, USA}

\author[0000-0003-3893-854X]{Mark R. Krumholz}
\affiliation{Research School of Astronomy and Astrophysics, Australian National University, 233 Mount Stromlo Road, Stromlo ACT 2611, Australia}

\author[0000-0001-8490-6632]{Thomas S.-Y. Lai}
\affiliation{IPAC, California Institute of Technology, 1200 East California Boulevard, Pasadena, CA 91125, USA}

\author[0009-0009-5509-4706]{Drew Lapeer}
\affiliation{Department of Astronomy, University of Massachusetts Amherst, 710 North Pleasant Street, Amherst, MA 01003, USA}

\author[0000-0002-1000-6081]{Sean T. Linden}
\affiliation{Department of Astronomy and Steward Observatory, University of Arizona, Tucson, AZ 85721, USA}

\author[0000-0003-1427-2456]{Matteo Messa}
\affiliation{INAF - Osservatorio di Astrofisica e Scienza dello Spazio di Bologna, Via Gobetti 93/3, I-40129 Bologna, Italy }

\author[0000-0002-3005-1349]{G\"{o}ran \"{Os}tlin}
\affiliation{Department of Astronomy, The Oskar Klein Centre, Stockholm University, AlbaNova, SE-10691 Stockholm, Sweden}

\author[0000-0002-8222-8986]{Alex Pedrini}
\affiliation{Department of Astronomy, The Oskar Klein Centre, Stockholm University, AlbaNova, SE-10691 Stockholm, Sweden}

\author[0000-0003-1682-1148]{M\`onica Rela\~no}
\affiliation{Dept.F\`{i}sica Te\`{o}rica y del Cosmos, Universidad de Granada, Campus de Fuentenueva, E-18071 Granada, Spain}
\affiliation{Instituto Universitario Carlos I de F\`{i}sica Te\'{o}rica y Computacional, Universidad de Granada, 18071, Granada, Spain}

\author[0000-0003-2954-7643]{Elena Sabbi}
\affiliation{Gemini Observatory, NOIRLab, 950 N. Cherry Ave., Tucson, AZ 85719, USA }

\author[0000-0002-3933-7677]{Eva Schinnerer}
\affiliation{Max Planck Institut f\"ur Astronomie, K\"onigstuhl 17, D-69117 Heidelberg, Germany}

\author[0000-0003-0605-8732]{Evan Skillman}
\affiliation{School of Physics and Astronomy, University of Minnesota, 116 Church Street, Minneapolis, MN 55455 }

\author[0000-0002-0806-168X]{Linda J. Smith}
\affiliation{Space Telescope  Science Institute, 3700 San Martin Drive, Baltimore, MD 21218, USA}

\author[0000-0002-0986-4759]{Monica Tosi}
\affiliation{INAF - Osservatorio di Astrofisica e Scienza dello Spazio di Bologna, Via Gobetti 93/3, I-40129 Bologna, Italy }

\author[0000-0003-4793-7880]{Fabian Walter}
\affiliation{Max Planck Institut f\"ur Astronomie, K\"onigstuhl 17, D-69117 Heidelberg, Germany}

\author[0009-0005-8923-558X]{Tony D. Weinbeck}
\affiliation{Department of Physics and Astronomy, University of Wyoming, Laramie, WY 82071, USA}

\begin{abstract}
We combine James Webb Space Telescope images of the nearby galaxy NGC\,5194  in the hydrogen recombination line Pa$\alpha$ ($\lambda$1.8756~$\mu$m) from the Cycle 1 program JWST--FEAST with 21~$\mu$m dust continuum images from the Cycle 2 Treasury program JWGT to quantify the difference in the calibration of mid--infrared star formation rates (SFR) between HII regions and galaxies. We use the archival HST H$\alpha$ image to correct the Pa$\alpha$ emission for the effects of dust attenuation. Our data confirm previous results that the dust--corrected Pa$\alpha$ flux is tightly correlated with the 21~$\mu$m emission at the scales of HII regions. When combined with published JWST data for the HII regions of the galaxy NGC\,628 and Spitzer 24~$\mu$m data for whole galaxies and for kpc--size galaxy regions, we show that the L(24)--L(Pa$\alpha$) correlation has exponent $>$1 across six decades in luminosity. In addition, the hybrid 24~$\mu$m+H$\alpha$ SFR indicator has a scaling constant about 4.4 times higher for HII regions than for whole galaxies, also in agreement with previous results. Models of stellar populations with a range of star formation histories reveal that the observed trends can be entirely ascribed to and quantified with the contribution to the IR emission by stellar populations older than $\sim$5--6~Myr. Based on the models' results, we provide: (1) a calibration for the infrared SFR across six orders of magnitude in L(24), from HII regions to luminous galaxies, and (2) a prescription for the scaling constant of the hybrid infrared SFR indicators as a function of the star formation timescale.

 \end{abstract}


\section{Introduction} \label{sec:intro}
Much of the light from star formation is absorbed by dust and re--emitted in the infrared (IR, $\gtrsim$3~$\mu$m), affecting galaxy populations at all redshifts and 
especially in the range z=1--4, where  dust--obscured star formation dominates by factors 2--4 over the unobscured component, with non--negligible effects in galaxies out to z$\sim$8 \citep[e.g.,][]{MadauDickinson2014,Casey+2018,Bouwens+2020,Zavala+2021,Dayal+2022, Liu+2025}. Locally, galaxies show a wide range of IR properties, as they include different mixes of highly obscured, IR--bright regions and almost--transparent, IR--faint ones. 

The efforts to calibrate IR--based indicators of the dust--obscured star formation span several decades in time, starting with the galaxy surveys from the IRAS satellite,
and continuing through recent years, as sensitivity and resolution increased. For as long, it has been clear that the IR emission is at best an imperfect tracer of 
the dust--obscured star formation. While stars form in dusty environments, at least in the local Universe, and young stars can therefore be considered prime sources of dust heating, the  IR emission we observe in galaxies and in galaxy regions is from dust heated by stellar populations of all ages. 

Because of the above, models predict that the calibration of the IR emission into a SFR is age--dependent, with lower calibration constants for population mixes that span longer star formation timescales  \citep{Calzetti2013}. The duration $\tau$ of star formation, and the star formation history in general, affects the IR SFR calibrations by changing the bolometric luminosity, the shape of the UV/optical spectral energy distribution (SED) and the dust optical depth of galaxies and regions over time; for monochromatic IR SFR indicators, calibrations are also affected by the dependence of the effective dust temperature on mean stellar population age. Observationally, this has been handled in indirect ways, through either spectral decomposition of the IR SED \citep[][]{Helou1986, LonsdaleHelou1987, Buat+1988, RowanRobinson+1989, Hunter+1989,  Sauvage+1992, Walterbos+1996, Buat+1996, Boselli+2004, Dale+2012} or spatial decomposition, when spatial information is available \citep{Calzetti+2005, Calzetti+2007, Calzetti+2010, Bendo+2012, SmithDunne+2012, Li+2013, Magnelli+2014, Boquien+2016, Tomicic+2019, Gregg+2022, Leroy+2023}. The results yield a large range, $\approx$20\%--to--70\%, for the contribution of young star--forming regions to the IR emission in local galaxies, possibly linked to differences in the constituents of individual samples \citep{Liu+2011, Li+2013, Boquien+2016, Leroy+2023, Calzetti+2024}.

While the studies listed above have recognized the age--dependence of IR SFR indicators and proposed ways to take them into account, explicit calibrations of the IR SFR as a function of star formation timescales have 
been scant \citep{Li+2013, Calzetti+2024}. 
This has  mainly been driven by limitations in the observations, especially in spatial resolution, since star formation is hierarchical and younger ages are associated with smaller scales in star forming galaxies \citep[e.g.,][]{Efremov+1998,Delafuente+2009,Elmegreen+2011,Grasha+2015,Gouliermis+2015,Grasha+2017a,Grasha+2017,Gouliermis+2018,Elmegreen+2018, Shashank+2025}. Direct evidence for the existence of an age--dependence in the IR SFR calibration exists in the Milky Way. Several authors have demonstrated that different SFR indicators yield a consistent value $\sim$2~M$_{\odot}$~yr$^{-1}$ for the Milky Way \citep{Chomiuk+2011, Mendigutia+2018, Elia+2022, Soler+2023}; however, it is also clear that each indicator can be used exclusively in the regime it applies. \citet{Elia+2025} demonstrate that applying an IR SFR indicator calibrated for $\sim$kpc--sized galaxy regions to the IR luminosity of Milky Way clumps yields a more-than-one-order-of-magnitude underestimate of the Galaxy's SFR. This is understood because the galaxy--wide calibrations include (and correct for) the diffuse IR emission from the entire stellar population and not from just the $<$1~Myr old clumps. Now that the synergy between JWST and ALMA is providing samples of distant galaxies with both UV--optical and infrared information, exploring the explicit age dependency of IR SFR indicators is becoming more necessary and urgent.

In this paper, we expand on the investigation presented in \citet{Calzetti+2024} to quantify the relation between the 21~$\mu$m dust emission and the SFR at the scale of HII regions in two nearby galaxies: NGC\,5194 and NGC\,628. We also expand our analysis to include results, from Spitzer MIPS/24~$\mu$m \citep{Rieke+2004} and other data, on larger, kpc--sized, regions and whole galaxies;  we finally use models to reconcile the different IR SFR calibrations into a single, coherent picture. 

Our first step is to focus on HII regions because the ionized gas traces young, massive stars and are usually easily identifiable units of recent star formation: hydrogen recombination lines probe ages $<$6--7~Myr \citep{Leitherer+1999}. The ionizing photon rate mainly probes the masses of these star formation units, but when averaged over a large number of HII regions of different masses, it becomes an accurate SFR
indicator over that same timescale of $\sim$6 Myr. This sets the shortest timescale we can probe with our investigation, which it is sufficient to detect differences in the IR emission when compared with kpc--sized galaxy regions or whole galaxies, since the latter probe crossing times of $\approx$100~Myr to Gyrs.  

\citet{Calzetti+2024} already analyzed the HII regions in NGC\,628, and we will be adding here those of NGC\,5194. The main properties of NGC\,5194 used in this paper are listed in Table~\ref{tab:properties}, together with those of NGC\,628. 
The two galaxies present both similarities and differences. NGC\,5194 is slightly closer than NGC\,628, and has twice higher overall SFR, implying that it has a higher density of HII regions\footnote{The two galaxies have comparable areas as measured from R$_{25}$, see Table~\ref{tab:properties}}, and is twice as massive. 
The two galaxies are both located a factor of a few above the main sequence of star formation \citep{Cook+2014, Renzini+2015}, but their dust--to--stellar mass ratios fall along the local galaxies sequence \citep{Dale+2023}.
Both galaxies have a little over solar oxygen abundance in their center, and modest radial gradients \citep{Berg+2020}. The region in common among the JWST mosaics in different bands of NGC\,5194 extends to about 187$^{\prime\prime}$ from the center (0.556~R$_{25}$ or 6.8~kpc), implying that the metallicity has decreased by only 0.15~dex at that point. We will, therefore, model the HII regions in NGC\,5194 as having a single, $\sim$solar, metallicity value.

We concentrate on 21~$\mu$m (24~$\mu$m) as opposed to the shorter--$\lambda$ dust emission features, as the latter are also sensitive to metallicity and ionization variations and dust carrier destruction \citep{Egorov+2023, Pedrini+2024}. The hydrogen recombination line Pa$\alpha$ ($\lambda$1.8756~$\mu$m) is used as an unbiased SFR indicator, after it is corrected for dust attenuation using the archival HST H$\alpha$ ($\lambda$0.6563~$\mu$m) imaging. As in \citet{Calzetti+2024}, we select HII~regions that are bright enough, and thus massive enough, to minimize the effects of stochastic (random) sampling of the stellar initial mass function (IMF). This approach is different from the one utilized by \citet{Belfiore+2023}, which analyzed $\sim$20,000 HII regions in 19 nearby galaxies, but included faint, stochastic HII regions and did not account for diffuse IR emission in their analysis. 

This paper is organized as follows: Section~\ref{sec:data} presents the data used in this analysis, Section~\ref{sec:selection} describes the source identification and measured  properties, Section~\ref{sec:results} presents the main results for NGC\,5194 first and then combined with the HII regions of NGC\,628. Section~\ref{sec:discussion} discusses the results, also in comparison with published samples of galaxies and of $\sim$kpc--sized galaxy regions, providing HII regions--to--galaxies calibrations for IR SFRs. Conclusions and  recommendations on the use of these IR SFR calibrations are given in Section~\ref{sec:conclusions}.

\begin{deluxetable*}{llrcr}
\tablecaption{Properties of the Galaxies\label{tab:properties}}
\tablewidth{0pt}
\tablehead{
\colhead{Parameter} & \colhead{Units} &\colhead{NGC\,5194} &\colhead{References$^1$}& \colhead{NGC\,628$^2$} \\
}
\decimalcolnumbers
\startdata
Distance & Mpc & 7.55 & (a) & 9.3\\
Inclination & degrees & 22 & (b) & 9\\
R$_{25}$ & arcsec (kpc) & 336.6 (12.32) & (c) & 314.2 (14.15)\\
E(B--V)$_{MW}^3$ & mag & 0.031 & (d) & 0.06 \\
M$_{star}$ & M$_{\odot}$ & 2.3$\times$10$^{10}$ & (e) & 9.9$\times$10$^9$\\
SFR & M$_{\odot}$~yr$^{-1}$  & 6.7 & (e) & 3.2\\
12+Log(O/H)$^4$ & & 8.75 & (f) & 8.71\\
Gradient$^5$ & R$_{25}$ & $-$0.27 & (f) & $-$0.40\\
M$_{dust}$/M$_{star}$ &    & 4.5$\times$10$^{-3}$ & (g) & 4.2$\times$10$^{-3}$ \\
\enddata
$^1$  References for the NGC\, 5194 parameters:  (a) \citet{Sabbi+2018}, using TRGBs  -- similar results are obtained by \citet{Csornyei+2023} and a slightly larger distance, 8.58~Mpc, by \citet{McQuinn+2016} --;  (b) \citet{Colombo+2014}, using CO emission; (c) \citet{deVaucouleurs+1991}; (d) \citet{Schlafly+2011}; (e) \citet{Calzetti+2015a}; (f) \citet{Berg+2020}; (g) \citet{Dale+2023}.\\
$^2$ The parameters for NGC\,628 are the same reported in \citet{Calzetti+2024}. We refer the reader to that work for the relevant references. The M$_{dust}$/M$_{star}$ value for this galaxy is from \citet{Dale+2023}.\\
$^3$  Foreground Milky Way extinction.\\
$^4$ Central oxygen abundance. We adopt a solar oxygen abundance of 12+Log(O/H)=8.69, \citet{Asplund+2009}.\\
$^5$ Metallicity gradient as a function of galactocentric radius in units of R$_{25}$.\\
\end{deluxetable*}

\section{NGC\,5194 Imaging Data and Processing} \label{sec:data}

The galaxy NGC\,5194 is one of six targets of the Cycle 1 JWST program \# 1783 (Feedback in Emerging extrAgalactic Star clusTers, JWST--FEAST, P.I.: A. Adamo) and the target of the Cycle 2 JWST Treasury program \# 3435 (The JWST Whirpool Galaxy Treasury, JWGT, P.Is.: K. Sandstrom \& D. Dale). Both programs have observed the galaxy with both NIRCam \citep{Rieke+2005, Rieke+2023} and MIRI  \citep{Rieke+2015}, obtaining mosaics in complementary filters across the two instruments, and covering the wavelength range 1.1--21~$\mu$m. Mosaics in NIRCam were processed through the JWST pipeline version 1.12.5 (Dec 2023 release) using the Calibration Reference Data System (CRDS) context ``jwst\_1174.pmap'', while the MIRI mosaics were processed through the JWST pipeline version 1.13.4 (Feb 2024 release) using the CRDS context ``jwst\_1241.pmap'' \footnote{\href{https://jwst-pipeline.readthedocs.io/en/latest/jwst/user\_documentation/reference\_files\_crds.html}{https://jwst-pipeline.readthedocs.io/en/latest/jwst/user\_documentation/reference\_files\_crds.html}}. The mosaics utilized in this work are listed in Table~\ref{tab:images} together with the program number they originated from: we use the NIRCam mosaics from JWST--FEAST and one MIRI (21~$\mu$m) mosaic from JWGT. A small rotation between the FoVs of the NIRCam mosaics  and the MIRI mosaic, both centered on the galaxy's center, reduces the effective overlap area to $\sim$2$^{\prime}\times$5.7$^{\prime}$, or 4.4$\times$12.6~kpc$^2$. Furthermore, the NIRCam mosaics have common pixel scale of 0.04$^{\prime\prime}$/pix and are in units of Jy/pix, while the MIRI mosaic has pixel scale of 0.11$^{\prime\prime}$/pix and is in units of MJy/sr. 

HST Wide Field of the Advanced Camera for Surveys (ACS/WFC) imaging of NGC\,5194 was retrieved from the MAST Archive\footnote{MAST: Mikulski Archive for Space Telescopes at the Space Telescope Science Institute; \href{https://archive.stsci.edu/}{https://archive.stsci.edu/}.}. The HST imaging is a 2$\times$3 pointings mosaic covering the entire bright area of NGC\,5194 and its companion NGC\,5195 obtained by HST--GO--10452 (P.I. S. Beckwith), as part of the HST image releases program by the Hubble Heritage Team. The HST mosaic coverage is sufficiently extended to completely include the JWST mosaics. The JWST--FEAST team reprocessed the archival images, resampling them to a pixel scale of 0.04$^{\prime\prime}$/pix. Flux  calibration is in units of counts/s, which we convert to physical units using the PHOTFLAM image header keywords. Of the four ACS/WFC filters available for this galaxy from the Heritage program, we 
only utilize three: F555W, F658N and F814W, as will be discussed below. See Table~\ref{tab:images} for details on the telescope/instrument/filters for the HST mosaics.

The F2100W mosaic has the lowest resolution among all images used in this work: the MIRI/F2100W Point Spread Function (PSF) has Full Width at Half Maximum (FWHM)=0$^{\prime\prime}$.674, which subtends about 25~pc at the distance of NGC\,5194. This is comparable to or larger than the size of a single HII region (see next Section); thus, we will be relying on the shorter wavelength, higher angular resolution mosaics to identify HII regions that are ionized by individual star clusters. The Short Wavelength NIRCam mosaic in the F150W filter has a PSF FWHM=0.05$^{\prime\prime}$; the PSF of F200W has similar FWHM=0.066$^{\prime\prime}$ and the F187N has FWHM=0.061$^{\prime\prime}$. At the distance of NGC\,5194, the Short Wavelength NIRCam PSF corresponds to 1.8--2.4~pc, implying that compact star clusters, which have characteristic radius $\sim$3~pc \citep{Ryon+2015,Ryon+2017,Brown+2021}, are marginally resolved. 

Emission line maps are derived from the narrow--band mosaics: the NIRCam F187N centered on the Pa$\alpha$  line emission ($\lambda$=1.8789~$\mu$m at the fiducial redshift z=0.001745\footnote{From NED, the NASA Extragalactic Database.}) and the ACS/WFC F658N centered on the H$\alpha$+[NII] doublet  line emission ($\lambda$=0.6559,0.6574,0.6595~$\mu$m at z=0.001745). The interpolation between F150W and F200W is used to produce a stellar continuum image to subtract from the F187N. 
Since the F200W contains the Pa$\alpha$ emission, we iteratively subtract the line from this filter, using the procedure described in \citet{Messa+2021} and \citet{Calzetti+2024}, until differences between two subsequent iterations are $\lesssim$0.1\% in flux. The interpolation between F555W and F814W is used to produce the stellar continuum image to subtract from the F658N. The F555W filter includes the [OIII]($\lambda$0.5007~$\mu$m) line emission, but \citet{Calzetti+2024} showed that in metal--rich galaxies this contribution is small, affecting the interpolated stellar continuum by $\lesssim$1.5\%. The F658N line emission is then corrected for the [NII] contribution, using [NII]/H$\alpha$=0.6 for the sum of the two [NII] components \citep{Moustakas+2006, Kennicutt+2008}. As discussed in section~\ref{sec:intro} (see, also, Table~\ref{tab:properties}), the metallicity gradient is  $\le$0.15~dex within the JWST footprint; we thus, neglect [NII]/H$\alpha$  variations and only include the 6\% uncertainty on the [NII]/H$\alpha$ measurement in our error propagation \citep{Moustakas+2006}. The final line flux maps in H$\alpha$ and Pa$\alpha$ are derived by multiplying the continuum--subtracted mosaics by the respective bandwidths (0.00875~$\mu$m for F658N\footnote{\href{https://etc.stsci.edu/etcstatic/users\_guide/appendix\_b\_acs.html}{https://etc.stsci.edu/etcstatic/users\_guide/appendix\_b\_acs.html}} and 0.024~$\mu$m for F187N\footnote{\href{https://jwst-docs.stsci.edu/jwst-near-infrared-camera/nircam-instrumentation/nircam-filters}{https://jwst-docs.stsci.edu/jwst-near-infrared-camera/nircam-instrumentation/nircam-filters}}) and by correcting for the filter transmission curve value at the galaxy's redshift.

\begin{deluxetable*}{llll}
\tablecaption{NGC\,5194 Imaging Data Sources\label{tab:images}}
\tablewidth{0pt}
\tablehead{
\colhead{Telescope$^1$} & \colhead{Instrument$^2$} & \colhead{Filters$^3$} & \colhead{Proposal ID$^4$}\\
}
\decimalcolnumbers
\startdata
JWST & NIRCam S+L   &  F150W, F187N, F200W, F444W  & 1783\\
JWST & MIRI   &  F2100W  & 3435 \\
HST  & ACS/WFC   &  F555W, F658N, F814W  & 10452\\
\enddata
$^1$  JWST=James Webb Space Telescope \citep{Gardner+2023, Rigby+2023}; HST=Hubble Space Telescope\\
$^2$ NIRCam S+L = Near Infrared Camera, Short and Long Wavelength Channels \citep{Rieke+2005, Rieke+2023}; MIRI=Mid--Infrared Instrument \citep{Rieke+2015}; ACS/WFC= Advanced Camera for Surveys Wide Field Channel \citep{Sirianni+2005}.\\
$^3$ Filter names. The F187N and the F658N narrow--band filters are centered on the hydrogen recombination lines Pa$\alpha$($\lambda$1.8756~$\mu$m) and H$\alpha$($\lambda$0.6563~$\mu$m) respectively.\\
$^4$ Identification of the GO program that obtained the images: JWST/GO--1783 (JWST--FEAST), PI: Adamo; JWST/GO--3435 (JWGT), PIs: Sandstrom \& Dale;  HST/GO--10452, PI: Beckwith, Hubble Heritage Team.\\
\end{deluxetable*}

\section{Source Selection, Photometry, and Physical Quantities} \label{sec:selection}

\begin{figure}
\plotone{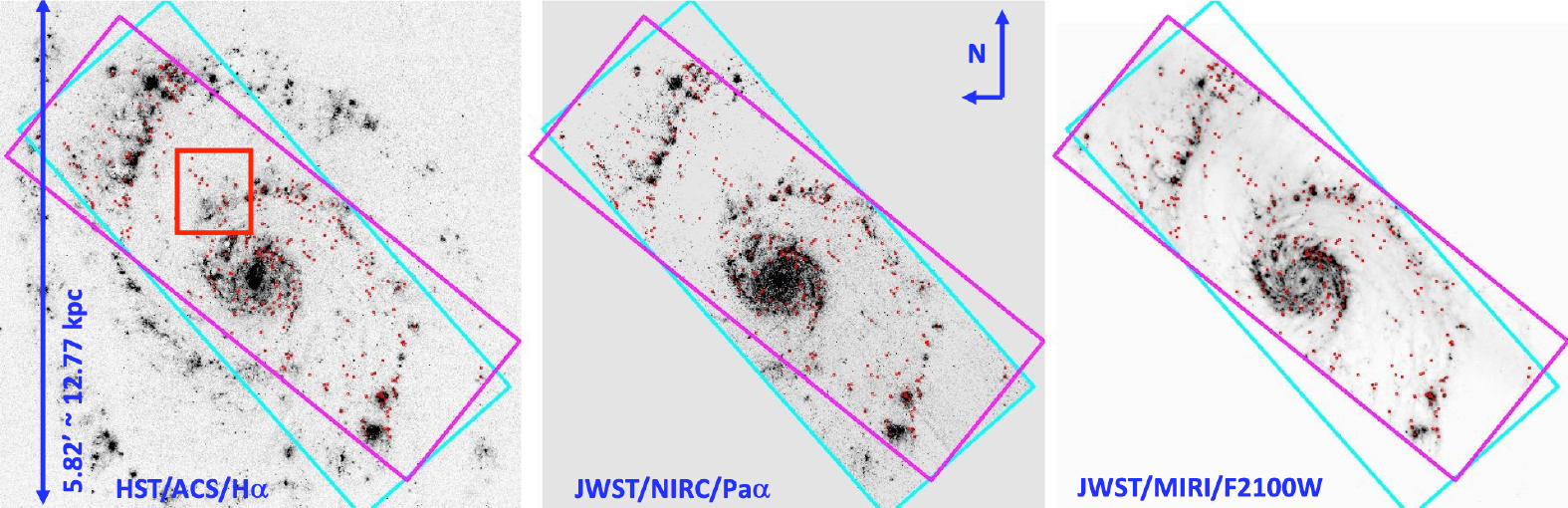}
\plotone{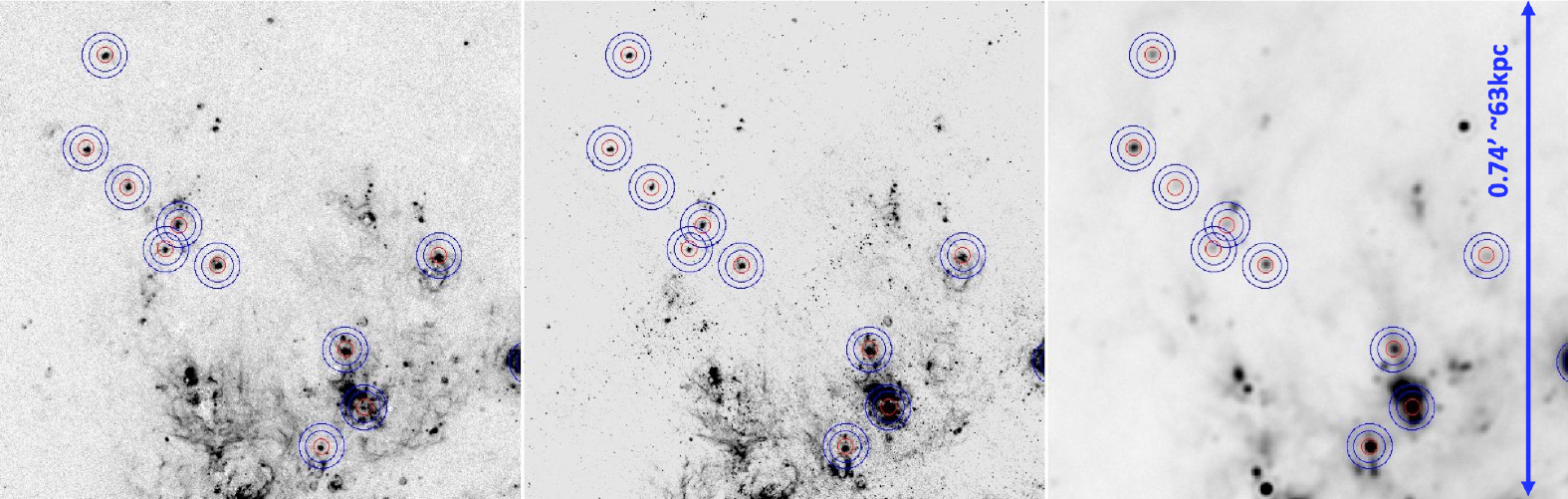}
\caption{{\bf Top Row:}: The 254 sources emitting in H$\alpha$, Pa$\alpha$ and 21~$\mu$m are identified with red circles on the JWST and HST mosaics of NGC\,5194 in the stellar--continuum subtracted HST/ACS/H$\alpha$ (left), JWST/NIRCam/Pa$\alpha$ (center), and JWST/MIRI/F2100W (right), see Section~\ref{sec:data}. The NIRCam footprint and the MIRI footprint are shown as cyan and magenta rectangles, respectively, on all three panels. The radius of the circles matches the photometric aperture used in this study, 0$^{\prime\prime}.7$, or $\sim$26~pc.  North  is up, East is left. {\bf (Bottom Row:)} A detail of the images above, in the same order (H$\alpha$, Pa$\alpha$, 21~$\mu$m).  The double blue circle around each region shows the size of the annulus used for the local background subtraction (1.4$^{\prime\prime}$ inner radius with 0.6$^{\prime\prime}$ width). The location of this region is drawn on the top--row H$\alpha$ image with a red rectangle. }
 \label{fig:sources}
\end{figure}

For the source selection, we adopt the same approach and criteria as \citet{Calzetti+2024}. The Pa$\alpha$ and 21~$\mu$m images are inspected visually to isolate sources that are compact both in Pa$\alpha$ and in stellar continuum (from the F814W mosaic), are detected with  S/N$\gtrsim$5 at 21~$\mu$m and S/N$\gtrsim$3 in both Pa$\alpha$ and H$\alpha$, and are spatially coincident in the two bands to better than the 21~$\mu$m PSF FWHM. In this context, a `compact source'  is a source that displays a single peak in the line emission image within the photometric aperture (see below) and  has a colocated  stellar continuum source that shows either a single peak or clustered emission; the latter are sources showing multiple peaks immersed in a common, fainter emission region (likely low--mass, unresolved stars).  At the resolution of our images (section~\ref{sec:data}) these sources are likely to be individual star clusters or compact associations  \citep{Adamo+2017}. All sources that appear extended or shell--like in the Pa$\alpha$ image are excluded from further consideration, as these are 
expected to correspond to evolved (older than $\sim$6~Myr) HII regions \citep{Whitmore+2011}. We further require that the sources are sufficiently separated from each other that adjacent photometric apertures do not overlap more than 10\% to ensure independent photometric measurements. As will be discussed later, requiring {\em both} Pa$\alpha$ and H$\alpha$ to be detected with S/N$\gtrsim$3 imposes a luminosity--dependent upper limit on the highest value of dust extinction a region can have; however, this upper limit is sufficiently above the bulk of the sources' measurements to not represent a limitation to this analysis. Our selection is incomplete for  sources that are faint in all three Pa$\alpha$, H$\alpha$ and 21~$\mu$m images; this will not affect our analysis, because we implement a luminosity cut at the faint end to remove sources that are affected by stochastic sampling of the stellar IMF (see Section~\ref{sec:results}). A total of 254 sources meeting the above criteria are found across the JWST footprint of NGC 5194 (Figure~\ref{fig:sources}).

For the photometry, we deviate from \citet{Calzetti+2024} in that we select a smaller aperture radius, 0$^{\prime\prime}$.7 which corresponds to about 26~pc at the distance of NGC\,5194, instead of the radius 1.4$^{\prime\prime}$ that we used for NGC\,628 (corresponding to a physical scale of 63~pc in this galaxy). The reason for the different choice is that HII regions are more densely packed in NGC\,5194 than in NGC\,628, requiring a smaller aperture size to separate adjacent regions. Although the de--projected aperture corresponds to an ellipse with semi--major and semi--minor axes of  28pc$\times$26pc, we will continue to treat the apertures as round, under the assumption that HII regions can be basically considered spherically symmetric. A concern is whether the small aperture may be missing significant portions of the ionized gas emission surrounding each star cluster. For our sources, the largest Str\"omgren radius is about 26--27~pc \citep{Osterbrock+2006} for uniform density, when using the electron densities from  \citet{Croxall+2015}, n$_e\sim$70--200~cm$^{-3}$ along the JWST mosaic's strip. Thus, our apertures are well matched to the largest Str\"omgren sphere we sample, alleviating any concern of missed ionizing flux. 

The photometric apertures are centered on the centroid of the 21~$\mu$m sources. In a few cases, the wings of the nebular emission are sufficiently bright that it is desirable to center the aperture on the Pa$\alpha$ peak, so to capture as much as possible the source's ionized gas emission. In an additional two dozen cases, the regions are sufficiently crowded that the apertures need to be moved slightly to avoid $>$10\% overlap between adjacent apertures. These can result in apertures' centers that are  slightly offset relative to the 21~$\mu$m peak. For all off--center cases, we measure the impact of the de--centering on the 21~$\mu$m photometry and find that, except for six cases, the impact is smaller than the measurement uncertainty. The six impacted sources (all to $<$20\% level) are located in crowded regions and their photometry may be affected by the wings of neighboring sources; for these cases, we carry additional uncertainty in the 21~$\mu$m photometry. Photometry is obtained with local background subtraction, with the background measured in an annulus with inner radius=1.4$^{\prime\prime}$ and outer radius=2.0$^{\prime\prime}$ (blue circles in Figure~\ref{fig:sources}, bottom panels). The size of the background annulus is twice as large as the radius of the photometric aperture, to exclude the majority of the emission in the wings of the 21~$\mu$m PSF: 1.4$^{\prime\prime}$ corresponds to 5$\sigma$ for this PSF, and only 17\% of the 21~$\mu$m flux is contained outside of it. The value of the local background around each source is the mode of the pixel value distribution after iterative $\sigma$ clipping. 

Aperture corrections are calculated for Pa$\alpha$, H$\alpha$, 21~$\mu$m and F444W, the latter used to asses the stellar continuum contribution to the 21~$\mu$m band. Simulated flight PSFs are retrieved from the STScI Box repository\footnote{\href{https://stsci.app.box.com/v/jwst-simulated-psf-library}{https://stsci.app.box.com/v/jwst-simulated-psf-library}}, and growth curves are generated to emulate our photometric approach. Aperture corrections to an infinite aperture are: 7\% for F187N (we adopt the same for F658N, as the PSFs are similar), 9\% for F444W, and 41\% for 21~$\mu$m. The 21~$\mu$m correction, which is the largest applied to our photometry, agrees with that derived from the in--flight PSF to within 1\% \citep{Libralato+2024, Dicken+2024}. The aperture corrections are applied to the narrow--band filters photometry after removal of the underlying stellar continuum. 

Stellar continuum is removed from the 21~$\mu$m measurements by rescaling the F444W photometry to the central wavelength of the MIRI/F2100W filter and then subtracting it from the 21~$\mu$m flux, using the approach described by \citet{Calzetti+2024}:  f(21)$_{dust}$=f(21)-0.046 f(444), where the flux densities are in units of Jy. This method assumes that the F444W filter is dominated by stellar emission, and both dust attenuation and emission are negligible. The latter is not strictly correct, as the dust emission contribution to the F444W can be substantial, especially in star--forming regions \citep{Meidt+2012}. However, even under the strict assumption that the F444W is entirely due to stellar emission, the stellar contribution to the 21~$\mu$m emission is $<$2\% for all selected sources. Thus, the 21~$\mu$m emission is, for all practical purposes, not affected by stellar contamination. Local background subtraction removes the diffuse dust emission from the general stellar population in the galaxy, thus eliminating contributions that are unrelated to the dust heating by the HII region. The flux density at 21~$\mu$m is then multiplied by the frequency 
to convert it to flux, which is  standard for SFR measurements, and then converted to a luminosity using our adopted distance for NGC\,5194 (Table~\ref{tab:properties}). The 5$\sigma$ limit for Log[L(21)] is 37.62\footnote{Log=log$_{10}$ in this work.}, in units of erg~s$^{-1}$.

Line fluxes for the nebular emission are derived by applying  the procedure described in Section~\ref{sec:data} to the photometric measurements. For the HII regions, the stellar continuum subtraction corrects also for the underlying stellar absorption, which can lead to underestimates of the true line emission if not removed, while the local 
background subtraction removes the diffuse contribution from photon leakage and from scattered light coming from other regions. Both H$\alpha$ and Pa$\alpha$ fluxes are corrected for foreground Milky Way extinction 
(Table~\ref{tab:properties}) and then converted to luminosities. The 3$\sigma$ detections limits are 35.65 and 35.54 in erg~s$^{-1}$ for Log[L(H$\alpha$)] and Log[L(Pa$\alpha$)], respectively.

Additional quantities used in this work are the equivalent width (EW) of Pa$\alpha$ and the color excess E(B$-$V). The EW is the ratio of the line luminosity to the luminosity density of the interpolated stellar continuum, in units of \AA. The color excess is derived from the observed line ratio H$\alpha$/Pa$\alpha$, adopting Case B recombination and an intrinsic ratio of 7.82. This ratio is appropriate for metal rich sources \citep[T$_e$=7,000~K and n$_e$=100~cm$^{-3}$,][]{Osterbrock+2006}. We use the extinction 
curve extended to the JWST filters by \citet{Fahrion+2023} for 30~Doradus: $\kappa($H$\alpha$)=2.53 and $\kappa$(Pa$\alpha$)=0.678, where $\kappa(\lambda)$=A($\lambda$)/E(B--V) is the ratio of the attenuation to the color excess. Although our sources are in a more metal--rich environment than the LMC, they are all HII regions like 30~Doradus, thus we assume the curve by \citet{Fahrion+2023} to be appropriate for our case; however, we adopt R$_V$=3.1 as appropriate for metal--rich galaxies like the Milky Way \citep{Fitzpatrick+2019}. Dust attenuation corrected luminosities in the nebular lines are derived under the assumption of foreground dust, as: $L(\lambda)_{corr} = L(\lambda) 10^{0.4 E(B-V) \kappa(\lambda)}$, where {\it L}($\lambda$) and {\it L}($\lambda$)$_{corr}$ are the observed and attenuation--corrected luminosities in units of erg~s$^{-1}$, respectively, and $\kappa$($\lambda$) is the dust attenuation curve.  Given the small spatial scale sampled by our measurements, we assume the same dust attenuation values for emission lines and stellar continuum, implying that equivalent widths are not affected by attenuation corrections. Table~\ref{tab:sources} lists for each source: ID, location on the sky in RA(2000) and DEC(2000), observed luminosity in H$\alpha$, Pa$\alpha$ and 21~$\mu$m, the equivalent width in Pa$\alpha$, and the color excess E(B$-$V).

\section{Analysis and Results} \label{sec:results}

\subsection{Properties of the Emission Line Regions}\label{subsec:properties}

\begin{figure}
\plottwo{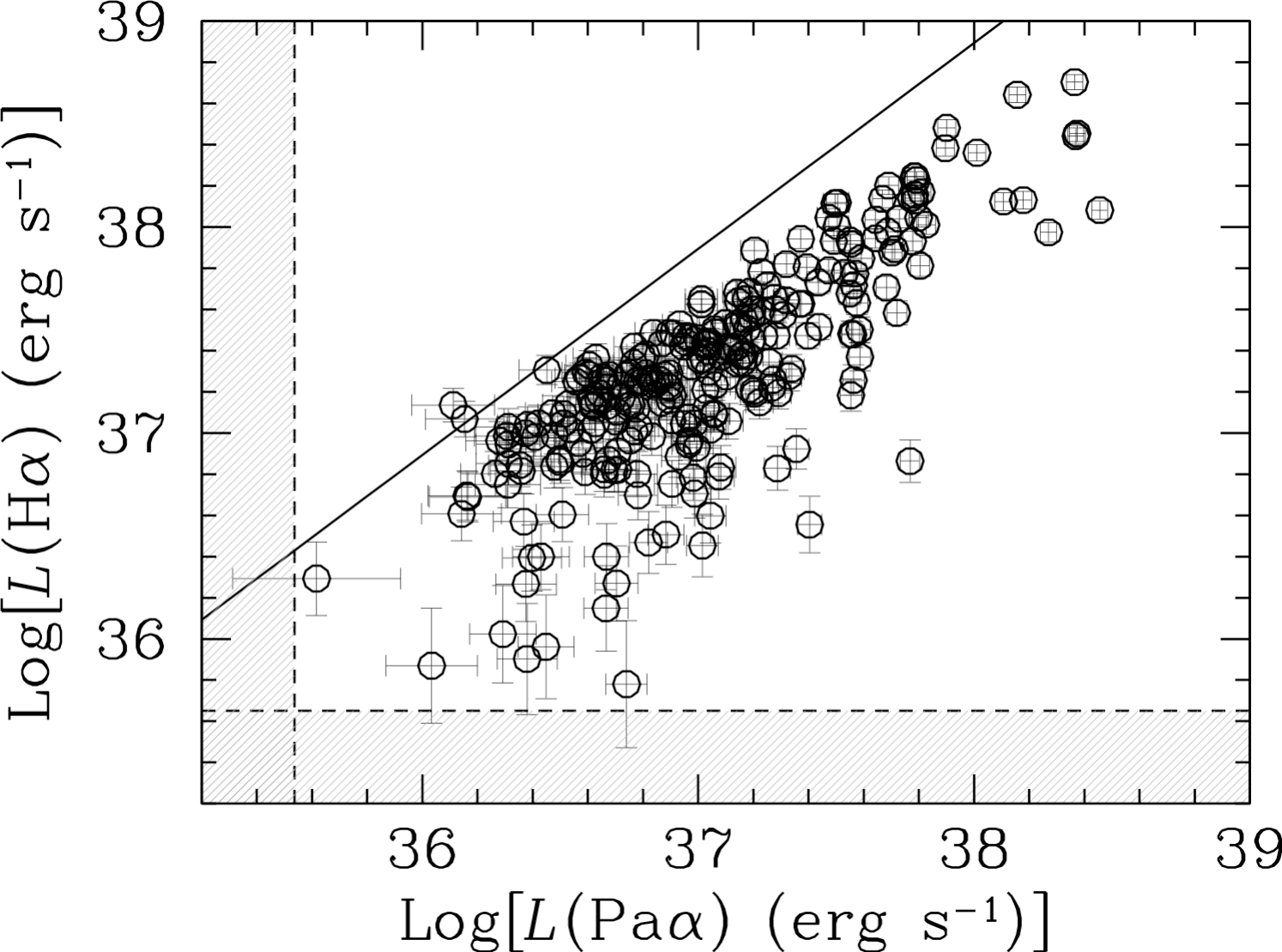}{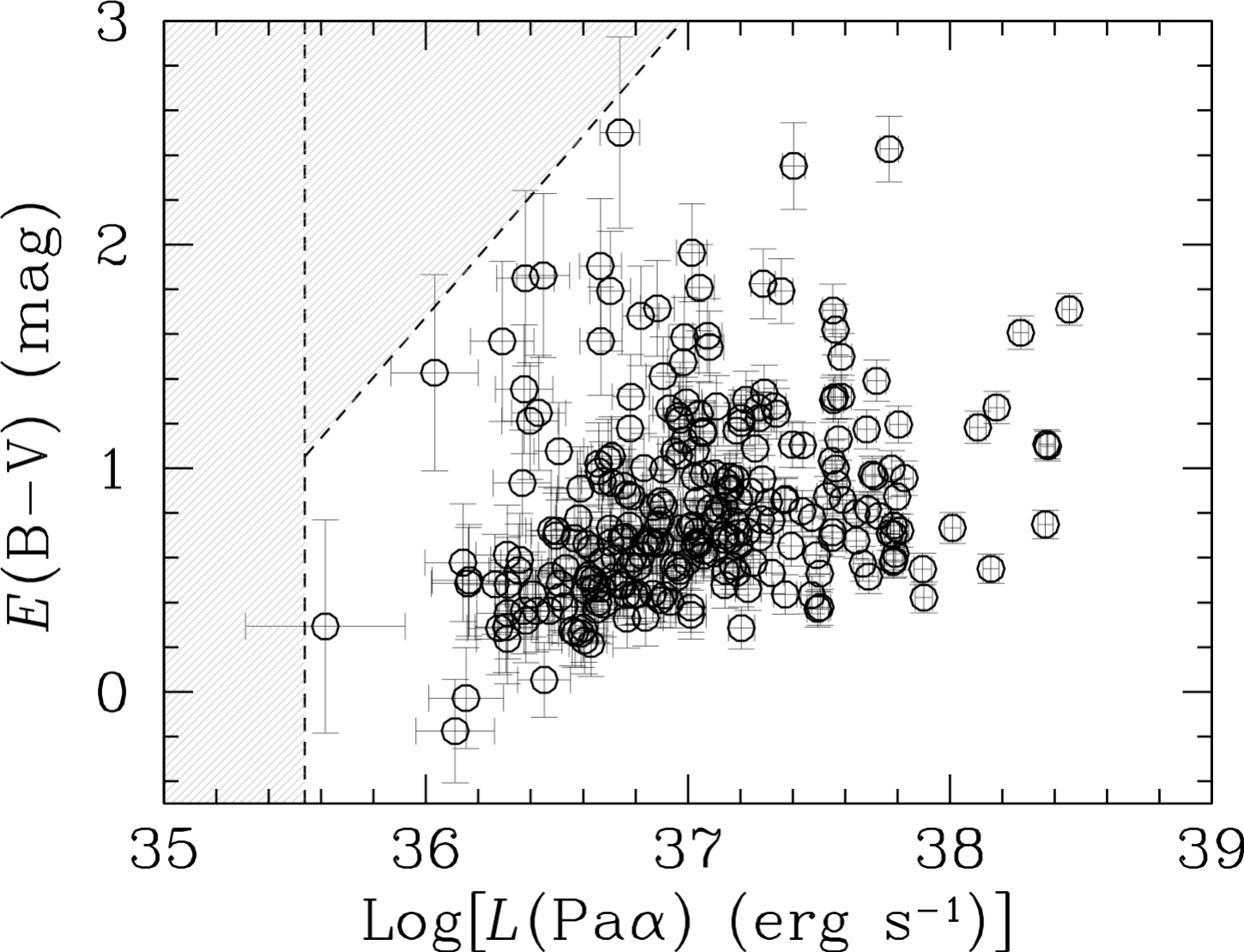}
\caption{The observed luminosity in H$\alpha$ (left panel) and the color excess E(B$-$V) derived from the observed nebular luminosities (right panel) as a function of the observed Pa$\alpha$ luminosity for the 254 regions identified in the JWST+HST mosaics of NGC\,5194. All quantities are corrected for the foreground extinction from the Milky Way (Table~\ref{tab:properties}). The data are shown as black circles with their 1$\sigma$ uncertainties. The continuous black line in the left panel marks the expected location of luminosities of the two emission lines for our adopted line ratio of 7.82 and with zero dust attenuation. The grey--shaded regions mark the areas below 3$\sigma$ detection.} 
\label{fig:obs_lum}
\end{figure}

The HII regions in NGC\,5194 show many characteristics that are similar to the HII regions in NGC\,628 \citep{Calzetti+2024}. The observed H$\alpha$ is underluminous relative to the Pa$\alpha$ when using as reference the intrinsic ratio L(H$\alpha$)/L(Pa$\alpha$)=7.82 discussed in the previous Section, implying that almost all regions have some dust, with the only exception of the faintest ones (Figure~\ref{fig:obs_lum}, left). In addition, the lower envelope to the data for the color excess E(B-V) increases for increasing luminosity (Figure~\ref{fig:obs_lum}, right), i.e., more luminous HII regions tend to be dustier than less luminous ones, with a minimum E(B-V)$_{min}\sim$0.5~mag at L(Pa$\alpha$)$\sim$10$^{38}$~erg~s$^{-1}$. Visual inspection of the H$\alpha$ image indicates that this is not the result of biased selection: no regions are missing from our sample across the area in common between the HST and JWST mosaics that are at the same time bright in H$\alpha$ and low in dust attenuation, while also satisfying both our Pa$\alpha$ and 21~$\mu$m selection criteria. As already mentioned above, our 3$\sigma$ limit in line detection introduces a luminosity--dependent upper limit to E(B-V), shown as a grey slanted area in Figure~\ref{fig:obs_lum} (right panel). The majority of our regions are not affected by this limit, although we cannot exclude the presence of even dustier regions. However, since the distribution of the selected regions cluster around a locus well below the upper limit in E(B-V), we will assume that our sample is not missing a significant number of very dusty regions.

\begin{figure}
\plottwo{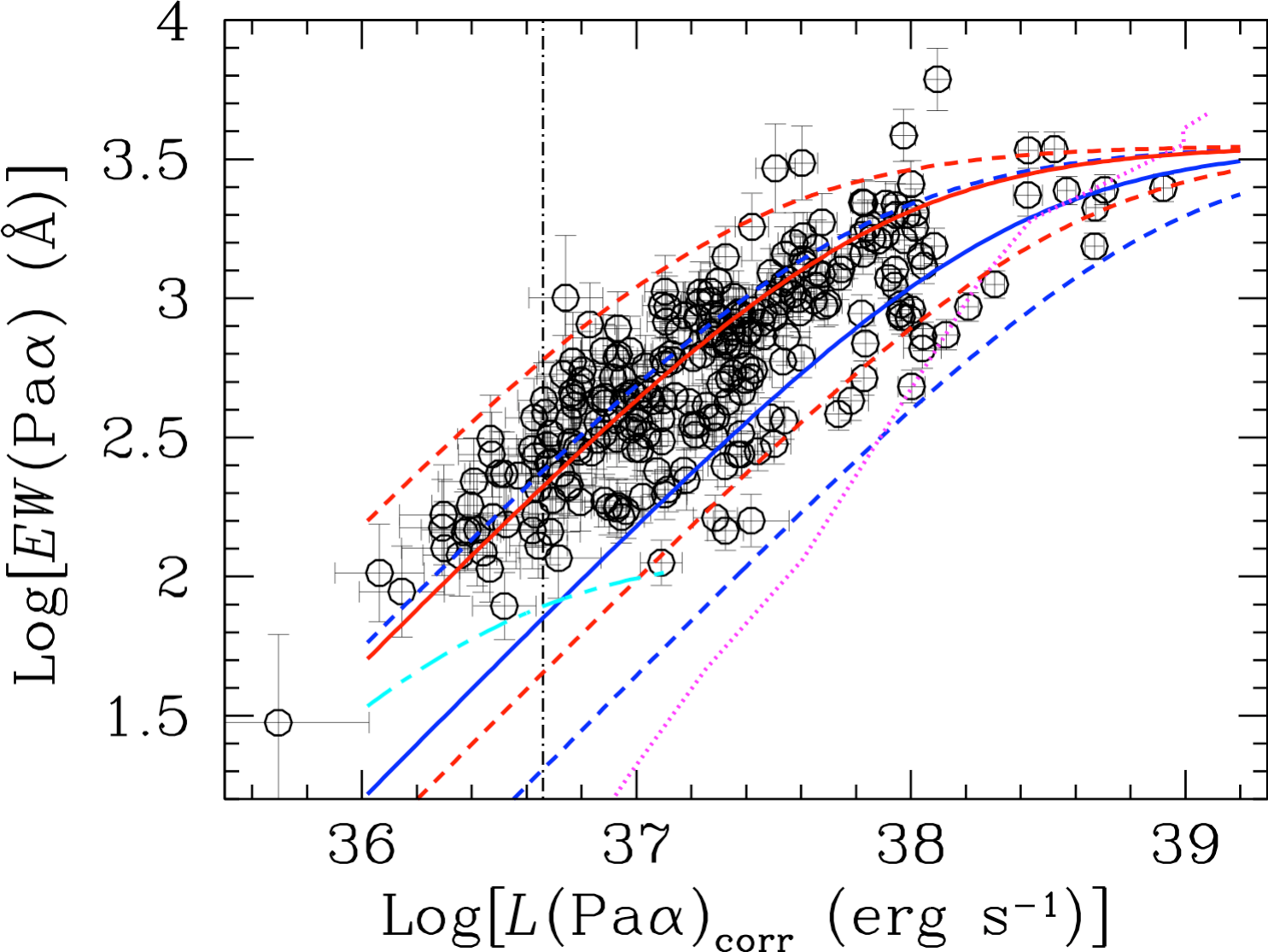}{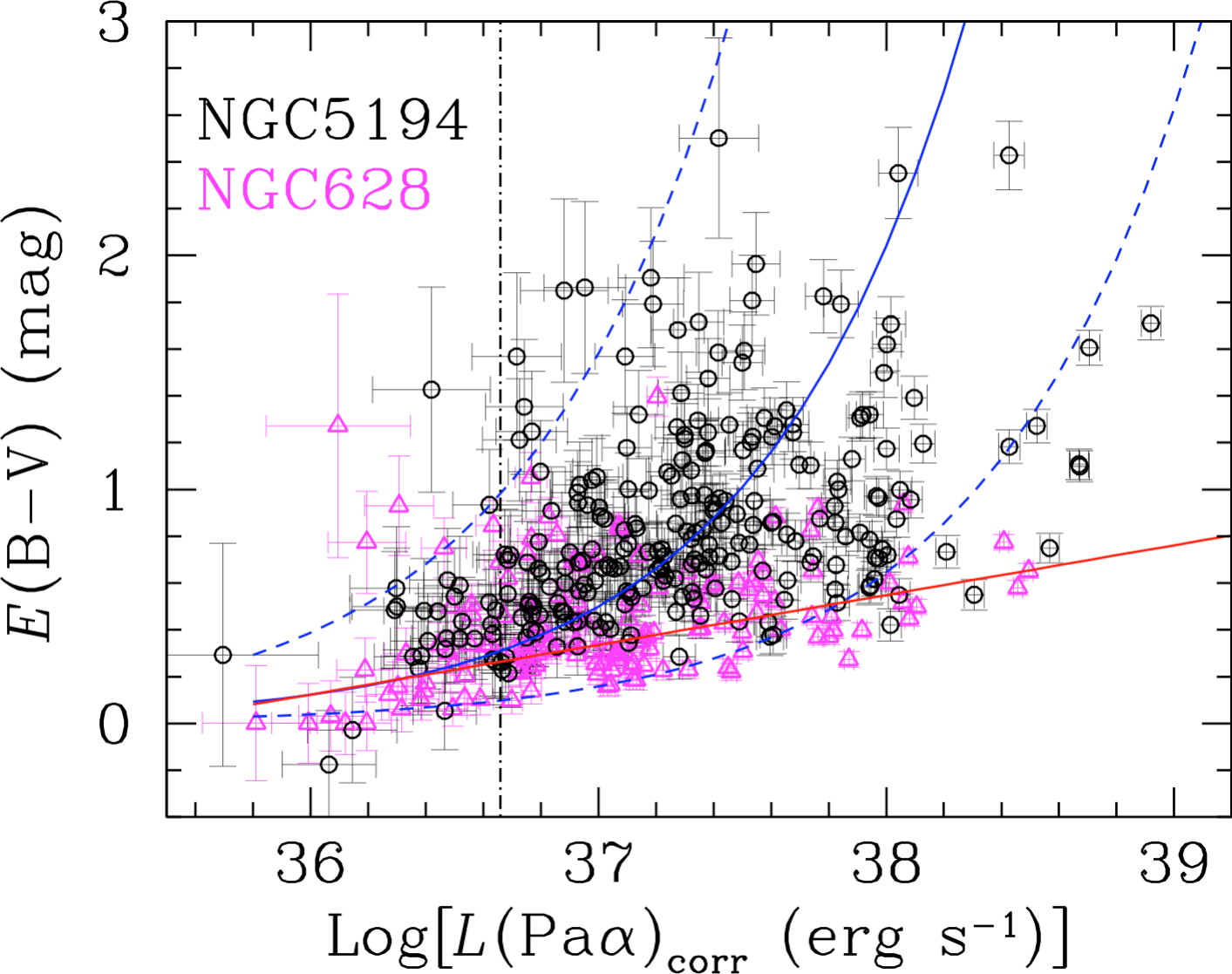}
\caption{The EW(Pa$\alpha$) (left panel) and the color excess E(B$-$V) (right panel) as a function of the attenuation--corrected luminosity in Pa$\alpha$ for the HII regions in NGC\,5194. The data are shown as black circles, with their 1$\sigma$ uncertainties. The right panel includes the data for NGC\,628 as magenta triangles, with their 1$\sigma$ uncertainties. Both panels show as a vertical black dot--dashed line the expected Pa$\alpha$ luminosity of a 4~Myr old, 3,000~M$_{\odot}$ cluster, the lower-end luminosity limit in our analysis to mitigate stochastic IMF sampling effects \citep{Cervino+2002}. {\bf (Left):} The magenta dotted line shows the EW(Pa$\alpha$) model track of an HII region of constant mass $\sim$2$\times$10$^5$~M$_{\odot}$ with increasing age from 1~Myr (top--right) to 10~Myr (bottom--left). The red continuous line is the model track for a constant age, 3~Myr old, HII region of decreasing mass, from $\sim$3$\times$10$^5$~M$_{\odot}$ down to $\sim$2$\times$10$^2$~M$_{\odot}$,  immersed in a constant luminosity non--ionizing stellar field, with a scatter of factors $-$3/$+$5 (higher/lower red dashed lines) about the mean trend. The cyan line is the same track, but for a 6~Myr old HII region. The blue lines are the tracks for NGC\,628,  from \citet{Calzetti+2024}, using the same model but with a larger photometric aperture 
(continuous line for the mean trend and dashed lines for the scatter). {\bf (Right):} The two empirical relations shown are from \citet{Calzetti+2007} for $\sim$0.5~kpc  star forming regions (blue solid line for the mean trend and dashed lines for the 90--percentile) and from \citet{Garn+2010} for galaxies (red continuous line). } 
\label{fig:corr_lum}
\end{figure}

The EW(Pa$\alpha$) marks a clear trend of decreasing values for decreasing luminosity of the attenuation--corrected Pa$\alpha$ (Figure~\ref{fig:corr_lum}, left). The maximum EW and L(Pa$\alpha$)$_{corr}$ is consistent with the expected values for an HII region powered by a $\sim$3~Myr old, 3$\times$10$^5$~M$_{\odot}$ star cluster. However, the trend is overall inconsistent with the expected track of aging HII regions (magenta dotted line in Figure~\ref{fig:corr_lum}, left), since the majority of the data are located above this track. The model that best describes the observations is a modification of the model used to explain a similar trend in NGC\,628 \citep{Calzetti+2024}: the clusters powering the HII regions are all roughly the same age, $\lesssim$5~Myr, but have decreasing mass when going from high to low EW and are immersed in a constant--luminosity, non--ionizing background stellar field. 

The models used for NGC\,628 are shown as blue lines in Figure~\ref{fig:corr_lum}  (left), and they underestimate the EW of the HII regions in NGC\,5194 at any given luminosity, with the exception of the brightest regions. The reason for this discrepancy is that there are differences in the properties and treatment of the two galaxies: the photometric aperture used for NGC\,5194 is four times smaller in angular area than that used in NGC\,628 and the median stellar field surrounding HII regions in NGC\,5194 has about 30\% higher  
surface brightness than in NGC\,628. When these two differences are included, the resulting model accounts reasonably well for the mean observed trend (red continuous line) of EW versus L(Pa$\alpha$)$_{corr}$. The observed scatter in the background stellar field, corresponding to factors $-$3/$+$5 (red dashed lines), is also consistent with the width of the distribution of the field's luminosity values. All regions have EWs consistent with HII regions younger than 6~Myr: the 6~Myr model track (cyan line) is located below the locus of the data, and therefore underestimates the observations.  The youth of our HII regions is consistent with our selection criteria of regions with compact ionized gas emission, in agreement with \citet{Whitmore+2011} and \citet{Hannon+2022}.

The seven brightest HII regions, with L(Pa$\alpha$)$_{corr}\gtrsim$10$^{38.4}$~erg~s$^{-1}$, mark the asymptotic behavior of the EW with L(Pa$\alpha$) at high luminosity. Their values, Log(EW)$\sim$3.19--3.55, are consistent with either a 3--4~Myr old massive cluster (see above) with no ionizing photon loss or a younger, $\sim$1-2~Myr, cluster about four times less massive and with $\approx$50\% ionizing photon loss. Recent results (Pedrini et al., in prep.) indicate that current models may not account completely for the presence of pre--Main--Sequence stars, which leads to an underestimate of the true stellar continuum around the Pa$\alpha$ emission line by 0.12--0.2~dex. The higher continuum corresponds to an identical downward shift in EW(Pa$\alpha$) and the ages of the brightest clusters would be closer to $\sim$2~Myr. If we assume $\lesssim$50\% of ionizing photon loss, there are at least two possible sources for this, as already discussed by \citet{Calzetti+2024}. One possibility is photon leakage from the HII regions, which \citet{Pellegrini+2012} determined to correlate with source luminosity, being larger for brighter sources. 
The second possibility is direct absorption of ionizing photons (pre--recombination cascade) by dust, which is also consistent with our data, as more luminous HII regions are also dustier on average (Figure~\ref{fig:obs_lum}, right, and Figure~\ref{fig:corr_lum}, right). According to the models of \citet{Krumholz+2009} and \citet{Draine+2011}, there is a maximum theoretical value of 20\%--70\% ionizing photons that can be directly absorbed by dust in the brightest and dustiest HII regions depending on physical conditions \citep[see, also,][]{Inoue+2001, Dopita+2003}. Our brightest regions, with L(Pa$\alpha$)$_{corr}\gtrsim$10$^{38.4}$~erg~s$^{-1}$, which corresponds to an ionizing photon rate Q$_o\gtrsim$1.4$\times$10$^{51}$~s$^{-1}$, and E(B--V)$\gtrsim$1~mag, satisfy the conditions for such high level of direct dust absorption. 

The color excess E(B--V) shows a general trend of higher values for larger Pa$\alpha_{corr}$ luminosities (Figure~\ref{fig:corr_lum}, right). For reference, E(B--V)=1~mag suppresses the H$\alpha$ emission by a factor $\sim$10 and the Pa$\alpha$ by a factor $\sim$1.9. The observed trend is very similar to the one \citet{Calzetti+2024} observed in NGC\,628 (magenta points in Figure~\ref{fig:corr_lum}, right): there appears to be a lower envelope to the color excess of HII regions, with a lower limit that increases with luminosity. The  E(B--V) lower envelope is well described by the model of  \citet{Garn+2010} (red line), which these authors derived for the mean extinction of galaxies, but seems to apply well to describe the lower boundary to the extinction of HII regions as a function of luminosity. The trend, however, is not limited to the lower envelope: brighter regions have larger values of the color excess on average, although the scatter increases with luminosity. \citet{Calzetti+2007} derived an empirical fit for $\sim$kpc galaxy regions together with the 90\% scatter about the mean relation (see, also, equation~\ref{equa:newebv} in Appendix~\ref{sec:appendixB}); those authors' fit is consistent with the trend observed for our HII regions and their 90\%--tile lines bracket our data reasonably well (blue continuous and dashed lines). 

Both panels in Figure~\ref{fig:corr_lum} mark the expected Pa$\alpha$  luminosity of a 4~Myr old cluster with mass 3,000~M$_{\odot}$ as a vertical dot--dashed line, corresponding to L(Pa$\alpha$)$_{corr}$=10$^{36.66}$~erg~s$^{-1}$. We elect this luminosity as the lower-limit we consider in our analysis, since HII regions powered by lower mass clusters are affected by stochastic sampling of the IMF \citep{Cervino+2002}. For those low--mass clusters, measurements of the ionizing photon rate is no longer a faithful representation of the massive star end of the IMF: the IMF is not fully sampled and the number of massive stars is randomly drawn from its distribution. With this luminosity limit, the remaining sample of 225 HII regions spans a little over two orders of magnitude in luminosity which is sufficient to relate it to the 21~$\mu$m emission.

\subsection{Properties of the Mid--Infrared Emission and Calibration of SFR Indicators}\label{subsec:IRSFR}

\begin{figure}
\plottwo{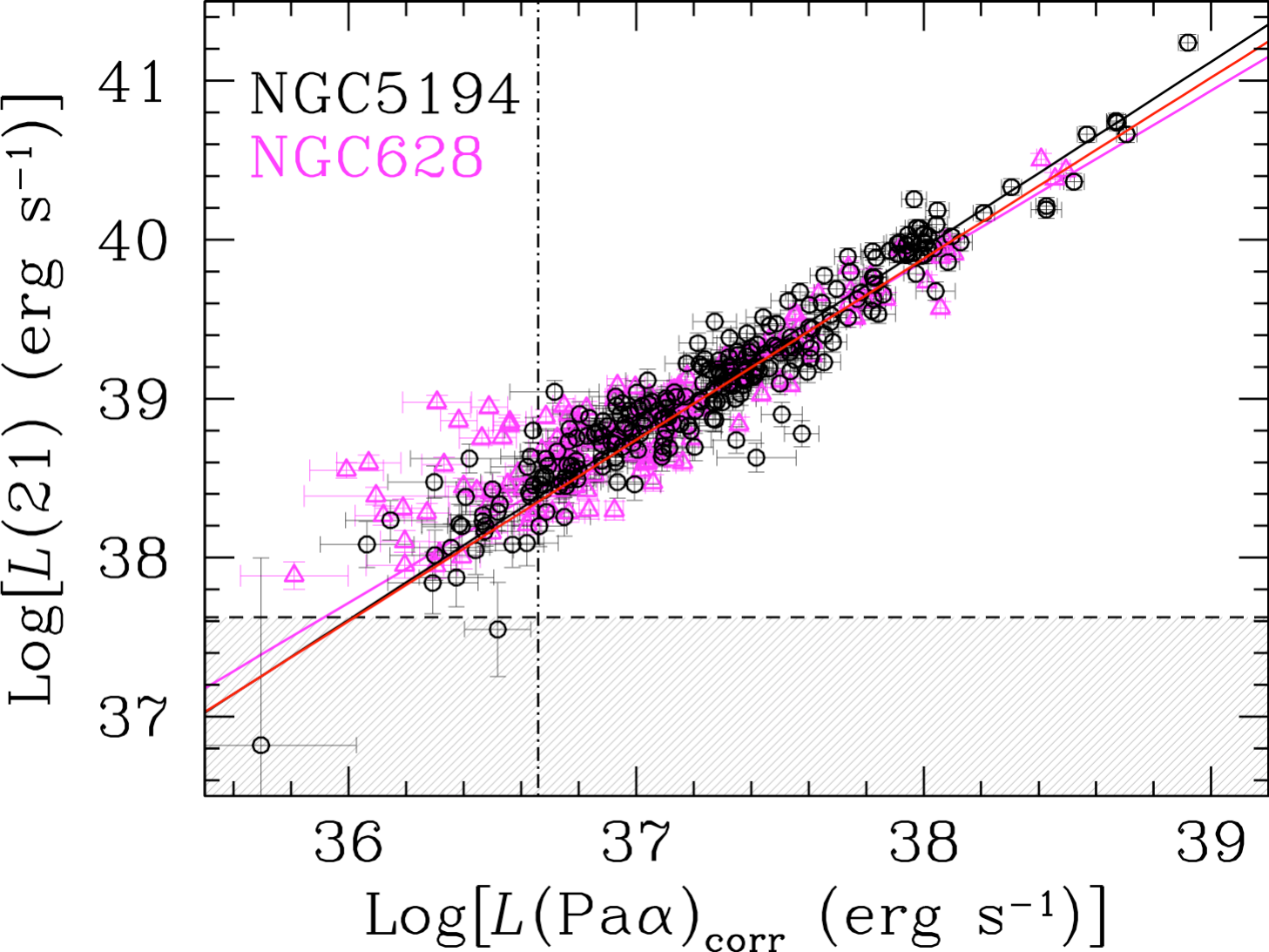}{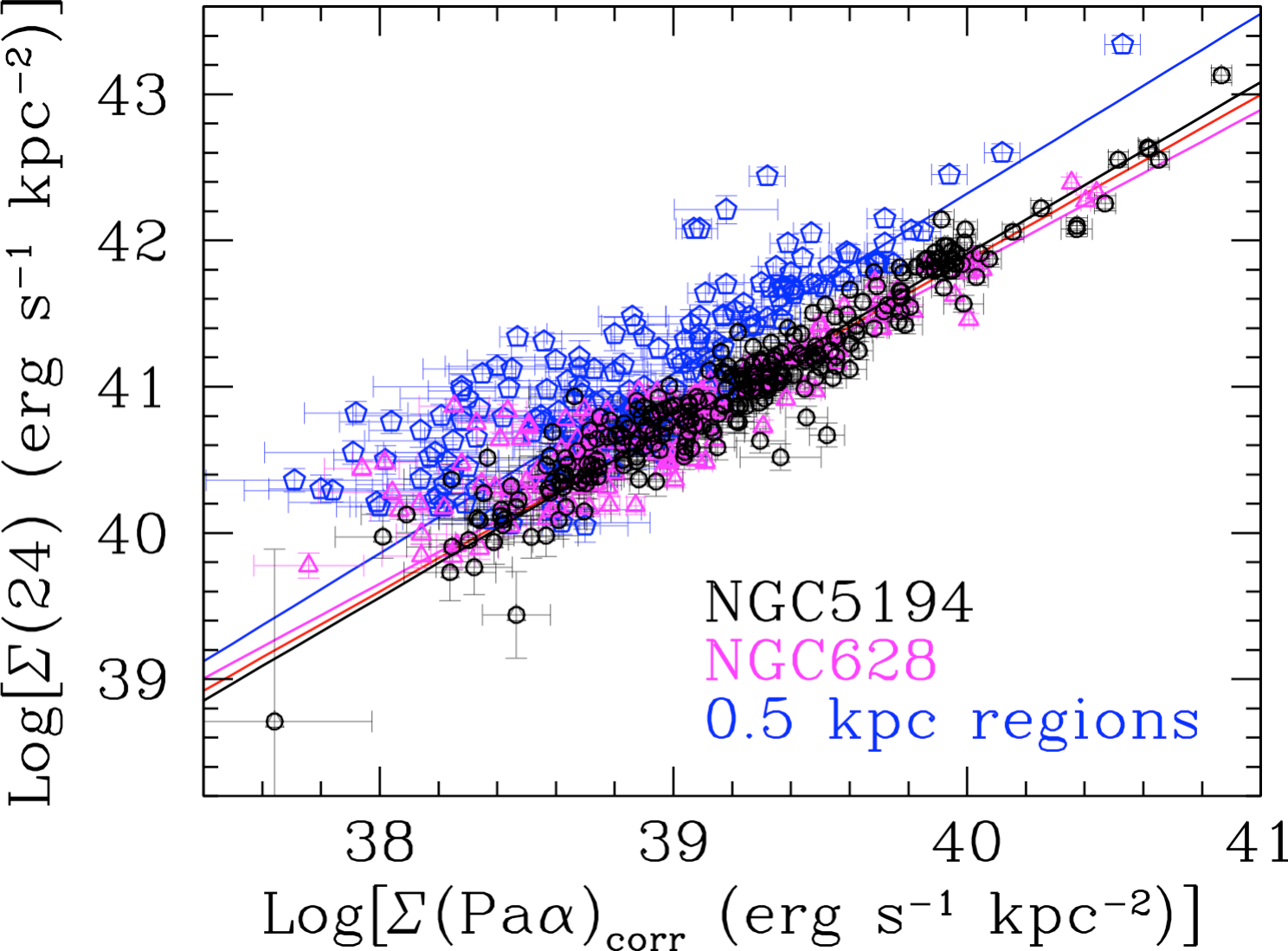}
\caption{{\bf (Left:)} The luminosity at 21~$\mu$m, L(21), as a function of the attenuation--corrected luminosity at Pa$\alpha$ for the 254 line--emitting regions in NGC\,5194 (black circles, this work) and the 143 regions in NGC\,628 (magenta triangles, \citet{Calzetti+2024}, with 1$\sigma$ uncertainties. The horizontal dashed line and grey region show the location of the 5$\sigma$ threshold at 21~$\mu$m in NGC\,5194. The black and magenta lines show the best fits through the data with Pa$\alpha$ luminosity above the IMF sampling limit (vertical dot--dashed line) for the two galaxies separately, while the red line is for the data from the two galaxies combined. {\bf (Right:)} The luminosity surface density at 24~$\mu$m, $\Sigma$(24), as a function of the attenuation--corrected luminosity surface density at Pa$\alpha$ for the same HII regions in NGC\,5194 (black circles) and NGC\,628 (magenta triangles), also with 1$\sigma$ uncertainties, after converting the luminosity at 21~$\mu$m to 24~$\mu$m. The solid lines indicate the fits through the HII regions using the same color scheme as the left panel.  The data for the luminosity surface densities of 160 $\sim$0.5~kpc regions at $\sim$solar metallicity from \citet{Calzetti+2007} are shown as blue pentagons with the best linear fit from those authors marked as a blue line.} 
\label{fig:l21_vs_lpa}
\end{figure}

The 21~$\mu$m luminosity of the HII regions in NGC\,5194 is tightly correlated with the extinction--corrected Pa$\alpha$ luminosity, similarly to what was found in NGC\,628 \citep{Calzetti+2024} and other galaxies \citep{Belfiore+2023}. The left panel of Figure~\ref{fig:l21_vs_lpa} shows L(21) as a function of L(Pa$\alpha$)$_{corr}$ for the HII regions in NGC\,5194 and NGC\,628, with the best fits to the two sets shown both separately and combined together. The best fit line to the logarithm of the luminosities for NGC\,5194, for the 225 regions which are above the stochastic IMF sampling limit L(Pa$\alpha$)$_{corr}$=10$^{36.66}$~erg~s$^{-1}$, is:
\begin{equation}
 Log[L(21)] = (1.17\pm 0.02) Log[L(Pa\alpha)_{corr}] - (4.51\pm 0.90),
\label{equa:l21lpa_5194}
\end{equation} 
 with scatter=0.09~dex. We use the LINMIX package\footnote{The python version of LINMIX used in this work can be found at: \href{https://github.com/jmeyers314/linmix}{https://github.com/jmeyers314/linmix}.}, which applies a hierarchical Bayesian approach to linear regression \citep{Kelly2007}, for all fits in this work. The reported best fit parameters are the average between the y--vs--x and x--vs--y relations, to mitigate the impact of the L(Pa$\alpha$)$_{corr}$=10$^{36.66}$~erg~s$^{-1}$ limit. Still--used deterministic  approaches to linear fitting, like the FITEXY \citep{Press+1992} and the Ordinary Least-Square bisector \citep{Isobe+1990} algorithms, yield a similar slope, $1.18$, but with significant smaller uncertainties, which have been shown to underestimate the true scatter in the data. 
The super--linear slope (slope$>$1 in log--log space) can be attributed to several mechanisms, two already discussed in Section~\ref{subsec:properties}: photon leakage out of HII regions and direct dust absorption of ionizing photons, both of which affect bright regions more than faint ones. A third mechanism, which is likely to account for the majority of the effect is the fact that lower--luminosity HII regions tend to be less extincted than brighter ones (Figure~\ref{fig:corr_lum}, right), implying that a larger fraction of the ionizing and non--ionizing photons at low luminosities emerges unaffected by dust attenuation and is not captured by the 21~$\mu$m luminosity. A fourth mechanism postulates that higher luminosity HII regions produce `hotter' infrared SEDs, which would increase the fraction of 21~$\mu$m emission relative to the total infrared emission. This has been observed, with large scatter, both in HII regions \citep{Relano+2013} and in galaxies \citep{Rieke+2009, Calzetti+2010}. 

The slope of 1.17 for the HII regions in NGC\,5194 (equation~\ref{equa:l21lpa_5194}) is slightly steeper than the one found for those in NGC\,628 of 1.07$\pm$0.03. The small, $<$3$\sigma$  difference between the two slopes can be accounted for by the presence of brighter HII regions in NGC\,5194, which extend to a factor $\sim$3 higher luminosities than in NGC\,628 and, for the reasons discussed above, steepens the slope. The difference in the slopes of the two sets of HII regions reduces to about 2~$\sigma$, when fit over the same range of luminosities. In addition, the locus of the HII regions from the two galaxies largely overlaps  (Figure~\ref{fig:l21_vs_lpa}), and we derive a combined relation for the 21~$\mu$m luminosity  for the 336 HII regions (225 from NGC\,5194 and 111 from NGC\,628) that are above our adopted limit for  stochastic sampling of the IMF:
\begin{equation}
Log[L(21)] = (1.14\pm 0.02) Log[L(Pa\alpha)_{corr}] - (3.44\pm 0.76),
\label{equa:l21lpa}
\end{equation} 
with scatter=0.09~dex. This fit has, as expected, a slope that is intermediate between the individual ones of the two galaxies, as is shown in red in Figure~\ref{fig:l21_vs_lpa} (left). 

Using L(Pa$\alpha$)$_{corr}$ as a reference SFR indicator \citep{KennicuttEvans2012}, we derive a SFR calibration using the 21~$\mu$m luminosity:
\begin{equation}
SFR(21) (M_{\odot} yr^{-1}) = 4.44^{+2.91}_{-2.68}\times 10^{-38} L(21)^{0.877\pm 0.015} \ \ \ \ \ \ \ \ \ \ \ \ \ \ \ \ \ \ \ \ \ \ for \ \ \ \ \ 10^{38}\lesssim L(21) \lesssim 2\times10^{41}.
\label{equa:SFR21}
\end{equation}
The JWST 21~$\mu$m luminosities can be converted to Spitzer/MIPS 24~$\mu$m equivalents using the relation derived by \citet{Calzetti+2024} for NGC\,628: $Log L(24) = Log L(21) - 0.057$. Given the proximity in wavelength between the two infrared luminosities, we do not expect major variations to this relation between NGC\,628 and NGC\,5194 and small variations have negligible impact on the overall picture. We thus derive also a SFR calibration for the 24~$\mu$m emission: 
\begin{equation}
SFR(24) (M_{\odot} yr^{-1}) = 4.98^{+3.26}_{-3.01}\times 10^{-38} L(24)^{0.877\pm 0.015} \ \ \ \ \ \ \ \ \ \ \ \ \ \ \ \ \ \ \ \ \ \ for \ \ \ \ \ 10^{38}\lesssim L(24) \lesssim 2\times10^{41},
\label{equa:SFR24}
\end{equation}
with both luminosities in units of erg~s$^{-1}$. 
The slope is indistinguishable from the one derived by \citet{Calzetti+2010} for the luminosities of a sample of 160 $\sim$solar metallicity, $\sim$0.5~kpc size regions in 21 nearby galaxies, but the intercept is a factor $\sim$1.8 higher, although the two agree within the formal uncertainty. This shift becomes obvious when comparing the luminosity  surface densities (luminosity/area), as opposed to the luminosities, of the HII regions and the 0.5~kpc regions, shown in Figure~\ref{fig:l21_vs_lpa} (right). Although the HII regions in NGC\,5194 have been measured employing photometric apertures with 40\% of the physical radius of the measurements in NGC\,628, the derived luminosities for both cases are total HII region luminosities, implying that the photometric aperture is not the regions' characteristic size (i.e.: larger or smaller apertures would not yield larger or smaller luminosities, because all luminosities are corrected for aperture losses). The choice of area is therefore somewhat arbitrary, and we choose 60~pc as the radius of the region to normalize the HII region luminosities and derive surface densities, as this is the size of the aperture radius in the more distant NGC\,628 (1$^{\prime\prime}$.4=63~pc). This is the assumption used to construct the right panel of  Figure~\ref{fig:l21_vs_lpa}, where we have neglected the small inclination corrections for the two galaxies. Using a different assumption for the region size, e.g., the NGC\,5194 photometric aperture ($\sim$26~pc radius), causes the HII regions locus to shift away from the 0.5~kpc regions by 0.05~dex, exacerbating the discrepancy between the two sets of regions. Similarly, if the HII regions in our sample are experiencing ionizing photon leakage, the discrepancy would be further magnified: the locus of the HII regions would shift away from the 0.5~kpc regions locus by $\sim$0.05~dex for a 50\% ionizing photon loss.  The best fits to the HII regions shown in Figure~\ref{fig:l21_vs_lpa} (right panel, solid black, magenta and blue lines)  are derived using the regions that are above the stochastic IMF sampling limit in $\Sigma$(Pa$\alpha$)$_{corr}$=10$^{38.56}$ erg~s$^{-1}$~kpc$^{-2}$ when normalizing to the area of a circle with 60~pc radius. We revisit the shift between HII regions and 0.5~kpc regions  luminosity surface densities   in Section~\ref{sec:discussion}, when we discuss our findings within the context of both previous results and models.

The existence of similar super--linear relations for both HII regions and larger 0.5~kpc regions lends support to the interpretation that, in both cases, the super--linear slope is mainly driven by lower--luminosity regions being more transparent to dust  than more luminous regions, together with a potential increase of the mid--IR emission relative to the total IR (hotter dust) in brighter regions. In fact, neither ionizing photon leakage nor direct dust absorption of ionizing photons apply to the larger regions, as these are extended enough to include the leaked photons in their area and to encompass random collections of HII regions at a range of luminosities. 

\subsection{The Calibration of Hybrid SFR Indicators}\label{subsec:hybridSFR}

\begin{figure}
\plottwo{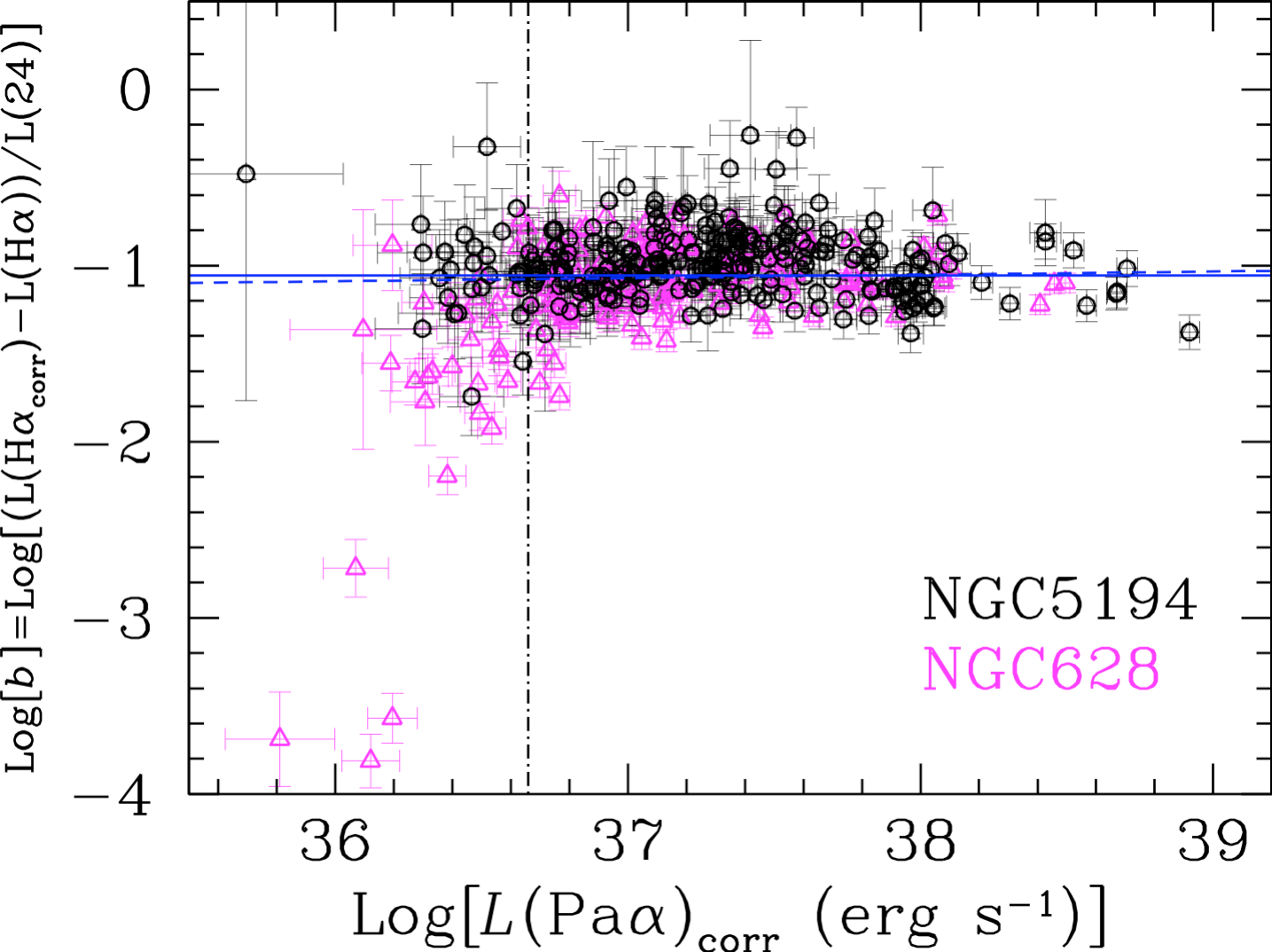}{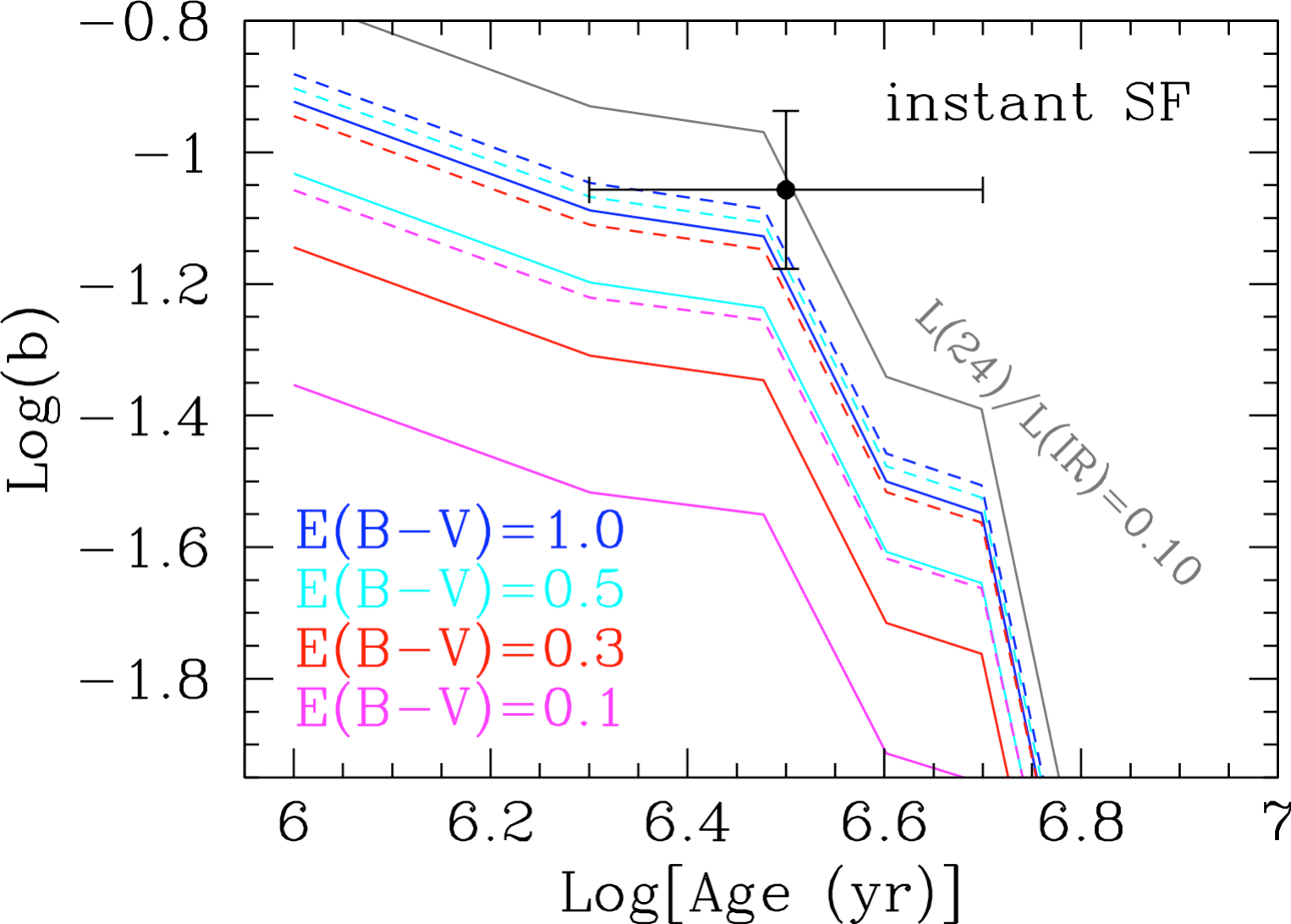}
\caption{{\bf (Left):} The ratio of the nebular hydrogen emission absorbed by dust at H$\alpha$ and the dust emission at 24~$\mu$m as a function of the attenuation--corrected Pa$\alpha$ luminosity for the HII regions in NGC\,5194 (black circles) and NGC\,628 (magenta triangles) with 1$\sigma$ uncertainties. The best linear fit, in log--log scale, and the best constant value fit for Pa$\alpha$ luminosities above the stochastic IMF sampling limit (vertical dot--dashed black line) are shown as dashed and continuous blue lines, respectively; the value of the scaling factor $b$ in this regime is practically independent of the Pa$\alpha$ luminosity. {\bf (Right):} The best--fit value of $b$ from the left panel (filled circle with error bars), compared with models of instantaneous burst populations of increasing age. The observed $b$ is assigned an age of 3~Myr with $-$1~Myr,+2~Myr uncertainty. Continuous [dashed] lines show models with E(B--V)$_{gas}$=E(B--V) [E(B--V)$_{gas}$=2.27 E(B--V)], with different E(B--V) values identified with colors (see legend). In all cases, $f$=L(24)/L(IR)=0.14, except for the continuous grey line where $f$=0.10 is applied to the case E(B--V)=0.5 and  E(B--V)$_{gas}$=2.27 E(B--V) as example. } 
\label{fig:b_vs_Pa}
\end{figure}

Following several previous studies \citep[e.g.,][]{Wang+1996, Buat+1999, Meurer+1999, Hirashita+2003, Treyer+2010, Hao+2011, Liu+2011, Calzetti+2007, Kennicutt+2007, Kennicutt+2009, Belfiore+2023, Calzetti+2024}, we combine the IR emission with a shorter--wavelength SFR indicator, to derive a full census of the star formation energy budget in our regions. The IR probes the dust--reprocessed SFR while the short--wavelength indicator, typically UV or a nebular emission line, probes the portion of the SFR that emerges unaffected by dust. Adding the unattenuated SFR indicator should `correct' for the non--linearity of the IR SFR indicator (see previous Section). Within our study, we adopt the hybrid optical--IR SFR indicator:
\begin{equation}
SFR (M_{\odot} yr^{-1}) = 5.45\times 10^{-42} L(H\alpha_{corr}) = 5.45\times 10^{-42} [L(H\alpha) + b  L(24)],
\label{equa:mixSFR}
\end{equation}
where L(H$\alpha$) is the {\em observed} H$\alpha$ luminosity and the calibration constant $5.45\times 10^{-42}$ has been derived by many of authors using models \citep[e.g.,][and references therein]{KennicuttEvans2012, Calzetti2013}. All luminosities are in units of erg~s$^{-1}$. We use L(24) instead of L(21) for uniformity with previous derivations that leveraged the 24~$\mu$m band of the Spitzer Space Telescope. 
The scaling factor $b$ can be derived from the data directly, as shown in Figure~\ref{fig:b_vs_Pa} (left), where the fits to the combined data of the HII regions in NGC\,5194 and NGC\,628 are virtually independent of the Pa$\alpha$ luminosity, yielding a best fit value:
\begin{equation}
SFR(H\alpha + 24) (M_{\odot} yr^{-1}) = 5.45\times 10^{-42} [L(H\alpha) + (0.088\pm 0.025)  L(24)],
\label{equa:mixSFR24}
\end{equation}
while the equivalent calibration for the 21~$\mu$m luminosity is:
\begin{equation}
SFR(H\alpha + 21) (M_{\odot} yr^{-1})  = 5.45\times 10^{-42} [L(H\alpha) + (0.077\pm 0.022)  L(21)].
\label{equa:mixSFR21}
\end{equation}
As before, the fits are performed on the 336 regions that have Pa$\alpha$ luminosity above our adopted stochastic IMF sampling limit.
The values of $b$ derived for the combined sample are, within the uncertainties, indistinguishable from the values derived for the HII regions in NGC\,628 alone, b(24)=0.095 and b(21)=0.083 \citep{Calzetti+2024}. The similarities are expected, as both galaxies have solar metallicity, implying comparable amounts of dust emission per `unit' of ionizing radiation. 

By definition:
\begin{equation}
b = {L(H\alpha_{corr}) - L(H\alpha) \over  L(24)} = {L(H\alpha)_{abs} \over f L(bol)_{abs}} 
\label{equa:b_definition}
\end{equation}
is the H$\alpha$ luminosity (ionized gas light) absorbed by dust divided by the stellar bolometric luminosity, L(bol), absorbed by dust and emerging in the infrared, and $f$=L(24)/L(IR) is the fraction of infrared light emerging at 24~$\mu$m. We adopt L(bol)$_{abs}$=L(IR) under assumption of energy balance.  
Thus, the scaling factor $b$ can be modeled using simplified assumptions for the star formation histories and the dust distribution, since HII regions can be effectively  considered instantaneous burst populations \citep{Wofford+2016} for which we assume foreground dust. 

We generate the SEDs of instantaneous burst populations in the age range 1--10~Myr using Starburst99 models \citep{Leitherer+1999,Vazquez+2005} with Padova AGB evolutionary tracks \citep{Girardi+2000},  metallicity Z=0.02 (solar) and a \citet{Kroupa2001} IMF in the stellar mass range 0.1--120~M$_{\odot}$. The stellar models do not include pre--Main--Sequence stars. The stellar  SEDs are attenuated by dust using a starburst attenuation curve \citep{Calzetti+2000} and attenuation values in the range E(B--V)=0.1--1~mag. The ionized gas attenuation E(B--V)$_{gas}$ is assumed to be the same as that of the stellar continuum or 2.27 times higher \citep{Calzetti+1994}, to test the impact of both scenarios. In all cases, the attenuation of the ionized gas follows the extinction curve of \citet{Fahrion+2023}, following the prescription of \citet{Calzetti+1994} and \citet{Calzetti2001}. The fraction of infrared luminosity emerging at 24~$\mu$m is assumed to be constant with 
age, $f$=0.14, and we also test the effects of a smaller value $f$=0.10 in one case. \citet{Relano+2013} studied the IR SEDs of HII regions in the galaxy M\,33, finding that L(24)/L(IR) spans the range $f\sim 0.07-0.3$ for the filled HII regions, which are ones from their sample most closely resembling the morphology of our HII regions. Although M\,33 has a lower metallicity (about 1/2 solar) than either NGC\,5194 or NGC\,628, the shapes of the IR SEDs of HII regions are not expected to be sensitive to these relatively small differences.  We, thus, take the value $f=0.14$ as representative of the IR SED shape of HII regions, which is also consistent with the value determined for star--forming galaxies \citep[e.g.,][]{Rieke+2009, Dale+2009}.

The models for $b$ are compared with the observed value in Figure~\ref{fig:b_vs_Pa} (right): we assume that the observations correspond to 3~Myr age for the HII regions (Figure~\ref{fig:corr_lum}, left), although we allow for a range in ages. The models show a general agreement with the observations for cases with significant dust (E(B--V)$\gtrsim$0.3~mag) and with higher attenuation in the emission lines relative to the stellar continuum. Variations in the L(24)/L(IR) ratio also have significant impact, in the direction of favoring smaller ratios (cooler IR SEDs). Assuming ages younger than 3~Myr  still maintains agreement between data and models, and  accommodates some ionizing photon loss as well.  For instance, 50\% leakage correction would increase $b$ by 0.3~dex, which would still agree with the models for average cluster ages $\sim$1~Myr.

\subsection{HII Regions, Giant HII Regions, and the Large Scale Emission}\label{subsec:largescale}

\begin{figure}
\plotone{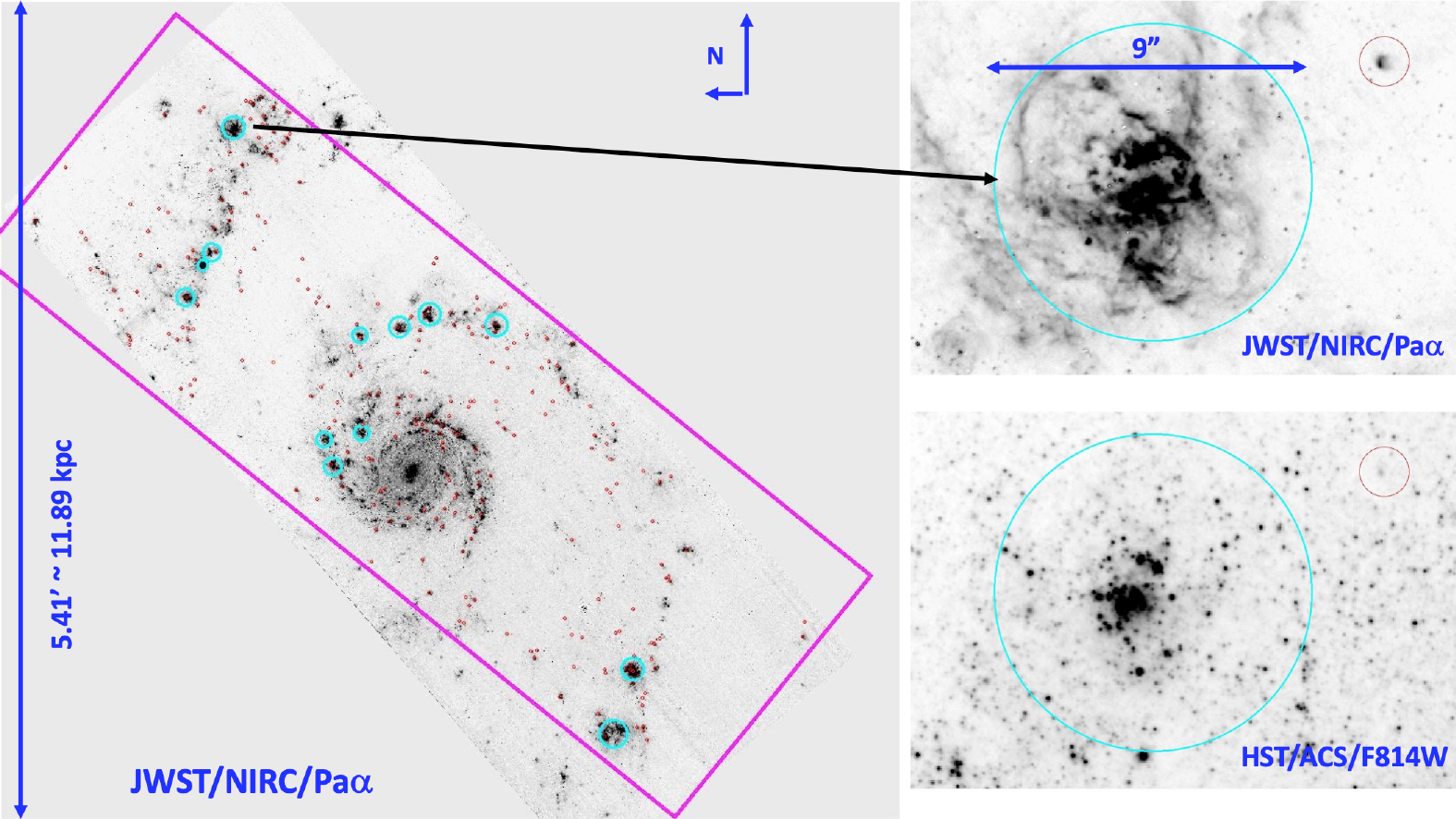}
\caption{{\bf (Left):} The JWST Pa$\alpha$ mosaic with the 13 Giant HII regions (cyan circles) together with the 254 HII regions (red circles). The radii of the circles match those of the apertures used for photometry, which for the Giant HII regions range from 2$^{\prime\prime}$ and 5$^{\prime\prime}$.5. North is up and East is left. The magenta rectangle shows the MIRI 21~$\mu$m footprint. {\bf (Right, Top and Bottom):} Detail showing one Giant HII region (cyan circle, its location marked by the black arrow) and one HII region (red circle) in the JWST/Pa$\alpha$ image (Top--Right) and the HST/ACS/F814W image (Bottom--Right). The diameter of the cyan circle, 9$^{\prime\prime}$, is indicated for reference. The Pa$\alpha$ image highlights the complexity of the ionized gas emission from the Giant HII region, while the F814W image shows the multiplicity of the continuum sources ionizing the gas; conversely, the HII region (red circle) is dominated by a single, compact source.} 
\label{fig:GiantPa}
\end{figure}

Our sample of HII regions, selected to be ionized by individual compact star clusters, leave out bright and extended emission line objects ionized by groups of closely clustered compact sources. These objects would have appeared only partially resolved in past ground--based imaging, and, because of their brightness and extent, were termed Giant HII regions \citep[e.g.,][]{Smith+1970, Terlevich+1981, Gallagher+1983, vanderHulst+1988}. We identify 13 Giant HII regions within the 4.4$\times$12.6~kpc$^2$ JWST mosaics footprint (Figure~\ref{fig:GiantPa}): nine of those regions include 15 of the single--source HII regions from our sample and the remaining four have no sources that can be easily isolated from the rest of the region. As the Giant HII regions provide significant contribution to the line emission encompassed by the mosaics, we measure their luminosities and check whether their inclusion would change any of the results from the previous Sections. We also calculate the fraction of the total emission within the JWST footprint captured by the single--source and giant HII regions, both in line emission and in the mid--IR.

The Giant HII regions require customized apertures for the photometry, as their sizes vary depending on the number and mix of ages and masses of the clusters that ionize them. We select aperture sizes to include the 21~$\mu$m emission out to within 3$\sigma$ of the local background, resulting in radii between 2$^{\prime\prime}$ and 5$^{\prime\prime}$.5 (73~pc to 200~pc), depending on the region (Table~\ref{tab:GiantHII}). The same quantities that were measured for the HII regions are measured for the Giant HII regions, including luminosities, equivalent widths, etc. Aperture corrections are not required for H$\alpha$ and Pa$\alpha$, but small aperture--dependent  corrections are applied to the 21~$\mu$m luminosity.  Table~\ref{tab:GiantHII} lists for each of the 13 Giant HII Regions: ID, location on the sky in RA(2000) and DEC(2000), the radius of the aperture used for photometry, the observed luminosity in H$\alpha$, Pa$\alpha$ and 21~$\mu$m, the equivalent width in Pa$\alpha$, and the color excess E(B$-$V). Both H$\alpha$ and Pa$\alpha$ are reported already corrected for the Milky Way foreground extinction. 

\begin{deluxetable}{lrrrrrrr}
\tablecolumns{8}
\tabletypesize{\small}
\tablecaption{Source Location, Luminosity and Derived Quantities for the Giant HII Regions in NGC\,5194\label{tab:GiantHII}}
\tablewidth{100pt}
\tablehead{
\colhead{ID} & \colhead{RA(2000),DEC(2000)}  &\colhead{Radius} & \colhead{Log[L(H$\alpha$)]} & \colhead{Log[L(Pa$\alpha$)]} & \colhead{Log[L(21)]} & \colhead{Log[EW(Pa$\alpha$)]} & \colhead{E(B$-$V)} 
\\
\colhead{(1)} & \colhead{(2)} & \colhead{(3)} & \colhead{(4)} & \colhead{(5)}  & \colhead{(6)} & \colhead{(7)} & \colhead{(8)} 
\\
}
\startdata
    G1 &13:29:55.7640,+47:11:44.489 & 3.5 &38.873$\pm$0.028 &38.724$\pm$0.018 &41.053$\pm$0.052 &2.816$\pm$0.027 &1.005$\pm$0.045\\
    G2 &13:29:54.6436,+47:11:57.720 & 3.0 &38.402$\pm$0.034 &38.153$\pm$0.024 &40.323$\pm$0.064 &2.353$\pm$0.032 &0.870$\pm$0.056\\
    G3 &13:29:54.6943,+47:12:36.450 & 3.0 &38.494$\pm$0.032 &38.237$\pm$0.023 &40.361$\pm$0.063 &2.462$\pm$0.031 &0.859$\pm$0.054\\
    G4 &13:29:53.1800,+47:12:39.760 & 4.0 &39.007$\pm$0.027 &38.587$\pm$0.019 &40.597$\pm$0.063 &2.743$\pm$0.028 &0.638$\pm$0.045\\
    G5 &13:29:52.0107,+47:12:44.770 & 4.5 &39.241$\pm$0.025 &38.806$\pm$0.018 &41.002$\pm$0.056 &2.747$\pm$0.025 &0.618$\pm$0.041\\
    G6 &13:29:49.3892,+47:12:40.510 & 4.5 &38.899$\pm$0.028 &38.734$\pm$0.018 &40.980$\pm$0.057 &2.836$\pm$0.027 &0.983$\pm$0.045\\
    G7 &13:29:44.0821,+47:10:23.813 & 4.5 &39.508$\pm$0.024 &39.076$\pm$0.016 &41.145$\pm$0.054 &2.971$\pm$0.024 &0.623$\pm$0.039\\
    G8 &13:29:44.8174,+47:09:58.070 & 5.5 &39.318$\pm$0.025 &38.911$\pm$0.017 &40.982$\pm$0.060 &2.581$\pm$0.023 &0.657$\pm$0.041\\
    G9 &13:29:59.6235,+47:13:59.019 & 4.5 &39.365$\pm$0.025 &38.824$\pm$0.017 &40.653$\pm$0.064 &2.807$\pm$0.025 &0.475$\pm$0.041\\
   G10 &13:30:01.4968,+47:12:51.728 & 3.5 &38.894$\pm$0.028 &38.782$\pm$0.018 &41.305$\pm$0.049 &2.744$\pm$0.025 &1.055$\pm$0.045\\
   G11 &13:30:00.4693,+47:13:09.574 & 3.5 &38.876$\pm$0.028 &38.424$\pm$0.021 &40.506$\pm$0.063 &2.662$\pm$0.030 &0.596$\pm$0.047\\
   G12 &13:29:56.1027,+47:11:55.183 & 3.0 &38.640$\pm$0.031 &38.255$\pm$0.023 &40.207$\pm$0.068 &2.322$\pm$0.030 &0.686$\pm$0.051\\
   G13 &13:30:00.8440,+47:13:04.439 & 2.0 &38.706$\pm$0.030 &38.355$\pm$0.021 &40.301$\pm$0.057 &2.924$\pm$0.034 &0.732$\pm$0.049\\
\hline
\enddata
\tablenotetext{}{(1) The identification number of the source.}
\tablenotetext{}{(2) Right Ascension  and Declination in J2000 coordinates.}
\tablenotetext{}{(3) Radius, in arcseconds, of the photometric aperture used for photometry.}
\tablenotetext{}{(4)--(6) Logarithm of the luminosity  of each source at the indicated wavelength, in units of erg~s$^{-1}$.  The H$\alpha$ and Pa$\alpha$ luminosities are corrected 
for the Milky Way foreground extinction.}
\tablenotetext{}{(7) The logarithm of the equivalent width (EW) of Pa$\alpha$, in \AA, calculated from the ratio of the emission line flux  to the stellar continuum flux density. }
\tablenotetext{}{(8) The color excess, E(B$-$V), in mag, derived from the H$\alpha$/Pa$\alpha$ luminosity ratio.}
\end{deluxetable}

\begin{figure}
\plottwo{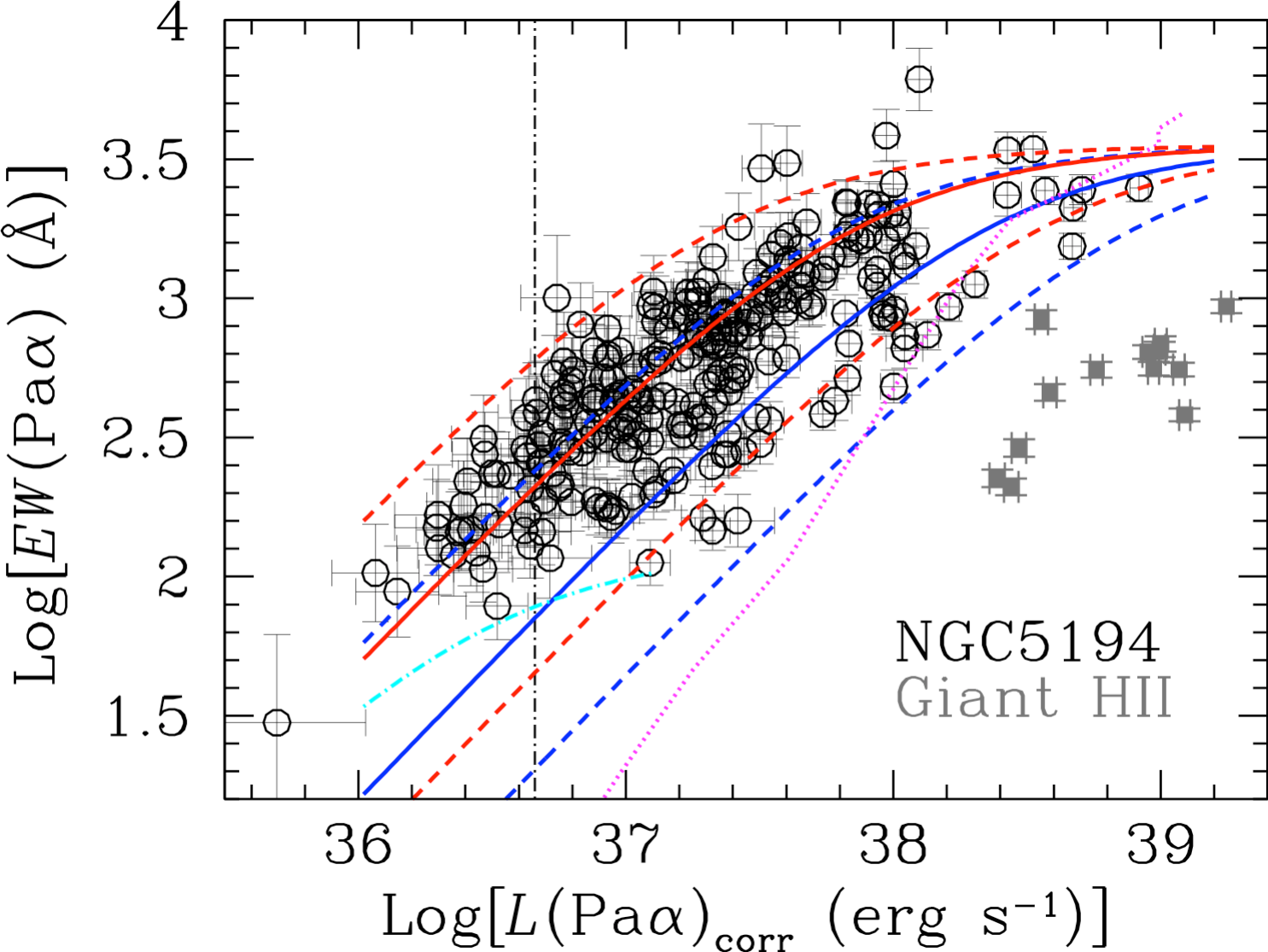}{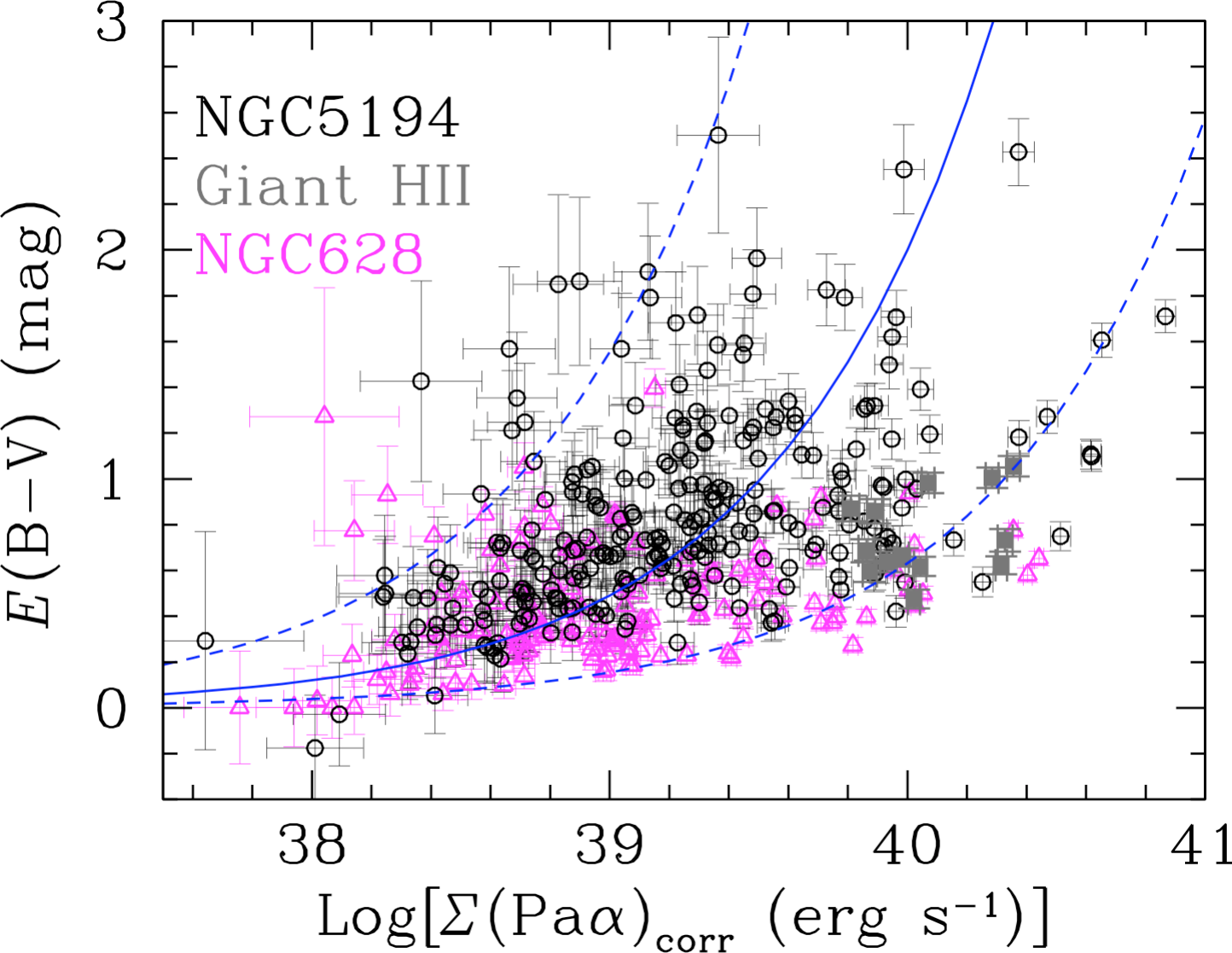}
\plottwo{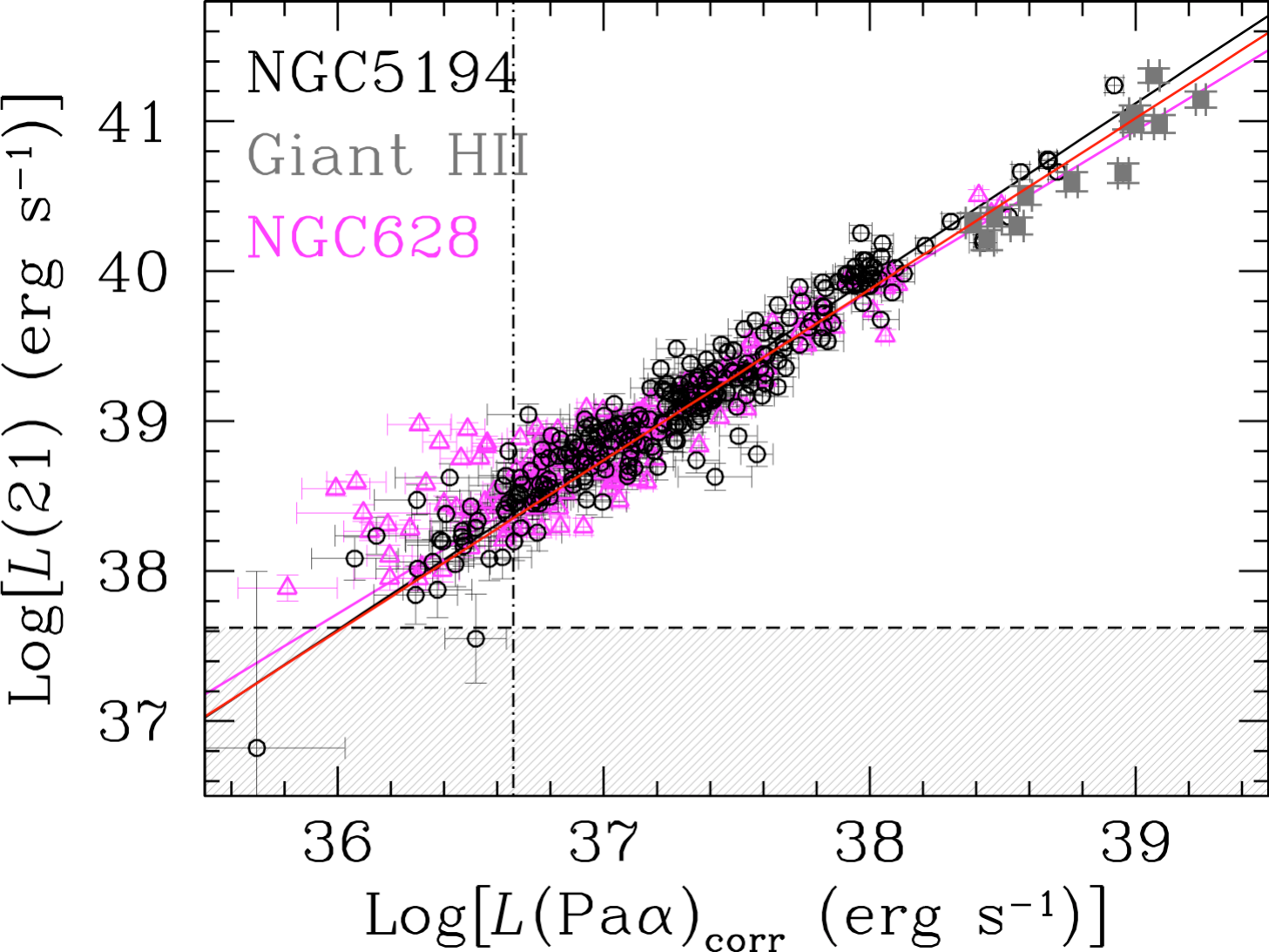}{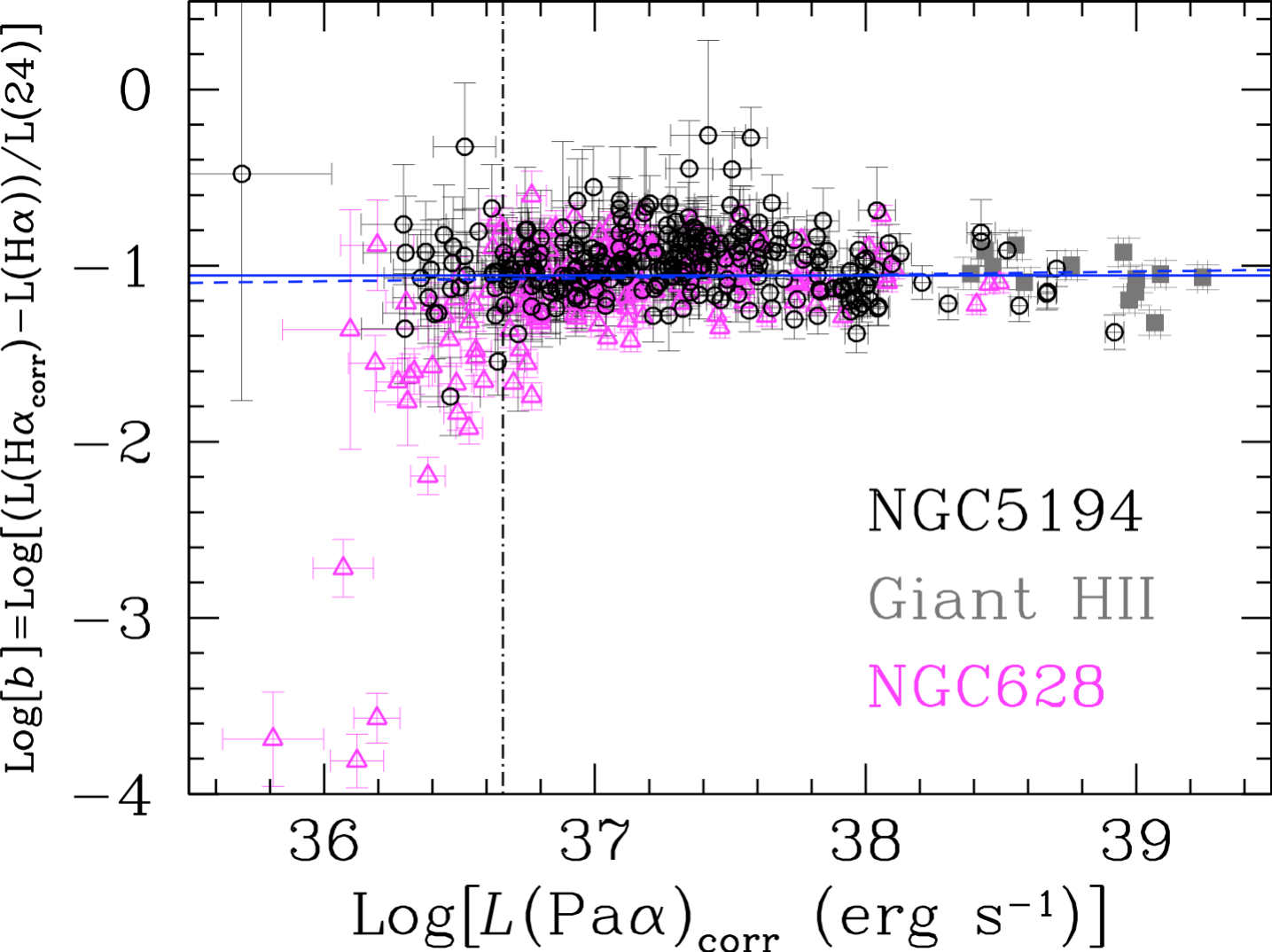}
\caption{The same plots as the left and right panels of Figure~\ref{fig:corr_lum} {\bf (Top--Left and Top--Right, respectively)}, the left panel of Figure~\ref{fig:l21_vs_lpa} {\bf (Bottom--Left)}, and the left panel of Figure~\ref{fig:b_vs_Pa} {\bf (Bottom--Right)}, with the Giant HII regions in NGC\,5194 added as grey filled squares. In the Top--Right panel, the luminosity surface density $\Sigma(Pa\alpha)_{corr}$ is used instead of the luminosity L(Pa$\alpha)_{corr}$. }
\label{fig:Giant_properties}
\end{figure}

As expected, the Giant HII Regions have larger luminosities, on average, than the HII regions, but also span a more restricted range of values both in EW(Pa$\alpha$) and E(B--V)  (Figure~\ref{fig:Giant_properties}, top two panels).  The values of the EW(Pa$\alpha$), which are lower than those of the HII regions at a given Pa$\alpha$ luminosity, can be explained with the contribution to the stellar continuum both by multiple clusters with different ages and by the larger apertures employed to measure the Giant HII regions, which admit a larger fraction of background stellar emission (see discussion in Section~\ref{subsec:properties}). Differential dust attenuation between nebular emission and stellar continuum, which is likely to occur when considering large spatial scales and stellar population mixes, can also contribute to decreasing the EW(Pa$\alpha$) values: for the observed range of E(B--V) the expected decrease can be as much as 0.08--0.18~dex. If the EWs are interpreted strictly in terms of mean ages, the Giant HII Regions are in the age range 3.5-5.5~Myr, only marginally older than the single--source HII regions. 

The color excess of the Giant HII regions is within the range spanned by the HII regions (Figure~\ref{fig:Giant_properties}, top right  panel), when using luminosity surface densities instead of luminosities, to account for the differences in area used to measure Giant HII and single--source HII regions. The models from \citet{Calzetti+2007} are shown as blue lines (Equation~\ref{equa:newebv}). 

The Giant HII region luminosities fill in the bright luminosity end of the distribution of the HII regions (Figure~\ref{fig:Giant_properties}, bottom--left), marking the same trend for L(21)--versus--L(Pa$\alpha$)$_{corr}$. A formal fit through the HII regions of NGC\,5194 and NGC\,628, including the Giant HII regions discussed in this Section, yields a slope of (1.12$\pm$0.02), consistent within 1$\sigma$ with the slope reported in equation~\ref{equa:l21lpa}. Not surprisingly, the Giant HII Regions distribute around the same value of the scaling factor $b$ as the HII regions (Figure~\ref{fig:Giant_properties}, bottom--right). In summary, the Giant HII Regions are consistent with being brighter versions of the single--source HII regions but similarly young, likely because their luminosity is dominated by the youngest members in the groups of clusters that ionize them.

The photometry of the Giant HII regions is measured in regions with sizes almost as large as those used by \citet{Calzetti+2007} for their measurements in this galaxy. The photometric sizes (6$^{\prime\prime}$.5 radii) used in that paper were driven by the low resolution of the Spitzer MIPS 24~$\mu$m channel \citep{Rieke+2004}, while our large apertures (Table~\ref{tab:GiantHII})  are driven by the extent of the 21~$\mu$m and Pa$\alpha$ emission. We thus discuss the level of overlap between our sample and the sample from \citet{Calzetti+2007}. Of the 39 regions in common between our footprint and the footprint of the 24~$\mu$m/Pa$\alpha$/H$\alpha$ observations used in \citet{Calzetti+2007}, only 7 are among our Giant HII regions. The remaining sources are broken down each into multiple single--source HII regions in our analysis. Thus, the 0.5~kpc regions identified by \citet{Calzetti+2007} as peaks of 24~$\mu$m emission contain a mix of sources, including Giant HII regions and individual HII regions with a variety of spatial distributions. Because of these differences, in the remainder of this paper we will treat the Giant HII regions as more luminous versions of HII regions, while the regions from \citet{Calzetti+2007} will be treated as general representatives of star formation over $\sim$0.5~kpc scales in galaxies.

We compare the sum total H$\alpha$, Pa$\alpha$ and 21~$\mu$m luminosity from HII regions and Giant HII regions with the luminosity captured by the entire JWST footprint, to evaluate whether our sample is representative 
of the full HII region population. Table~\ref{tab:totals} lists the observed and attenuation--corrected luminosities for the JWST footprint, for the sum of the HII regions above the stochastic IMF sampling limit of Log[L(Pa$\alpha$)$_{corr}$]=36.66 (after removing the 15 regions which are contained within the Giant HII regions) and for the sum of the Giant HII regions. Table~\ref{tab:totals} also reports the luminosity--weighted E(B--V) values for all three cases, showing that the JWST footprint suffers, as expected, lower dust attenuation than the HII regions, and the Giant HII regions are slightly less attenuated than the single--source HII regions. For reference, 
the single--source HII regions have median E(B--V)$\sim$0.8~mag and the Giant HII regions E(B--V)$\sim$0.7~mag, similar to the luminosity--weighted values in Table~\ref{tab:totals}. 

As discussed earlier,  the minimum luminosity for the HII regions to be considered above the stochastic IMF sampling limit corresponds to a star cluster with mass 3,000~M$_{\odot}$ and age 4~Myr. Roughly equal masses are in star clusters above and below 3,000~M$_{\odot}$ \citep{Adamo+2020}, implying that the HII regions include about twice as much luminosity as the one reported in Table~\ref{tab:totals}. We add the faint HII region luminosities under two assumptions: that their dust attenuation is similar to the one of the HII regions above the stochastic IMF sampling limit (Knutas et al. in prep.) and that their  attenuation is the same as the mean attenuation of the JWST footprint (Table~\ref{tab:totals}). These two assumptions yield slightly different values of the attenuation--corrected H$\alpha$ and Pa$\alpha$ luminosities for the JWST footprint, as listed in Table~\ref{tab:totals} (lines 2 and 3). The contribution of the Giant HII regions is added to the HII regions without extrapolations in their total luminosity, under the reasonable assumptions that the Giant HII regions include  clusters that sample the full range of masses within the age range of interest ($\sim$0--6~Myr).

Taking those calculations at face value, the H$\alpha$ luminosity of all HII regions (normal and Giant) represents 43\%--47\% of the total within the JWST footprint, depending on the assumption for the attenuation of the faint HII regions. For Pa$\alpha$, the fraction is about 51\%--54\%. For both, we find that the HII regions represent about half of the recovered ionizing photons, with the remaining half found in the diffused ionized medium, as typical of star--forming disks \citep{Oey+2007}. The difference between the H$\alpha$ and Pa$\alpha$ diffuse fractions, $\sim$55\% and $\sim$47\% respectively, is negligible within the uncertainties of our estimate, but, if confirmed, would support models where a small fraction, $\sim$20\%, of the H$\alpha$ diffuse light is due to reflection by dust \citep{Dong+2011, McCallum+2025}; the Pa$\alpha$ line would then be expected to have a smaller reflection component, since the wavelength of Pa$\alpha$ has a lower scattering cross--section, by a factor of several, than H$\alpha$ \citep{Weingartner+2001}. We refrain from speculating further on the nature of this difference as we do not control some of the systematics, such as small uncertainties in the stellar background removal, 
the small gradient in the Pa$\alpha$ mosaic (corresponding to 1$\sigma$ change from top to bottom), and faint residual 1/f noise in the NIRCam images\footnote{\href{https://jwst-docs.stsci.edu/known-issues-with-jwst-data/1-f-noise\#gsc.tab=0}{https://jwst-docs.stsci.edu/known-issues-with-jwst-data/1-f-noise\#gsc.tab=0}}. From the result of this analysis, we conclude that our sample  is representative of the HII region population within the JWST footprint in NGC\,5194. 

We have a better handle on the potential systematics for the 21~$\mu$m luminosity within the JWST footprint thanks to the Spitzer MIPS 24~$\mu$m image of the galaxy. Matching the JWST footprint to the MIPS image yields a flux density of 8.17~Jy at 24~$\mu$m, to be compared with 8.91~Jy at 21~$\mu$m, or 9\% lower, which can be ascribed to small uncertainties in matching the two mosaics. Thus, we conclude that the JWST mosaic includes the diffuse emission from the galaxy. When comparing the emission from the full population of HII regions (Giant and non) to the total, we find that the regions account for 23\% of the 21~$\mu$m emission in the JWST footprint, or about half of the fraction in the ionized gas. The nature of this difference is discussed in the next Section. 

\begin{deluxetable}{llrrrr}
\tablecolumns{5}
\tabletypesize{\small}
\tablecaption{Total Luminosities in NGC\,5194\label{tab:totals}}
\tablewidth{100pt}
\tablehead{
\colhead{Source} & \colhead{}   & \colhead{Log[L(H$\alpha$)]} & \colhead{Log[L(Pa$\alpha$)]} & \colhead{Log[L(21)]} & \colhead{E(B--V)} 
\\
\colhead{(1)} & \colhead{(2)}     & \colhead{(3)} & \colhead{(4)}  & \colhead{(5)} & \colhead{(6)} 
\\
}
\startdata
JWST Footprint\tablenotemark{a}   & Observed  & 41.21 & 40.52 & 42.95  & 0.27\\
                                                         & Corrected & 41.64 & 40.67 &  & \\
                                                         &                   & 41.60 & 40.65 &  &  \\
\hline
HII Regions ($>$stoch)\tablenotemark{b} & Observed  & 39.836$\pm$0.005 & 39.560$\pm$0.004 & 41.744$\pm$0.006  & 0.833$\pm$0.009\\
                           & Corrected & 40.729$\pm$0.005 & 39.836$\pm$0.013 & &  \\
\hline
Giant HII Regions \tablenotemark{c} & Observed  & 40.174$\pm$0.009 & 39.804$\pm$0.004  & 41.974$\pm$0.019 & 0.71$\pm$0.01\\
                             & Corrected & 40.898$\pm$0.007  & 40.005$\pm$0.016 &  & \\
\hline
\enddata
\tablenotetext{}{(1) The region's or the sum of the regions' luminosities.}
\tablenotetext{}{(2) For each source, the first line indicates the observed luminosities, the second line is for the attenuation corrected luminosities. For the JWST footprint, the second and third lines are for two separate assumption on the attenuation of the faint HII regions: the second line assume the faint regions have the same dust attenuation as the bright ones ($>$stoch) and  the third line assumes the faint regions have the same attenuation as the diffuse emission (see text for details).}
\tablenotetext{}{(3)--(5) Logarithm of the luminosity  of each source at the indicated wavelength, in units of erg~s$^{-1}$.  The H$\alpha$ and Pa$\alpha$ luminosities are corrected 
for the Milky Way foreground extinction.}
\tablenotetext{}{(6) Luminosity--weighted color excess, in mag.} 
\tablenotetext{a}{Luminosities for the JWST footprints are given without uncertainties. The formal uncertainties are small, but systematics are difficult to control.}
\tablenotetext{b}{Sum of the luminosities of the HII regions (after removal of the regions contained within the Giant HII regions) that are above the stochastic IMF sampling limit of Log[L(Pa$\alpha$)$_{corr}$]=36.66.}
\tablenotetext{c}{Sum of the luminosities of the Giant HII regions.}
\end{deluxetable}

\section{Discussion} \label{sec:discussion}

\subsection{Modeling The Age Dependence of Hybrid Infrared SFR Calibrations}\label{subsec:modeling}

The value of $b$ at 24~$\mu$m (Equation~\ref{equa:b_definition} and Figure~\ref{fig:b_vs_Pa}) derived  for the HII regions in the two galaxies NGC\,5194 and NGC\,628 can be compared with derivations of the same scaling factor by other authors. Here we expand on the discussion presented in \citet{Calzetti+2024} by adding stellar population models that take into account different possible star formation histories when comparing the scaling factor $b$ between regions of different sizes, including whole galaxies. Table~\ref{tab:b_parameter} summarizes previous results for ease of comparison. The HII region results of \cite{Belfiore+2023} are considered lower limits in this work, as those authors' fits include regions affected by stochastic IMF sampling, which lower the best--fit $b$ value \citep{Fumagalli+2011}. Despite the limited number of available measurements, it is clear, as already remarked in \citet{Calzetti+2024}, that $b$ decreases in value for increasing region sizes; in particular $b$ is $\sim$4.4 times larger in HII regions than in whole galaxies. 

To test how well these scaling factors work, we combine our sample of HII regions with the sample of $\sim$0.5~kpc regions discussed in Section~\ref{subsec:IRSFR} and other samples of kpc--sized regions and whole galaxies for which L(P$\alpha$), L(H$\alpha$) and L(24) are available, using data from the literature. We add: the 21 $\sim$2~kpc regions analyzed by \citet{Calzetti+2007},  integrated measures  for 46 local star--forming galaxies from the SINGS sample \citep{Kennicutt+2009}, 10 local starburst galaxies from \citet{Engelbracht+2008}, and a sample of 24 luminous infrared galaxies from \citet{Alonso+2006}. For the luminous infrared galaxies samples, only IRAS 25~$\mu$m data are available; \citet{Kennicutt+2009} showed that these are indistinguishable from the Spitzer 24~$\mu$m measurements. In addition, the luminous infrared galaxies lack H$\alpha$ measurements and we make the reasonable assumption that L(H$\alpha$) is negligible relative to L(24). The Pa$\alpha$ emission for the starbursts from the sample of \citet{Engelbracht+2008} is not corrected for dust attenuation due to the aperture mismatch between the H$\alpha$ data, which are the the whole galaxy, and the Pa$\alpha$ data, which are for the inner 50$^{\prime\prime}$ diameter region. Thus, the Pa$\alpha$ luminosity for this sample is underestimated, possibly by up to 65\%, depending on the amount of attenuation present in the galaxies. For the H$\alpha$ luminosity of the starbursts, from \citet{Engelbracht+2008} updated using the survey by \citet{Kennicutt+2008} where possible, we only use the observed values in what follows. For the SINGS sample of galaxies, only H$\alpha$, attenuation--corrected using H$\beta$($\lambda$ 0.4861~$\mu$m), is available; we assume that, in star forming galaxies, the ionized gas luminosity is dominated by emission from low--extinction regions, and the H$\alpha$/H$\beta$ ratio is sufficient to recover the intrinsic luminosity \citep{Kennicutt+2009, KennicuttEvans2012}. The attenuation corrected H$\alpha$ is converted to an equivalent Pa$\alpha$ by assuming H$\alpha$/Pa$\alpha$=7.82. The samples from \citet{Calzetti+2007} and \citet{Kennicutt+2009} are not independent: they are all drawn from the SINGS sample, although the two papers analyze different region sizes. 
In total, the combined sample of HII regions, galactic regions and galaxies has 658 data points. 

\begin{figure}
\plottwo{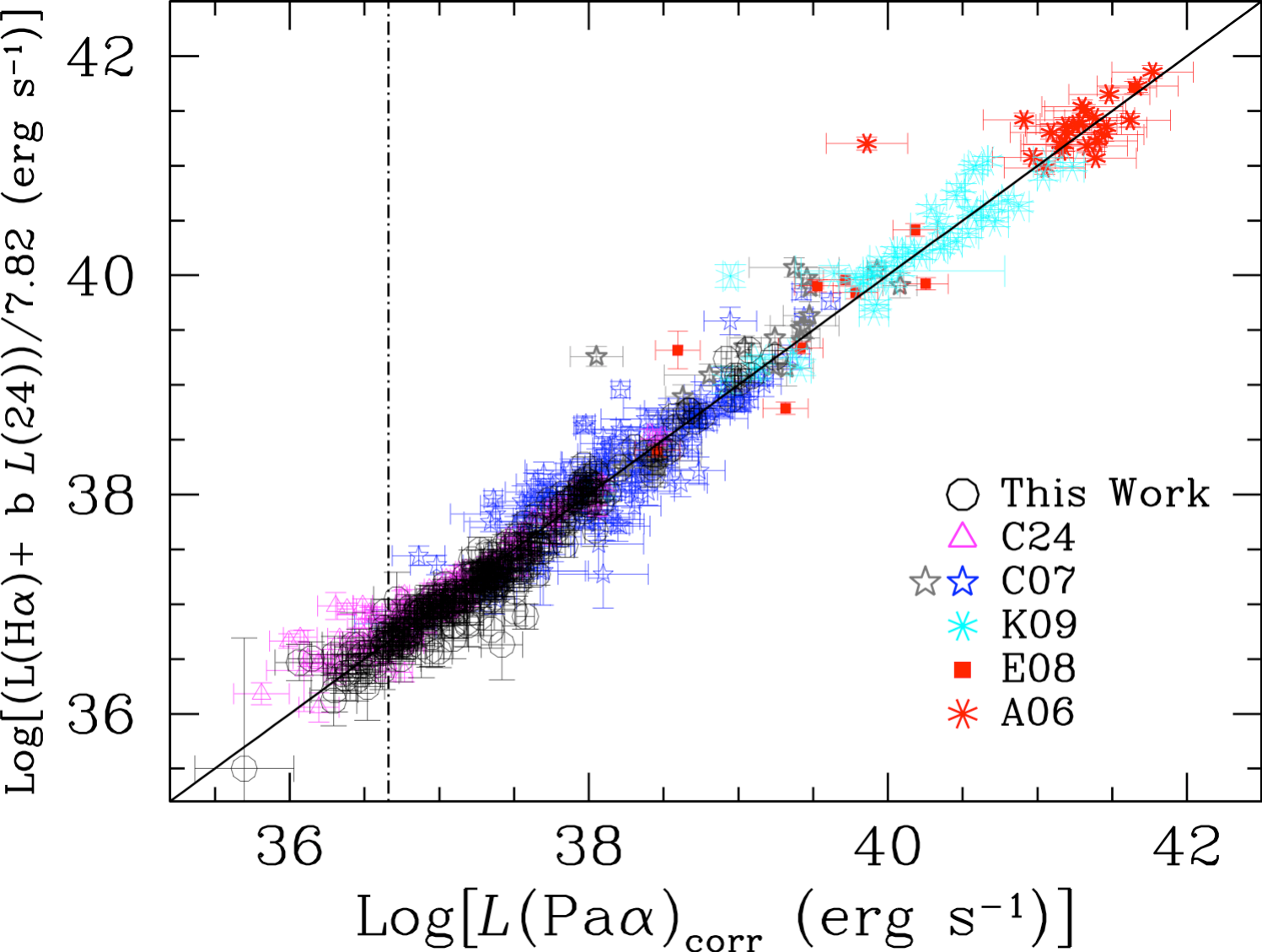}{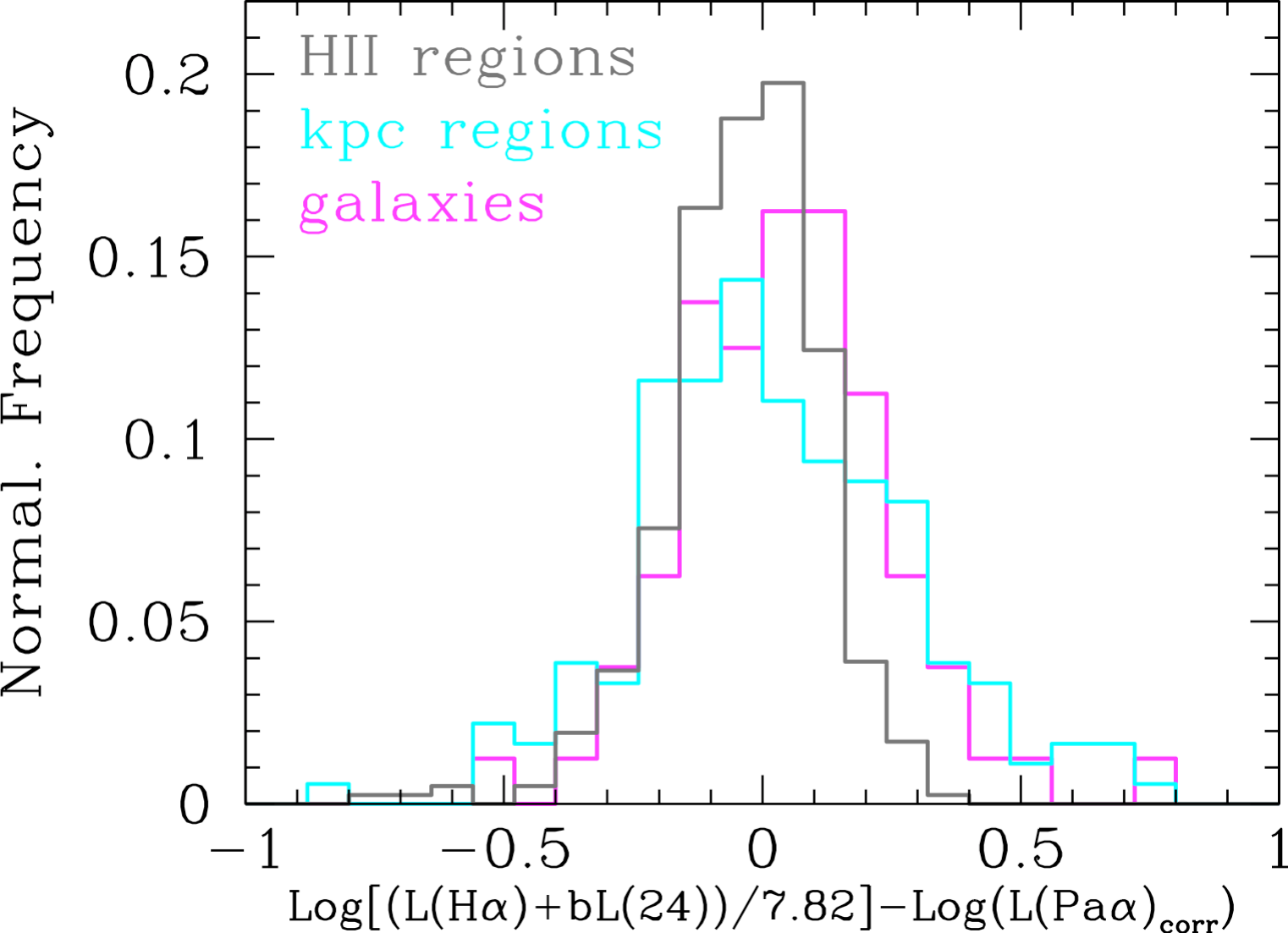}
\caption{{\bf (Left:)} The hybrid [L(H$\alpha$)+b L(24)] luminosity as a function of the attenuation--corrected Pa$\alpha$ luminosity for several samples: the HII regions in NGC\,5194 (black empty circles, This Work) and in NGC\,628 (magenta empty triangles, C24; \citet{Calzetti+2024}); the 0.5~kpc regions (blue stars)  and 2~kpc (grey stars) regions from \citet{Calzetti+2007} (C07); the integrated galaxy measurements of star--forming galaxies  by \citet{Kennicutt+2009} (cyan asterisks, K09), of  local starburst galaxies by \citet{Engelbracht+2008} (red filled squares, E08) and of luminous infrared galaxies  by \citet{Alonso+2006} (red asterisks, A06). All data are shown with 1$\sigma$ uncertainties. The hybrid luminosity is divided by 7.82 (the H$\alpha$/Pa$\alpha$ ratio) to bring the two axes to the same luminosity range. For each subsample: HII regions, kpc--size regions, and whole galaxies, the appropriate value of the scaling factor $b$ is adopted: 0.088, 0.031, and 0.020, respectively. The 1--to--1 relation is shown as a black solid line. The transition luminosity above which stochastic sampling of the IMF is mitigated is shown as a vertical dot--dashed line. {\bf (Right):} The histogram of the data in the Left panel relative to the 1--to--1 line, divided according to subsample: HII regions (grey histogram, samples from this work and \citet{Calzetti+2024}), kpc--size regions (cyan histogram, samples from \citet{Calzetti+2007}) and whole galaxies (magenta histogram, samples from \citet{Kennicutt+2009}, \citet{Engelbracht+2008}, and \citet{Alonso+2006}). 
}
\label{fig:Robplot}
\end{figure}

The data from the different samples are shown in Figure~\ref{fig:Robplot}, left, using the scaling factor value appropriate for each sample: b=0.088 for the HII regions samples of NGC\,628 and NGC\,5194; b=0.031 for the $\sim$kpc--sized regions from \citet{Calzetti+2007}; and b=0.020 for the galaxy samples of \citet{Kennicutt+2009}, \citet{Engelbracht+2008}, and \citet{Alonso+2006}. The 1--to--1 line (in log--log luminosity scale) is shown on the plot for comparison. The histogram of the distribution of the data relative to the 1--to--1 line for the three subsamples -- HII regions, galaxy regions, and whole galaxies, -- is shown in Figure~\ref{fig:Robplot}, right. The datapoints are reasonably well distributed around the 1--to--1 line, when using the b--value appropriate for each subsample. The source showing the most deviation from this trend in Figure~\ref{fig:Robplot}, left, is from the sample of \citet{Alonso+2006}. As noted by these authors, the galaxy, IC\,860, may include an Active Galactic Nucleus, which would explain its over--luminous 24~$\mu$m emission relative to the Pa$\alpha$.

The models of the scaling factor are developed in Appendix~\ref{sec:appendixA}, using the definition in equation~\ref{equa:b_definition} to derive $b$ from the SEDs of stellar populations; here we outline the basic approach. The H$\alpha$ emission is almost entirely contributed by photoionization in galaxies, with only a small contribution from other processes \citep[e.g., shocks,][]{Reynolds+1984, Reynolds+1990, Ferguson+1996, Hoopes+1996, Hoopes+2003, Voges+2006, Oey+2007, Zhang+2017}. This implies that populations younger than 6~Myr are required for modeling $L(H\alpha)_{abs}$ in equation~\ref{equa:b_definition}. When modeling galaxies or large ($\gg$100~pc) regions, star formation can be considered constant over these timescales. 

Conversely, the infrared emission is contributed by stellar populations of all ages, including Gyr--old ones \citep{LonsdaleHelou1987, Buat+1988, RowanRobinson+1989, Sauvage+1992, Walterbos+1996, Buat+1996, Calzetti+2010, Bendo+2012, Dale+2012, SmithDunne+2012, Magnelli+2014, Boquien+2016, Gregg+2022, Leroy+2023}. Modeling $L(24) = f L(bol)_{abs}$ (Equation~\ref{equa:b_definition}) requires analyzing the possible ranges of  $f$ and $L(bol)_{abs}$, separately. The fraction of infrared emission emerging at 24~$\mu$m, $f$, changes by less than a factor of three in galaxies and four in HII regions \citep{Rieke+2009, Calzetti+2010, Relano+2013}; however, this change is linked to surface brightness, not region size, and the surface brightness of HII regions has significant overlap with the surface brightness of large galaxy regions (e.g., Figure~\ref{fig:l21_vs_lpa}, right). We use $f$=0.14 as a reference value for the models, which is a mean value for both HII regions and star--forming galaxies, although variations to this assumption are also explored.

$L(bol)_{abs}$ increases with increasing E(B--V) and/or increasing bolometric output of the stellar population. At constant SFR, the latter can only occur by increasing the duration $\tau$ of the star formation, which progressively increases the population of low--mass, long--lived stars and the light they contribute to the non--ionizing SED.  A simple increase in E(B--V) cannot explain the decrease of $b$ with increasing region size, since, as shown in Table~\ref{tab:totals}, galaxies and large regions of galaxies are less extincted on average than the HII regions they contain. This has already been reported in the literature and is also seen in Figure~\ref{fig:corr_lum}, right, where the extinction--luminosity relation for galaxies by \citet{Garn+2010} marks the lower envelope of the same relation for HII regions. 
Appendix~\ref{sec:appendixA} presents the expected trend of the $b$ scaling factor as a function of the star formation duration $\tau$, for different assumptions of the star formation history (constant, exponentially decreasing and exponentially increasing), of the amount of dust attenuation in populations of different ages, and of the amount of dust attenuation in the ionized gas. 

In order to compare the models with the observed $b$ values, we need to relate star formation duration to region size. We leverage the fact that (1) our measurements are obtained in regions where the background emission from the galaxy has been removed, and (2) star formation is hierarchically clustered out to about 1~kpc or so and larger scales correspond to older mean ages for the stellar populations \citep[e.g.,][]{Efremov+1998,Delafuente+2009,Elmegreen+2011,Grasha+2015,Gouliermis+2015,Grasha+2017a,Grasha+2017,Gouliermis+2018,Elmegreen+2018, Shashank+2025}. Because of (1), the sub--galactic regions can be attributed the mean ages that are appropriate for their location in the hierarchy.  

For the HII regions, we adopt the mean age we derive from the EW-versus--luminosity trend (Figure~\ref{fig:corr_lum}, left), $\tau\sim$3~Myr, with some uncertainty attached to this value to span the range of possible HII region ages (see Table~\ref{tab:b_parameter}). For this discussion, we equate the mean $b$ value of the HII regions to the expected $b$ value resulting from constant star formation over $\tau\sim$~a~few~Myr.  
While this equivalence is generally not true, we argue that it is reasonable in our case for the following reason. 
Our sample of $>$300 HII regions is likely to distribute uniformly in age over the range $\sim$0--6~Myr (Section~\ref{subsec:properties}) and we select our regions to be massive enough to be above the limit of stochastic sampling for the stellar IMF, limiting the effects of biases in deriving average H$\alpha$ and infrared luminosities. Thus, this large sample is equivalent to sampling an event of constant star formation over the timescale range listed in Table~\ref{tab:b_parameter}.

For the $\sim$kpc regions, we use the crossing times of stars and star clusters. Cluster--to--cluster dispersion velocities range from a few km/s to $\sim$20 km/s  out to $\sim$1~kpc \citep{Whitmore+2005, Grasha+2017}. Gas dispersion velocities due to turbulence are relatively small, $\sim$1--2km/s \citep{Heyer+2004,Heyer+2009}, and the same small values are likely inherited by the stars; however, shear could increase the velocities by a factor of a few and shear rather than turbulence may be at the root of the observed hierarchy of star formation \citep{Elmegreen+2018}. With these velocities, we obtain a crossing time of 30--130~Myr for 0.5~kpc. This timescale agrees with the one measured from the age--separation relation of star clusters in several galaxies, which supports the turbulent star formation scenario \citep{Efremov+1998, Delafuente+2009, Grasha+2017}. For the specific case of NGC\,5194, \citet{Grasha+2017} derive a timescale of about 40--60~Myr at 0.5~kpc. As a caveat, these adopted timescales are subject to additional uncertainty, as crossing times are inversely proportional to the 1/4th power of pressure \citep{Elmegreen1989} and ISM pressure can change by over an order of magnitude within star--forming galaxies \citep{Querejeta+2023}, and more in starbursts and mergers. For whole galaxies, we assume that the duration of the star formation is $\gtrsim$10~Gyr.

\begin{figure}
\plottwo{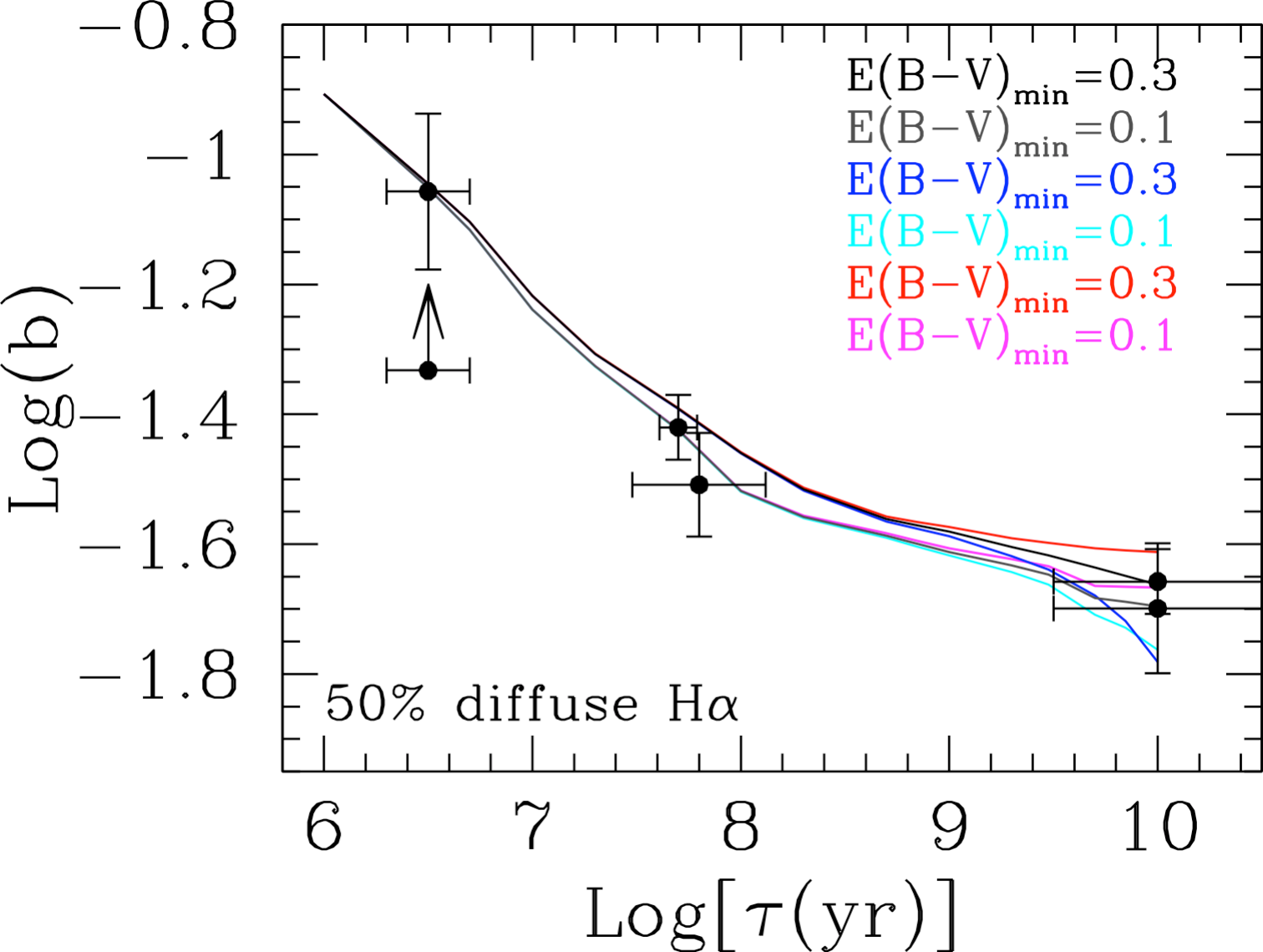}{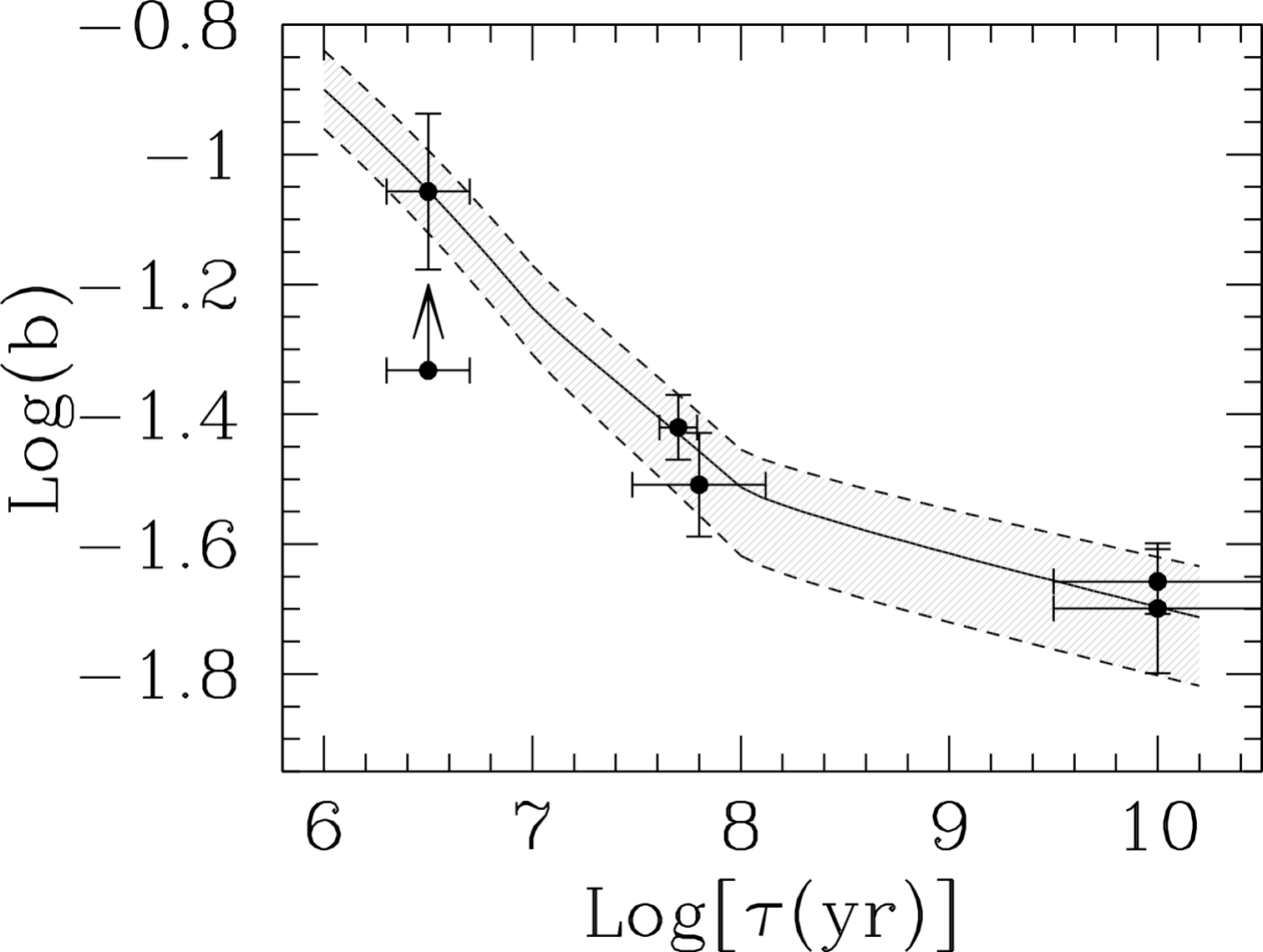}
\caption{{\bf (Left):} The scaling factor $b$ as a function of the duration of star formation $\tau$ for the data listed in Table~\ref{tab:b_parameter}, compared with model expectations. Three star formation histories are included, with decreasing color excess for stellar populations of increasing age and 50\% clustered--50\% diffuse H$\alpha$ emission, as described in Appendix~\ref{sec:appendixA}. $f$=0.14 for this model. The three SFHs are: constant (black and grey lines), decreasing (blue and cyan lines) and increasing (red and magenta) star formation with $\tau$, see details in Appendix~\ref{sec:appendixA}. {\bf (Right):} The smooth analytical representation of the trend of the scaling factor $b$ and its scatter as a function of star formation duration, as described in equation~\ref{equa:b_bestfit}, shown as black continuous and dashed lines, respectively.}
 \label{fig:b_model}
\end{figure}

The models from Appendix~\ref{sec:appendixA} 
are used here for comparison with the observed $b$ (Figure~\ref{fig:b_model}, left). The models are a good representation of the trend of the scaling factor $b$ with $\tau$, although we need to keep in mind that there is a non--negligible scatter in the models for different parameter choices, including, but not limited to, the dust attenuation in both the stellar populations and the ionized gas, the fraction of infrared emission emerging at 24~$\mu$m and, to a lesser extent, the star formation history. We provide an analytical expression for $b$ as a function of star formation duration $\tau$, 
by interpolating through the observational data, including their scatter:
\begin{equation}
SFR (M_{\odot} yr^{-1}) = 5.45\times 10^{-42} [L(H\alpha) + b(\tau)  L(24)],
\label{equa:b_tau}
\end{equation}
where $b(\tau)$ can be written as:
\begin{eqnarray} \label{equa:b_bestfit}
Log[ b(\tau)] &= 0.1^{+0.06}_{-0.06} - {\rm exp}\Bigl[ {(Log(\tau)-6.)\over 3.45^{+0.05}_{-0.10}}\Bigr]  \ \ \ \ \ \ \ \ \ \ \ \ \ \ \ \ \ \ \ \ \ \ for \ \ \ \ \ 6.\le Log(\tau)  < 7.,\nonumber\\
                    &= -0.236^{+0.065}_{-0.072} - {\rm exp}\Bigl[ {(Log(\tau)-7.)^{0.9} \over 4.10^{-0.10}_{-0.40}}\Bigr]  \ \ \ \ \ \ \ \ \ \ \ for \ \ \ \ \ 7.\le Log(\tau)  < 8.,\\
                    &= -0.512^{+0.058}_{-0.106} - {\rm exp}\Bigl[ {(Log(\tau)-8.)^{0.8} \over 10.3^{+1.1}_{-0.0}}\Bigr]  \ \ \ \ \ \ \ \ \ \ \ for \ \ \ \ \ 8.\le Log(\tau)  \le 10.,\nonumber
\end{eqnarray}
where the timescale $\tau$ is in years, and $b(\tau)$ is adimensional. Figure~\ref{fig:b_model}, right, shows the above analytic expression in relation to the observations. The superscript (subscript) for each coefficient in equation~\ref{equa:b_bestfit} marks the upper (lower) envelope in the figure. 

For all practical purposes, equations~\ref{equa:b_tau} and ~\ref{equa:b_bestfit} can be used to derive SFRs for systems with star formation durations between $\sim$1~Myr and $\sim$10~Gyr, with an accuracy that depends on the relative luminosity of the H$\alpha$ to the 24~$\mu$m emission, but is about 25\%--30\% at all ages. 

The above calibration has been derived for $\sim$solar metallicity systems and metallicity is expected to have second--order effects on $b(\tau)$. The absorbed portion of {\em both} L(H$\alpha$) and L(bol) decreases for decreasing dust content, and, therefore, decreasing metallicity, with a similar dependence on E(B--V) \citep{Calzetti+2024}. This implies that, to first order, the ratio L(H$\alpha$)/L(bol) does not change with E(B--V). However, the ionizing photon rate is dependent on metallicity, being higher at lower metal content  \citep{Chisholm+2019}, which affects the nebular lines luminosity, and can result in an increase of $b(\tau)$ for decreasing metallicity. Finally, $f$=L(24)/L(IR) can also depend on metallicity \citep[e.g.,][]{RemyRuyer+2015}.

\begin{deluxetable*}{llll}
\tablecaption{Infrared Scaling Factor for Hybrid SFR\label{tab:b_parameter}}
\tablewidth{0pt}
\tablehead{
\colhead{Region Diameter} & $\tau$(Myr)$^1$& \colhead{b$^2$} &\colhead{Reference} \\
}
\decimalcolnumbers
\startdata
50--120~pc & 2--5& 0.088$\pm$0.025 & This work, \citet{Calzetti+2024}\\
$\sim$100~pc & 2--5  & $>$0.035 & \citet{Belfiore+2023}$^3$\\ 
$\sim$500~pc  & 40--60   & 0.038$\pm0.005$ &\citet{Kennicutt+2007}$^4$\\
$\sim$500~pc  & 30--130   & 0.031$\pm0.006$ & \citet{Calzetti+2007}$^5$\\
$>$10~kpc       & $\sim$10,000 & 0.022$\pm0.002$ & \citet{Zhu+2008}$^6$\\
$>$10~kpc       & $\sim$10,000 & 0.020$\pm0.005$ & \citet{Kennicutt+2009}$^7$\\
\enddata
$^1$ Star formation duration attributed to the region, see text.\\
$^2$  Best fit L(24)/L(H$\alpha$) scaling constant for the hybrid H$\alpha$+24 SFR indicator.\\
$^3$ Derived from $\sim$20,000 HII regions in 19 galaxies, including regions affected by stochastic IMF sampling. The scaling parameter increases to b$\sim$0.08--0.09, when excluding such regions in the \citet{Belfiore+2023} sample.\\
$^4$ Derived from 42 regions in NGC\,5194.\\
$^5$ Derived from 160 regions in 21 $\sim$solar metallicity galaxies.\\
$^6$ Derived from 379 galaxies star--forming galaxies.\\
$^7$ Derived from 68 nearby ($<$140~Mpc) star--forming galaxies.\\
\end{deluxetable*}

\subsection{Calibration of SFR(24) from HII Regions to Galaxies}\label{subsec:SFR24model}

In Section~\ref{subsec:IRSFR} we briefly presented four mechanisms which may be the source of the super--linear correlation (in log--log scale)  between L(24) (or L(21)) and L(Pa$\alpha$)$_{corr}$, and which we summarize here again for clarity: 
\begin{enumerate}
\item lower--luminosity regions/galaxies tend to be less extincted than brighter ones, implying that a larger fraction of the ionizing and non--ionizing photons at low luminosities emerges  directly in the UV--optical and are not captured by the infrared emission; this depresses L(24) relative to L(Pa$\alpha$)$_{corr}$ for fainter regions relative to brighter ones;
\item leakage of ionizing photons out of HII regions affects bright regions more than faint ones \citep{Pellegrini+2012}, depressing L(Pa$\alpha$)$_{corr}$ at high luminosity more than at low luminosity, while leaving L(24) unaffected; 
\item direct absorption of ionizing photons by dust, which depresses the L(Pa$\alpha$)$_{corr}$ of bright (dustier) regions proportionally more than fainter (less dusty) regions and which may accompany the first mechanism for sufficiently high dust content \citep{Inoue+2001, Dopita+2003, Krumholz+2009, Draine+2011}; and 
\item higher luminosity regions produce `hotter' infrared SEDs, which increase the fraction of 24~$\mu$m emission relative to the total infrared emission \citep{Relano+2013, Rieke+2009, Calzetti+2010}. Although we do not discuss the exact nature of `hotter' IR SEDs, we note that dust temperature may not be the only cause of enhanced L(24)/L(IR) values; \citet{Relano+2018}, for instance, find that shattering of large dust grains into small dust grains in star forming regions can produce similar effects in the IR SED. 
\end{enumerate}
These four mechanisms are likely to affect, with various degrees of relevance, the behavior of the L(24)--L(Pa$\alpha$) correlation in regions drawn from uniform samples (e.g., HII regions). A fifth mechanism:
\begin{enumerate}
\setcounter{enumi}{4}
\item the contribution of stellar populations of all ages to the infrared emission as discussed and quantified in Appendix~\ref{sec:appendixA}; 
\end{enumerate}
becomes important when considering regions across different scales, e.g., samples that contain mixes of HII regions, galactic regions, and entire galaxies. 

\begin{figure}
\plotone{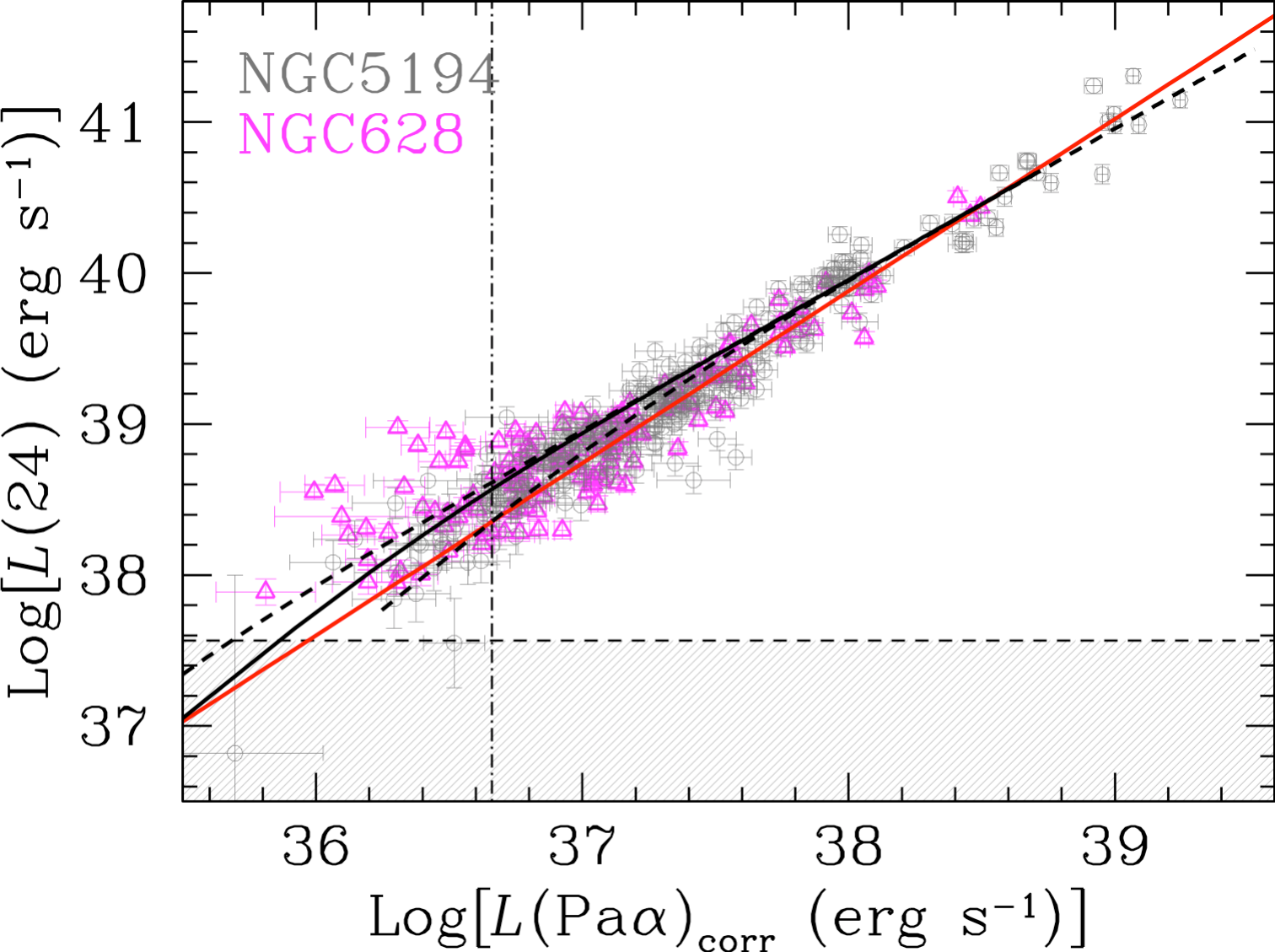}
\caption{The 24~$\mu$m luminosity as a function of the attenuation--corrected Pa$\alpha$ luminosity for the HII and Giant HII regions in NGC\,5194 (grey empty circles) and the HII regions in NGC\,628 (magenta empty triangles) with their 1$\sigma$ uncertainties. The horizontal dashed line and grey region show the location of the 5$\sigma$ threshold at 24~$\mu$m in NGC\,5194. The best fit line to the cumulative sample of both galaxies is shown in blue. The 3~Myr model from Appendix~\ref{sec:appendixB} is shown as a black solid line; larger values of both luminosities correspond to larger values of the color excess E(B--V), from 0.05~mag to 5~mag. Black dashed lines mark the 90\% scatter in the E(B--V)--L(Pa$\alpha$)$_{corr}$ relation from Figure~\ref{fig:corr_lum}. Limiting the maximum E(B--V) to 3~mag would decreases the maximum L(Pa$\alpha$)$_{corr}$ in each model by 0.37~dex.}
 \label{fig:SFR24_HIImodel}
\end{figure}

We use the models from Appendix~\ref{sec:appendixB} to compare the best fit through the  L(24)--L(Pa$\alpha$)$_{corr}$ measurements of HII regions in NGC\,5194 and NGC\,628 with the expected model trend from the first mechanism listed above. 
The model is shown in Figure~\ref{fig:SFR24_HIImodel} as the black lines, to be compared with the best fit line (in log--log space) through the data. The two, model and best fit, follow each other closely, well within the scatter in the data, suggesting that even a simplified model like the one in Appendix~\ref{sec:appendixB} is sufficient to explain the super--linear relation between L(24) and L(Pa$\alpha$)$_{corr}$. Although only suggestive, the data appear to continue along a trend with slope$>$1 at high luminosity, L(Pa$\alpha$)$_{corr}>$10$^{38}$~erg~s$^{-1}$, while the model converges to a slope=1. 
If confirmed by future HII region data at higher luminosity than currently available, some of the other mechanisms listed above, such as the increase in the L(24)/L(IR) fraction, direct dust absorption of ionizing photons and/or ionizing photon leakage, are required as secondary effects to keep the slope super--linear.  \citet{Relano+2013}, for instance, find a weak trend of increasing L(24)/L(IR) for increasing H$\alpha$ surface brightness in the HII regions of M33.

\begin{figure}
\plottwo{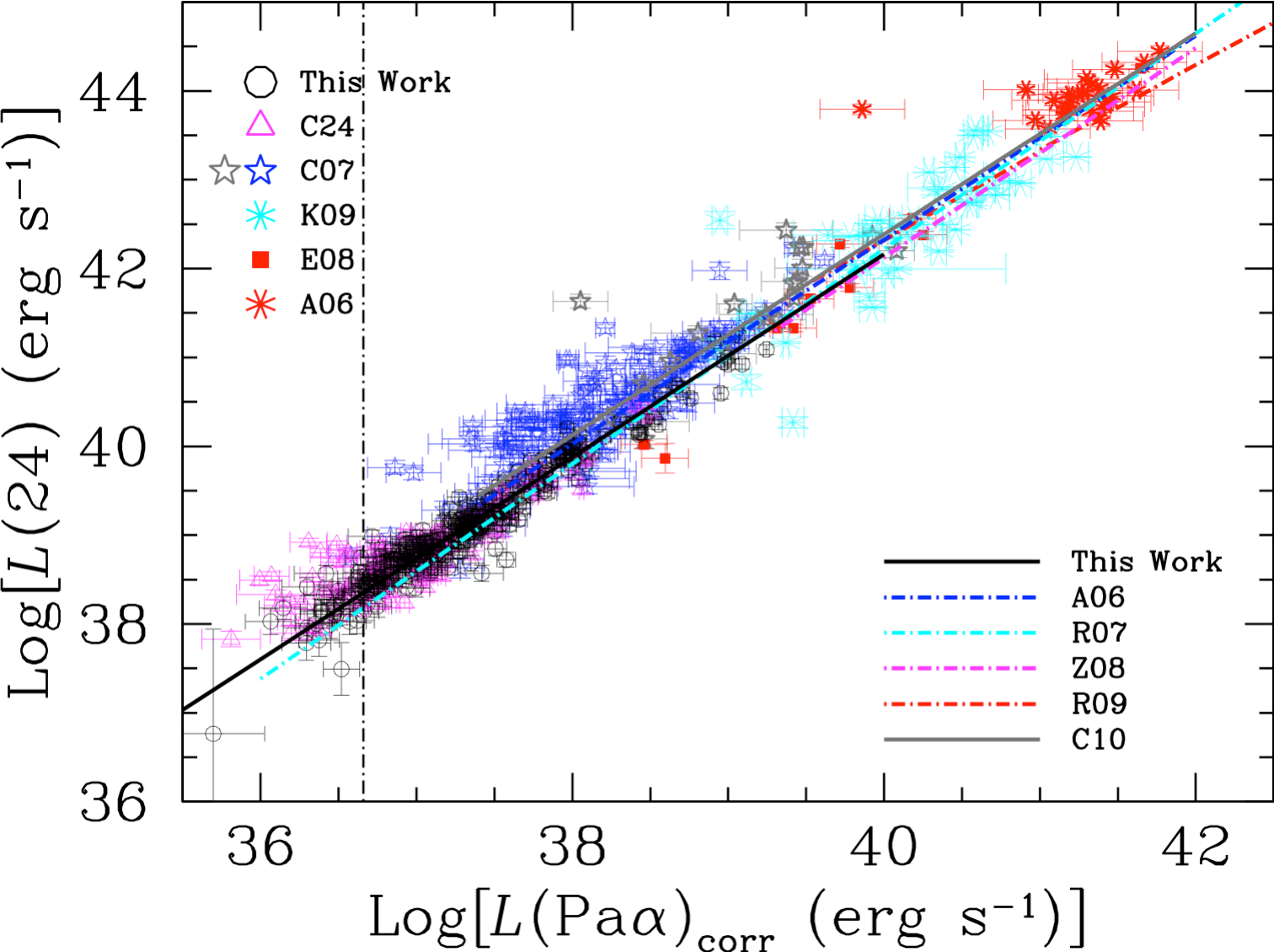}{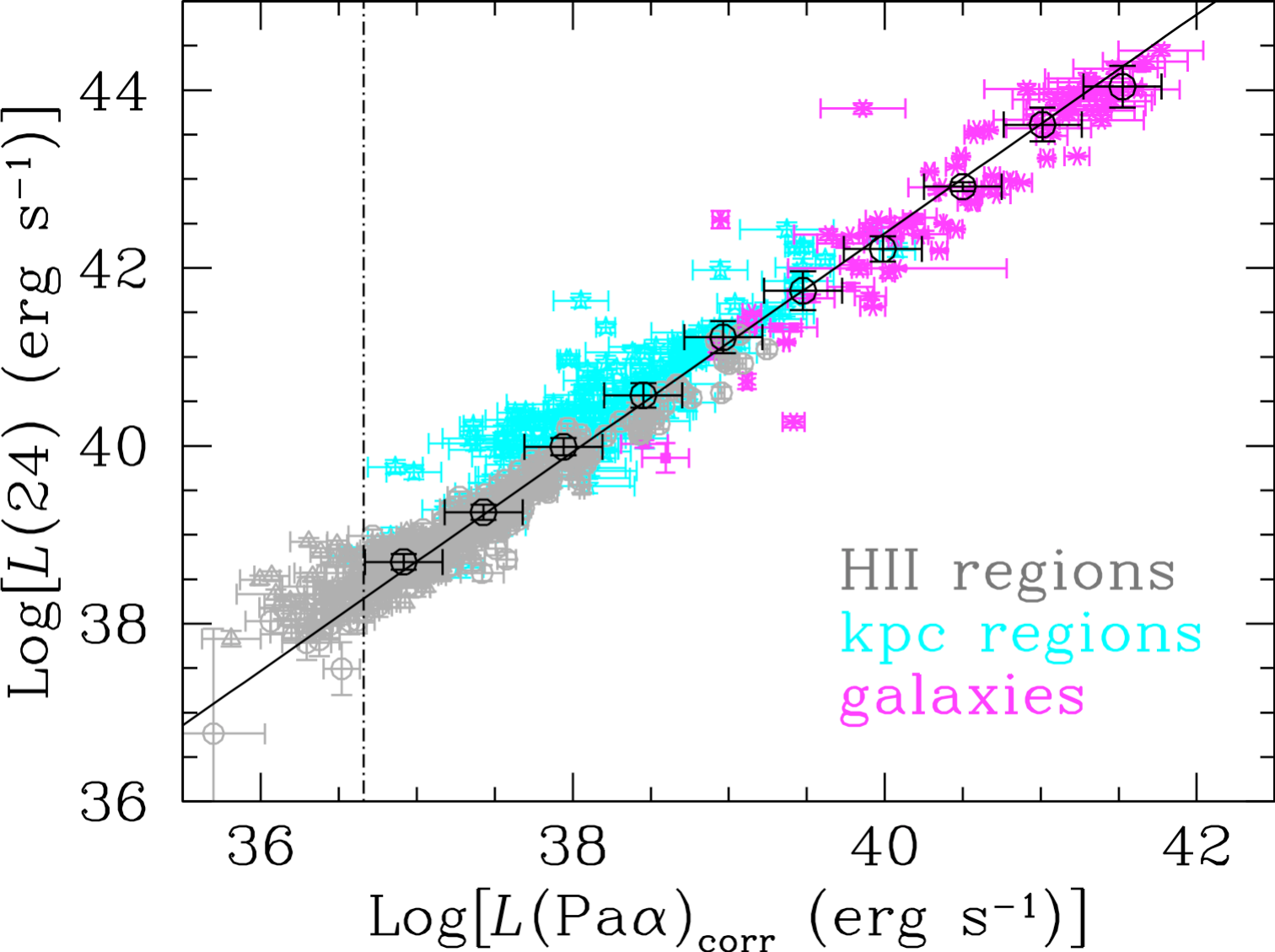}
\caption{{\bf (Left):}  The 24~$\mu$m luminosity as a function of the attenuation--corrected Pa$\alpha$ luminosity for the same samples as those of Figure~\ref{fig:Robplot}, left. 
All data are shown with 1$\sigma$ uncertainties. The transition luminosity above which stochastic sampling of the IMF is mitigated is shown as a vertical dot--dashed line. Several fits from previous authors 
are shown for comparison, using, for each relation, its range of validity: \citet{Alonso+2006} (blue dot--dashed line, A06), \citet{Relano+2007} (cyan dot--dashed line, R07), \citet{Zhu+2008} (magenta dot--dashed line, Z08), \citet{Rieke+2009} (red dot--dashed line, R09), \citet{Calzetti+2010} (grey solid line, C10), in addition to the one derived in this work (black solid line, This Work, equation~\ref{equa:l21lpa} with 21~$\mu$m replaced by 24~$\mu$m).  {\bf (Right):} The same data as in the Left panel, with the binned averages for all data above the stochastic IMF sampling limit (vertical dot--dashed line). Different colors identify sub--samples: HII regions (grey symbols), galactic regions (cyan symbols), and galaxies (magenta symbols). The binned averages are indicated as empty black circles with 1$\sigma$ uncertainties. The best linear fit for all samples in log--log space, across over five orders of magnitude in Pa$\alpha$ luminosity, is shown as a black line.}
 \label{fig:SFR24_all}
\end{figure}

The comparison between HII regions and $\sim$0.5~kpc regions discussed in Section~\ref{subsec:IRSFR} is extended here to include larger regions and whole galaxies, using the data from the literature introduced in Section~\ref{subsec:modeling}. 
Figure~\ref{fig:SFR24_all} shows that the combined data from all samples mark a fairly tight correlation across more than 5 orders of magnitude in Pa$\alpha$ luminosity and more than 6 orders of magnitude in the 24~$\mu$m luminosity. The correlation is not surprising since luminosities are always correlated with each other, especially when systems at different distances are included. However, it is interesting that the correlation remains super--linear. The left panel shows the individual samples with different colors and symbols, together with several fitting relations from the literature, as indicated in the caption.  The selected fitting relations are all non--linear (slope $>$1 in log--log scale), except for the linear relation by \citet{Rieke+2009}, which applies to galaxies with 24~$\mu$m luminosity $>$10$^{42.6}$~erg~s$^{-1}$ and becomes non--linear at L(24)$>$10$^{43.7}$~erg~s$^{-1}$. Most non--linear fits published in the literature do provide a reasonable representation of the data, provided they are kept within their range of validity (Figure~\ref{fig:SFR24_all}, left). 

A single line fit through all data above the stochastic IMF sampling limit (596 points) gives:
\begin{equation}
Log[L(24)] = (1.23\pm 0.01) Log[L(Pa\alpha_{corr})] - (6.81\pm 0.42),
\label{equa:l24_all}
\end{equation}
with scatter=0.14~dex, where both L(24) and L(Pa$\alpha$)$_{corr}$ are in units of erg~s$^{-1}$. Figure~\ref{fig:SFR24_all}, right, shows the best fit line in comparison with the data from  
the combined sample now separated into the same three sub--samples of Figure~\ref{fig:Robplot}, right: HII regions (grey), galactic regions (cyan) and whole galaxies (magenta). The fit aligns well with the binned averages along the sample, also shown in the figure. The best fit slope, with value 1.23, is 4~$\sigma$ higher than the one for HII regions alone (slope=1.14), also confirmed by a random drawing test (Appendix~\ref{sec:appendixE}), and is likely capturing the fact that larger regions and whole galaxies include infrared emission that is not due to young star forming regions. The added contribution to L(24) thus increases for increasing luminosity, producing the super--linear slope. 

\begin{figure}
\plottwo{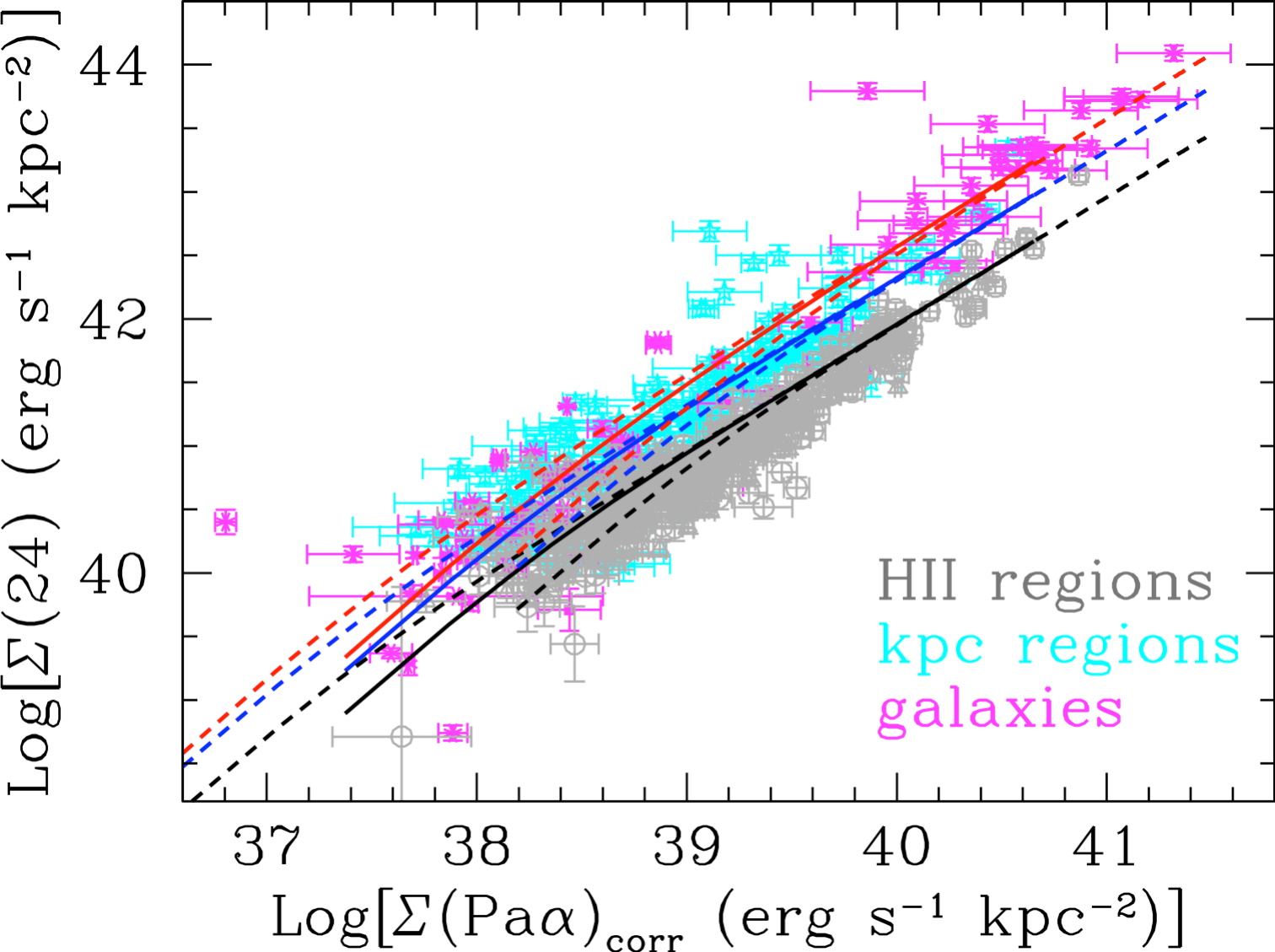}{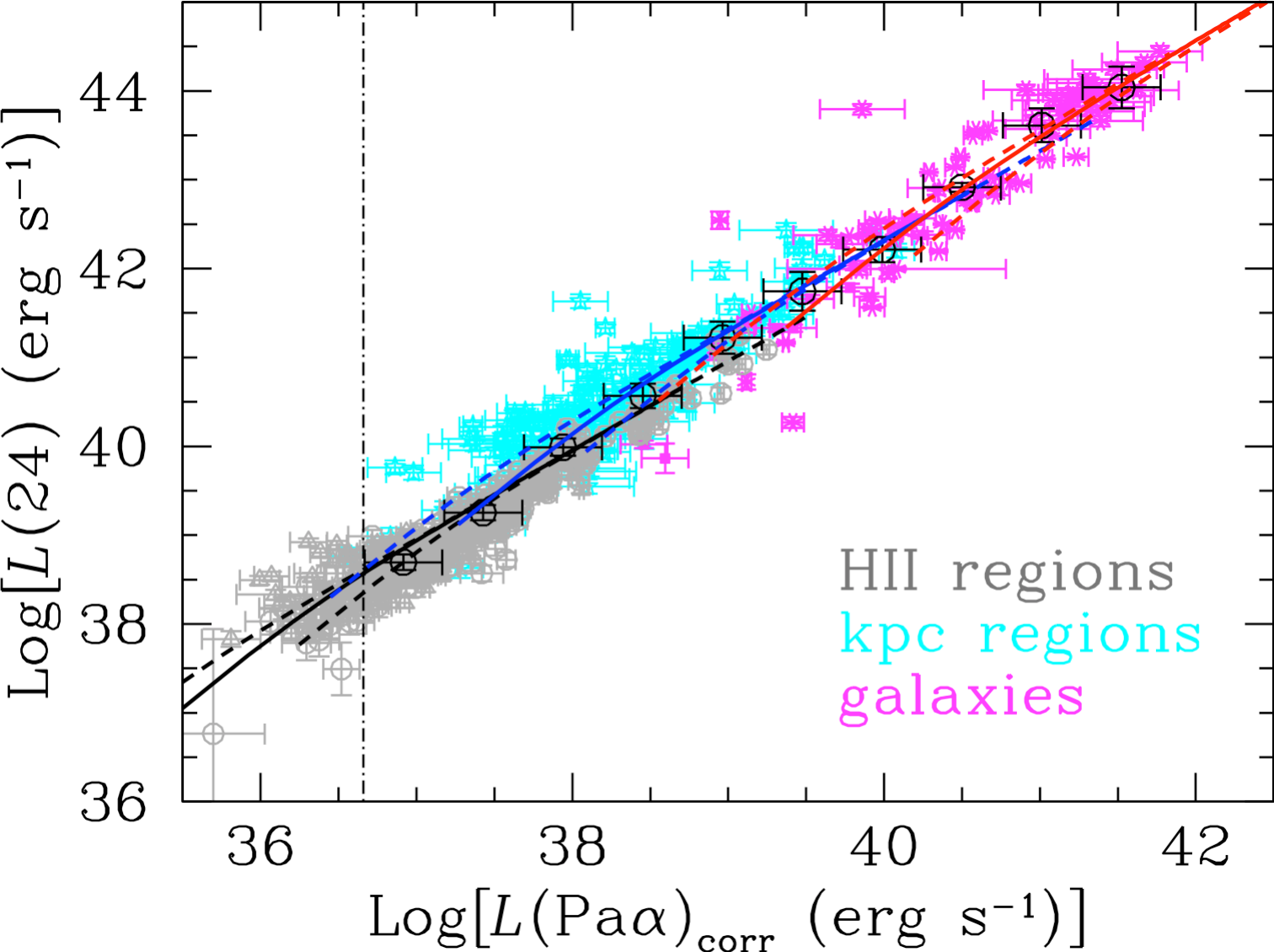}
\caption{{\bf (Left):}  The luminosity surface densities at 24~$\mu$m and at Pa$\alpha_{corr}$ for the three sub--samples of HII regions (grey symbols), galactic regions (cyan symbols) and galaxies (magenta symbols), with their 1$\sigma$ uncertainties, are compared with the models of Appendix~\ref{sec:appendixB}. The models are: a 3~Myr old instantaneous burst of star formation (black lines), and constant star formation with 100~Myr (blue lines) and 10~Gyr (red lines) durations, respectively. The dashed lines are the same as those in Figure~\ref{fig:SFR24_HIImodel}. Data and models both show similar increases in $\Sigma$(24) at constant $\Sigma$(Pa$\alpha$)$_{corr}$, as expected if HII regions, galactic regions and galaxies mark a progression in star formation duration. {\bf (Right):} The same as the left panel, but for luminosities. The model's luminosity surface densities are changed to luminosities using a characteristics spatial scale associated to each model (see text). The shifted models line up to mimic a linear relation with slope $>$1, as observed.}
 \label{fig:siglum_all}
\end{figure}

Comparisons with models of different ages/durations from Appendix~\ref{sec:appendixB} confirm this speculation: the models follow the data reasonably well, both in the case of luminosity surface densities and of luminosities (Figure~\ref{fig:siglum_all}, left and right panels, respectively). The models, a 3~Myr old instantaneous burst population, and two models for constant star formation, at 100~Myr and 10~Gyr, are taken as representative of $\Sigma$(24)--$\Sigma$(Pa$\alpha$)$_{corr}$ trends for the sub--samples of HII regions, galactic regions, and galaxies, respectively.  These models highlight that the differences observed in the data are real  (Figure~\ref{fig:l21_vs_lpa}, right, and Figure~\ref{fig:siglum_all}, left): galactic regions and galaxies have, on average, larger values of $\Sigma$(24) at a given $\Sigma$(Pa$\alpha$)$_{corr}$. Figure~\ref{fig:siglum_all}, right, shows the same models scaled by the characteristic size of each sub--sample: 0.1~kpc for HII regions, 1~kpc for galactic regions, and 10~kpc for galaxies, to compare them with luminosities. The models line up with the data in their respective sub--sample and with each other, resulting in a slope $>$1 and larger than the average slope of each individual model. In summary, one mechanism, the one for which low luminosity regions are more transparent than higher luminosity ones, can already explain much  of the non--linearity in the observed L(24)--L(Pa$\alpha$) relation. When linking together models of populations with different $\tau$, the contribution of the infrared emission from the low mass stars pushes the relation to even steeper slopes, in agreement with observations.

We refrain from drawing additional conclusions from those models, as they were created with a number of simplified assumptions. One example of their limitations  
 is the apparent agreement between the luminous infrared galaxies and the 10~Gyr model, both in $\Sigma$(24)--$\Sigma$(Pa$\alpha$)$_{corr}$ and L(24)--L(Pa$\alpha$)$_{corr}$. Luminous infrared galaxies have been shown to have younger average stellar populations than and different infrared SEDs from normal star forming galaxies \citep[e.g.,][]{Alonso+2006, Marcillac+2006, Rieke+2009, Howell+2010, Pereira+2015, Cortijo+2017, Paspaliaris+2021}. Thus, their agreement with the 10~Gyr model, generated to describe normal star forming galaxies, highlights the degeneracies intrinsic to the models. A future dedicated analysis to this and similar samples may reveal the source of this inconsistency. Despite discrepancies, however, the data distribute fairly symmetrically about the best fit line (equation~\ref{equa:l24_all}), yielding a Log(SFR) prediction with a 1$\sigma$ standard deviation of 0.19~dex across the full luminosity range (Appendix~\ref{sec:appendixC}).

Equation~\ref{equa:l24_all} translates into the following SFR calibration at 24~$\mu$m:
\begin{equation}
SFR(24) = (1.466\pm 0.508)\times10^{-35} L(24)^{(0.8130\pm 0.0066)}\ \ \ \ \ \ \ \ \ \ \ \ \ \ \ \ \ \ \ \ \ \ for \ \ \ \ \ 10^{38}\lesssim L(24) \lesssim 3\times10^{44}.
\label{equa:SFR24_all}
\end{equation}
The calibration of \citet{Relano+2007} is close to the one above, predicting about 20\% (46\%) higher SFRs at the lowest (highest) luminosity in the validity range. 
Given the approximately symmetric dispersion of the data about the best fit, the above calibration will generally yield accurate values of the SFR within 55\% (Appendix~\ref{sec:appendixC}), corresponding to a Half--Width at Half Maximum of 66\%, across the full 24~$\mu$m luminosity range from HII regions to entire galaxies at solar metallicity. 

Unlike $b(\tau)$, SFR(24) depends strongly on metallicity  \citep[e.g.,][]{Calzetti+2007, Calzetti+2010},  due to the fact that galaxies become more transparent for decreasing metal (and dust) content. Thus, use of IR SFR indicators are generally not recommended for use in sub--solar or low--dust systems. In these cases, a hybrid SFR indicator is preferable.

\section{Summary and Conclusions} \label{sec:conclusions}
We have combined new JWST/NIRCam and MIRI  observations in the light of the Pa$\alpha$ nebular line and the 21~$\mu$m dust continuum of the nearby galaxy NGC\,5194 to quantify the calibration of the mid--IR as a SFR indicator for HII regions, after correcting the Pa$\alpha$ emission for dust attenuation. To expand on our results, we have joined the 254 HII regions of NGC\,5194 with the 143 HII regions of NGC\,628 analyzed by \citet{Calzetti+2024}, finding that the Log[L(21)]--Log[L(Pa$\alpha$)] is linear with a slope $>$1. We have also found that the scaling factor $b$ for the hybrid [L(H$\alpha$)+b L(24)] SFR indicator is a factor $\sim$4.4 larger in HII regions than in galaxies. Our samples of HII regions explicitly choose regions ionized by single star clusters to ensure that a single (young) system contributes to both the ionized and infrared emission. However, we also find that adding Giant HII regions ionized by a group of star clusters to our HII region sample does not change the overall results. Applying models of stellar populations with different star formation histories to our HII regions sample as well as published samples of $\sim$kpc--sized galaxy regions and whole galaxies observed with the Spitzer Space Telescope, we determine:
\begin{itemize}
\item The relation Log[L(24)]--Log[L(Pa$\alpha$)]  steepens when expanding the sample from HII regions--only to HII regions+kpc regions+galaxies, with the slope changing from a value of 1.14 to 1.23 (Equations~\ref{equa:l21lpa} and \ref{equa:l24_all}). The models show this is consistent with two mechanisms acting on the luminosities: (1) lower luminosity regions are less attenuated by dust, with a larger fraction of their emission emerging directly at UV/optical wavelengths, driving the super--linear slope within homogeneous samples (e.g., HII regions--only); (2) larger physical regions correspond to longer durations of the star formation, which increases the contribution of progressively more evolved stars to the infrared emission, and produces a second steepening of the correlation when including mixed samples (HII regions+kpc regions+galaxies).
\item The result for the HII regions+kpc regions+galaxies sample translates  into a relation SFR$\propto$L(24)$^{\alpha}$ with the exponent $\alpha=(0.8130\pm 0.0066)$. The scatter of the data about the mean trend indicates that SFRs derived with equation~\ref{equa:SFR24_all} have a 1$\sigma$ standard deviation of 55\%. This calibration is recommended when using a single--band SFR indicator for metal--rich sources.
\item The scaling factor $b$ for the hybrid SFR indicator depends on the duration of the star formation $\tau$, if physically larger regions are associated with longer durations. The factor $b$ decreases for progressively longer durations, as expected if mechanism (2) above is taken into account. An analytical expression for $b$=$b(\tau)$ is given in equation~\ref{equa:b_bestfit}, and should be relatively insensitive to metallicity variations. Studies that derive SFRs from hybrid tracers should include timescale--dependent 
values of $b$ in their derivations.
\end{itemize}
These results cover six orders of magnitude in 24~$\mu$m luminosity, from 10$^{38}$~erg~s$^{-1}$ to 3$\times$10$^{44}$~erg~s$^{-1}$, with over 600 datapoints spanning HII regions, kpc--sized galaxy regions and whole galaxies, but are limited to solar metallicity systems. Future analyses will need to include sub--solar metallicity systems to generalize the results presented in this work. 

\begin{acknowledgments}

This work is based on observations made with the NASA/ESA/CSA James Webb Space Telescope and retrieved from the Mikulski Archive for Space Telescopes at the Space Telescope Science Institute, which is operated by the Association of Universities for Research in Astronomy, Inc., under NASA contract $ {\rm NAS} \ 5-03127$. These observations are associated with GO programs \# 1783 and  \#3435. The work also made use of archival data from the NASA/ESA Hubble Space Telescope obtained from the Space Telescope Science Institute, which is operated by the Association of Universities for Research in Astronomy, Inc., under NASA contract $ {\rm NAS} \ 5-26555$. HST data were retrieved from the Mikulski Archive for Space Telescopes at the Space Telescope Science Institute. Support for programs GO \# 1783 and  GO \#3435 was provided by NASA through grants from the Space Telescope Science Institute. 

VB and BG acknowledge partial support from grant JWST-GO-01783; DD and KS  acknowledge partial support from grant JWST-GO-03435. AA acknowledges support from the Swedish National Space Agency (SNSA) through grant 2021-00108. ADC acknowledges the support from the Royal Society University Research Fellowship URF/R1/191609. MRK acknowledges support from the Australian Research Council through Laureate Fellowship FL220100020. RSK acknowledges financial support from the European Research Council via the ERC Synergy Grant ``ECOGAL'' (project ID 855130),  from the German Excellence Strategy via the Heidelberg Cluster of Excellence (EXC 2181 - 390900948) ``STRUCTURES'', and from the German Ministry for Economic Affairs and Climate Action in project ``MAINN'' (funding ID 50OO2206). RSK also thanks the Harvard-Smithsonian Center for Astrophysics and the Radcliffe Institute for Advanced Studies for their hospitality during his sabbatical, and the 2024/25 Class of Radcliffe Fellows for highly interesting and stimulating discussions.
MR acknowledges support from project PID2023-150178NB-I00 financed by MCIU/AEI/10.13039/501100011033, and by FEDER, UE. 

The specific observations analyzed can be accessed via \dataset[10.17909/jp5q-s259] .

This research has made use of the NASA/IPAC Extragalactic Database (NED, \dataset[10.26132/NED1]\ ) which is operated by the Jet
Propulsion Laboratory, California Institute of Technology, under contract with the National Aeronautics and Space
Administration.

\end{acknowledgments}

\vspace{5mm}
\facilities{James Webb Space Telescope (NIRCam, MIRI)}
\software{JWST Calibration Pipeline \citep[][]{Bushouse+2022, Greenfield+2016}, Drizzlepac \citep[][and the STSCI Development Team]{Gonzaga+2012}, IRAF \citep{Tody1986, Tody1993}, SAOImage DS9 \citep{Joye+2003}, Fortran.}

\appendix

\section{Modeling the Light Absorbed by Dust in Stellar Populations}\label{sec:appendixA}

We model the scaling factor $b$ from equation~\ref{equa:b_definition} using stellar population models with extended star formation histories, as opposed to the instantaneous star formation used in Section~\ref{subsec:hybridSFR}. Extended star formation histories provide a better representation of the stellar populations in galaxies and large galaxy regions. Although they are not appropriate for HII regions, we argue in  Section~\ref{subsec:modeling} that the average behavior of large samples of HII regions across a range of ages is comparable to the behavior of constant star formation over durations $\tau\lesssim$6~Myr. With this similarity, we can connect our results for HII regions to the larger--regions results. 

We employ the Starburst99 models \citep{Leitherer+1999,Vazquez+2005}  to generate SEDs with three separate star formation histories (SFH): (1) constant star formation; (2) exponentially--decreasing star formation with an e--folding time of $\sim$4.8~Gyr to emulate the decrease of the cosmic SFR from redshift z$\sim$2 to today \citep{MadauDickinson2014}; and (3) exponentially--increasing star formation with e--folding time of 4.3~Gyr,  to represent galaxies like NGC\,5194 which has culminated in a burst of star formation about a Gyr ago \citep{Martinez+2018}. The SEDs are generated with star formation duration $\tau$ in the range 1~Myr--10~Gyr, using Padova AGB evolutionary tracks \citep{Girardi+2000} with metallicity Z=0.02 (solar) and a \citet{Kroupa2001} IMF in the stellar mass range 0.1--120~M$_{\odot}$. The attenuation curve of \citet{Calzetti+2000} is used to generate dust attenuated stellar SEDs. In order to apply different values of the dust color excess E(B--V) to populations of different ages, the SEDs are `sliced' in bins of time: 1--3~Myr, 3--5~Myr, 5--10~Myr, 10--20~Myr, 20--50~Myr, 50--100~Myr, 100--200~Myr, 200--500~Myr, 500--700~Myr, 700~Myr--1~Gyr, 1--2~Gyr, 2--3~Gyr, 3--5~Gyr, 5--7~Gyr, 7--10~Gyr.
The goal is to simulate a decrease in dust attenuation as a stellar population ages, while using high values  of the color excess (E(B--V)=1~mag) for the youngest stellar populations. The unattenuated and dust--attenuated SEDs are then subtracted from each other to calculate the amount of light absorbed by dust within each time bin. These contributions are finally added together between 0~Myr and a maximum age to derive L(IR) as a function of the duration $\tau$ of star formation. We assume energy balance, meaning that L(IR) is equal to the difference between the unattenuated and dust--attenuated SEDs. We also include the simplest case of constant star formation attenuated by a single value of E(B--V) at all ages for illustrative purposes.

We model the H$\alpha$ luminosity considering two cases: (1) the H$\alpha$ is attenuated by same color excess as the youngest ($\lesssim$5~Myr) stellar populations, i.e. E(B--V)=1.0~mag; and (2) 50\% of H$\alpha$ has the same attenuation of the young ($\lesssim$5~Myr) stellar population with the remaining 50\% assigned the attenuation of the oldest stellar population. The first model simulates the case that all the ionized gas emission is associated with the most recent star formation, while the second model simulates the presence of 50\% clustered lines emission and 50\% diffuse ionized gas \citep[e.g.,][]{Reynolds+1984, Reynolds+1990, Ferguson+1996, Hoopes+1996, Hoopes+2003, Voges+2006, Oey+2007, Zhang+2017}. 

\begin{figure}
\plottwo{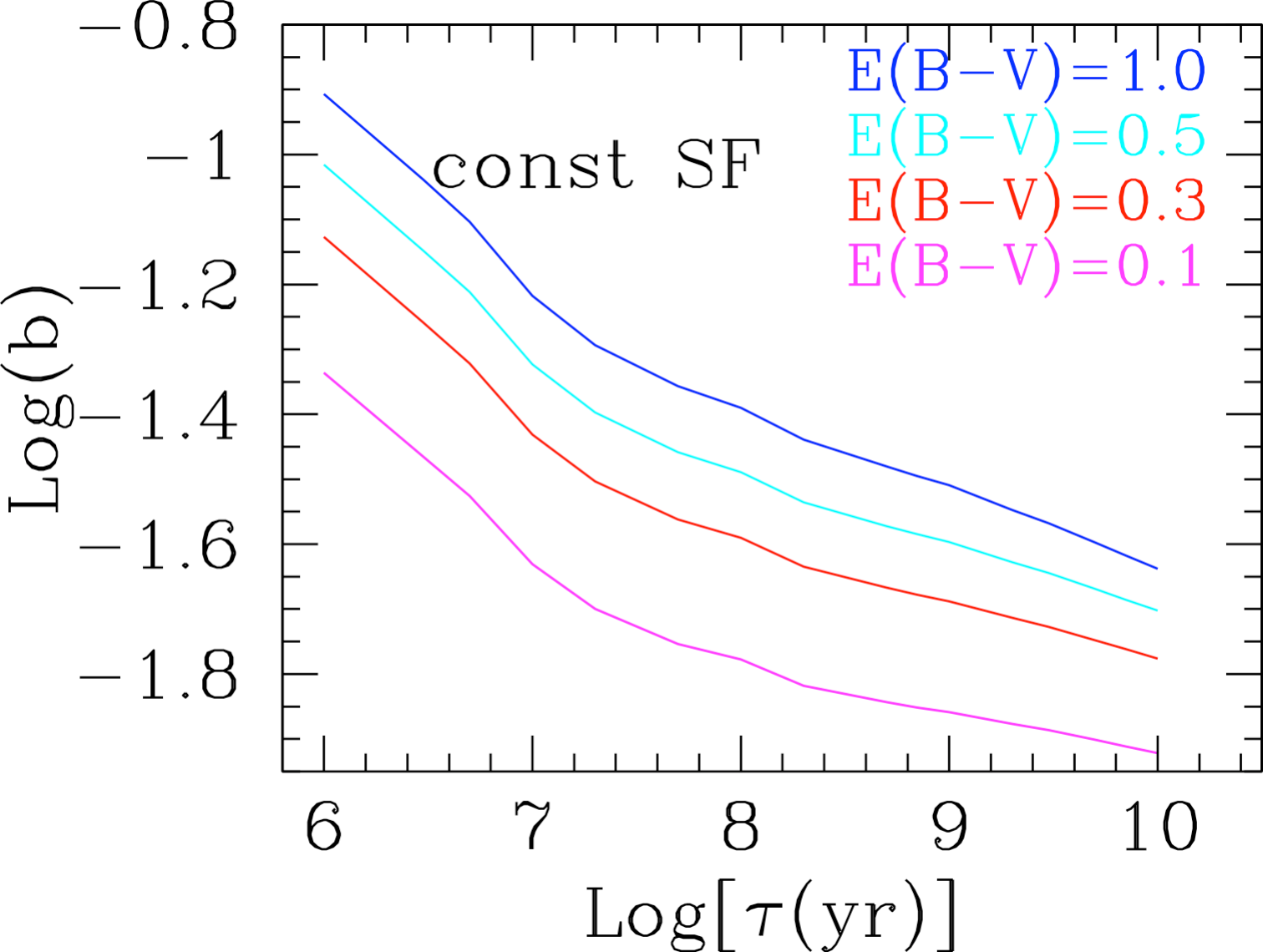}{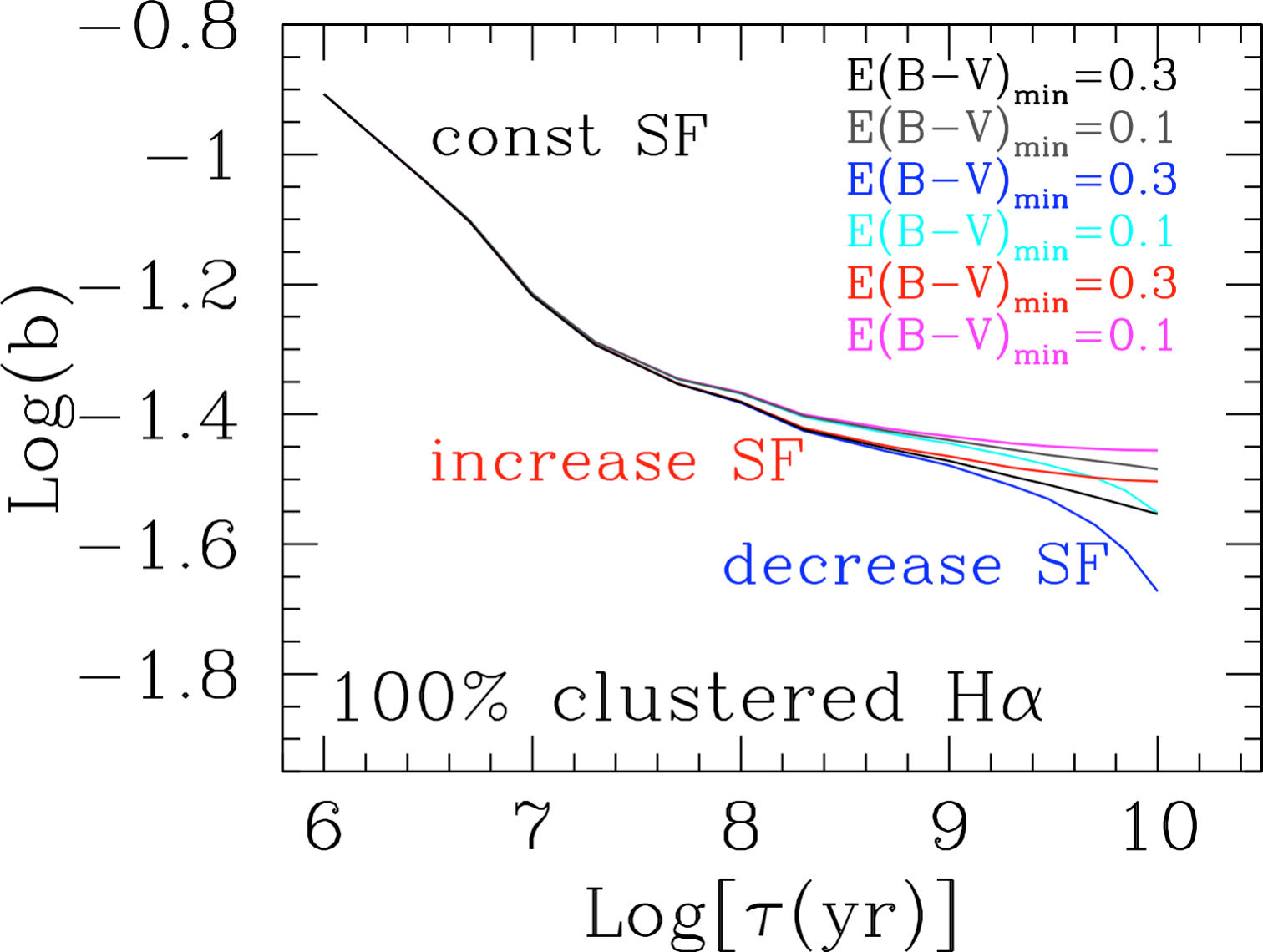}
\plottwo{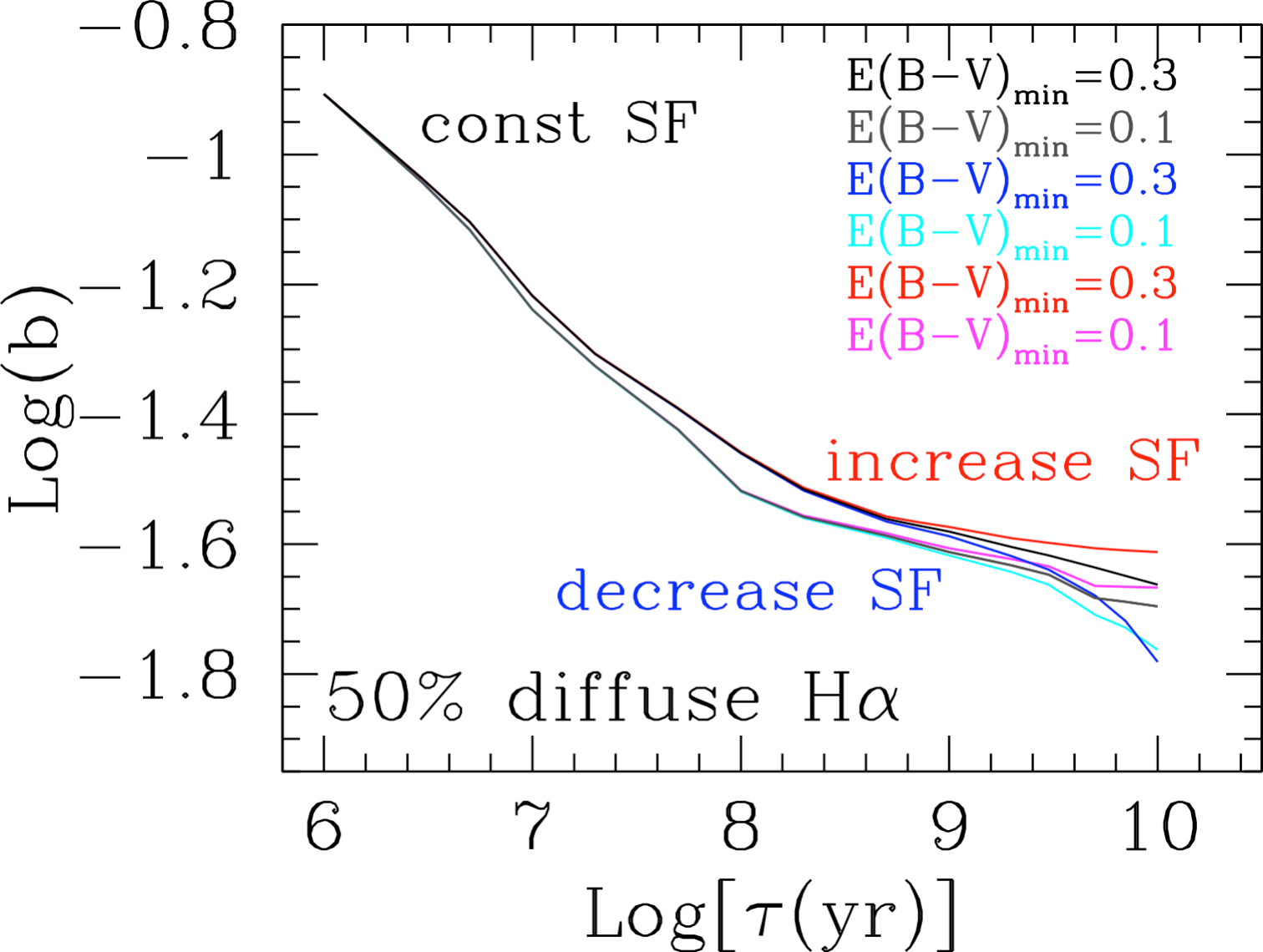}{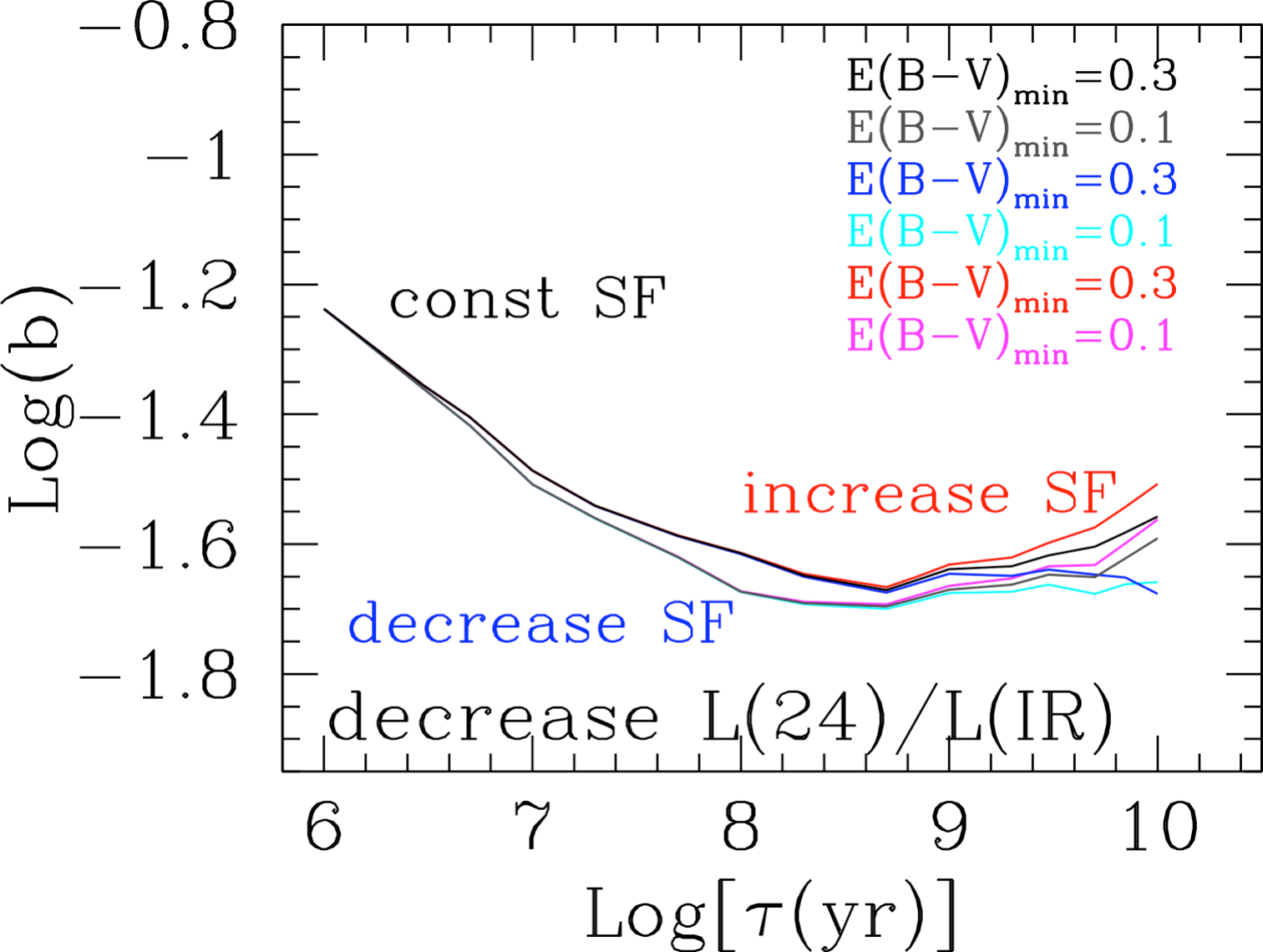}
\caption{{\bf (Top--Left:)} The scaling factor $b$ as a function of the star formation duration $\tau$, for constant star formation and constant value of the color excess at all ages (Table~\ref{tab:ir_model}); four values of E(B--V) are shown as examples. {\bf (Top--Right:)} The same as the top--left plot, but for the three SFHs of Table~\ref{tab:ir_model} and decreasing values of the color excess applied to population bins of increasing age; the H$\alpha$ emission is attenuated by a constant E(B--V)$_{gas}$=1~mag, the same color excess of the stellar populations $\lesssim$5~Myr. The SFHs are: constant (black and grey lines), decreasing (blue and cyan lines) and increasing (red and magenta lines) star formation. {\bf (Bottom--Left:)} Same models as the top--right panel, but now the H$\alpha$ attenuation is split into two components: 50\% of the line emission at attenuated with E(B--V)$_{gas}$=1.0~mag and the remaining 50\% has the color excess of the oldest population. {\bf (Bottom--Right:)} The same models as the bottom--left panel, but adding a decreasing fraction of the IR emission emerging at 24~$\mu$m for increasing age.}
 \label{fig:b_age}
\end{figure}

Finally, we also consider two cases for the fraction of L(IR) emerging at 24~$\mu$m: (1) a constant value, 0.14, at all ages \citep{Rieke+2009,Calzetti+2010}, and (2) a decreasing fraction from 0.3 at ages $<$5~Myr to 0.10 at ages$>$7~Gyr, thus covering the full range of both HII regions \citep{Relano+2013} and galaxies \citep{Rieke+2009, Calzetti+2010}. The latter model attributes hotter dust to younger stellar populations, which have both higher ionizing photon flux and non--ionizing UV emission, and therefore potentially a higher L(24)/L(IR) fraction  \citep{DaleHelou2002, Draine+2007}.  Table~\ref{tab:ir_model} lists the models used in this work and shown in Figure~\ref{fig:b_age}. 

These models are more sophisticated than the simple assumptions used in \citet{Calzetti+2024} in that they include three different star formation histories, decreasing extinctions for increasing ages, and a more flexible implementation of both the attenuation at H$\alpha$ and the fraction of IR emission emerging at 24~$\mu$m. Yet the basic result is similar to the findings by those authors: the scaling factor $b$ decreases for increasing duration of the star formation, for a large range of assumptions (Figure~\ref{fig:b_age}). The reason for this behavior is straightforward: while the H$\alpha$ emission is contributed by regions younger than $\sim$6~Myr, the IR emission is contributed by stellar populations of all ages, as discussed in the Introduction. As the stellar population ages, it accumulates low mass stars which contribute to the heating of the dust, but not to the ionizing photon flux. 

The panels in Figure~\ref{fig:b_age} describe increasingly complex assumptions from top--left to bottom--right, providing a sense for how $b$ changes for changing parameters, but 
also highlighting the many degeneracies that characterize such modeling. In fact, different assumptions can be combined together to reproduce the observed trend of $b$ with the duration of star formation. This implies that we cannot draw solid conclusions as to the star formation history, dust content, etc. of the galaxies under consideration by simply looking at the values of $b$. 

The results in Section~\ref{subsec:modeling} are discussed using as reference the models shown in the bottom--left panel of Figure~\ref{fig:b_age}. These models provide a representative description of the physical and geometrical conditions of nearby galaxies, in terms of the treatment of the attenuation of the stellar continuum and of the nebular lines. 

\begin{deluxetable*}{llll}
\tablecaption{Stellar Population Models with Dust Absorption\label{tab:ir_model}}
\tablewidth{0pt}
\tablehead{
\colhead{Parameter} & \colhead{Value} & \colhead{Range} & \colhead{Comments} \\
}
\decimalcolnumbers
\startdata
Star Form. History$^1$& Constant  & 1~Myr--10~Gyr & \\
			    & Decreasing & 1~Myr--10~Gyr & e--folding t=4.8~Gyr\\
			    & Increasing   & 1~Myr--10~Gyr & e--folding t=4.3~Gyr\\
\hline
E(B-V) (mag)$^2$ & Constant & 0.1, 0.3, 0.5, 1.0 & \\
		          & Decreasing & 1.0--0.3 & \\
		          & Decreasing & 1.0--0.1 & \\
\hline
L(H$\alpha$)$^3$ & 100\% clustered &       &\\
		    & 50\% clustered, 50\% diffuse & \\
\hline
L(24)/L(IR)$^4$ & constant & 0.14 &\\
		      & decreasing & 0.30--0.10 \\
\hline
\enddata

$^1$  The star formation histories (SFH) of the models include three cases: constant star formation (Constant), exponentially decreasing star formation with a 4.8~Gyr e--folding time (Decreasing), and exponentially--increasing star formation history with a 4.3~Gyr e--folding time (Increasing). The SF duration of each model spans the range 1~Myr--10~Gyr.  \\ 
$^2$ The color excess, E(B--V), applied to the population SEDs. Constant E(B--V) values at all ages are only applied to the Constant SFH model. E(B--V) values that decrease for increasing age of the stellar population are applied to all three SFHs; two values for the minimum E(B--V) are considered: 0.3~mag and 0.1~mag.\\
$^3$ The H$\alpha$ luminosity is attenuated by E(B--V)=1.0, i.e., the value of stellar populations younger than 10~Myr, for the 100\% clustered emission case; for the 50\% clustered, 50\% diffuse, half of the H$\alpha$ emission is attenuated by E(B--V)=1.0  and the remaining half has the same color excess of the oldest stellar population at the specified $\tau$.\\
$^4$ Fraction of IR emission emerging at 24~$\mu$m. This fraction is adopted to be either constant at all ages or slightly decreasing from 0.3 at ages $\le$~5~Myr to 0.10 at ages $>$7~Gyr.\\
\end{deluxetable*}

\section{Modeling the Correlation between Infrared and Ionized Gas Emission}\label{sec:appendixB}

The SED models introduced in the previous Appendix can be used also to model the observed correlation between L(24) and L(Pa$\alpha$). We restrict our attention to two of the five mechanisms listed in Section~\ref{subsec:SFR24model}: \# 1, regions of low luminosity are more transparent than high luminosity regions; and \# 5, the increasing contribution of low--mass stars to the infrared emission for increasing durations $\tau$ of the star formation, already discussed in Appendix~\ref{sec:appendixA}. 

\begin{figure}
\plottwo{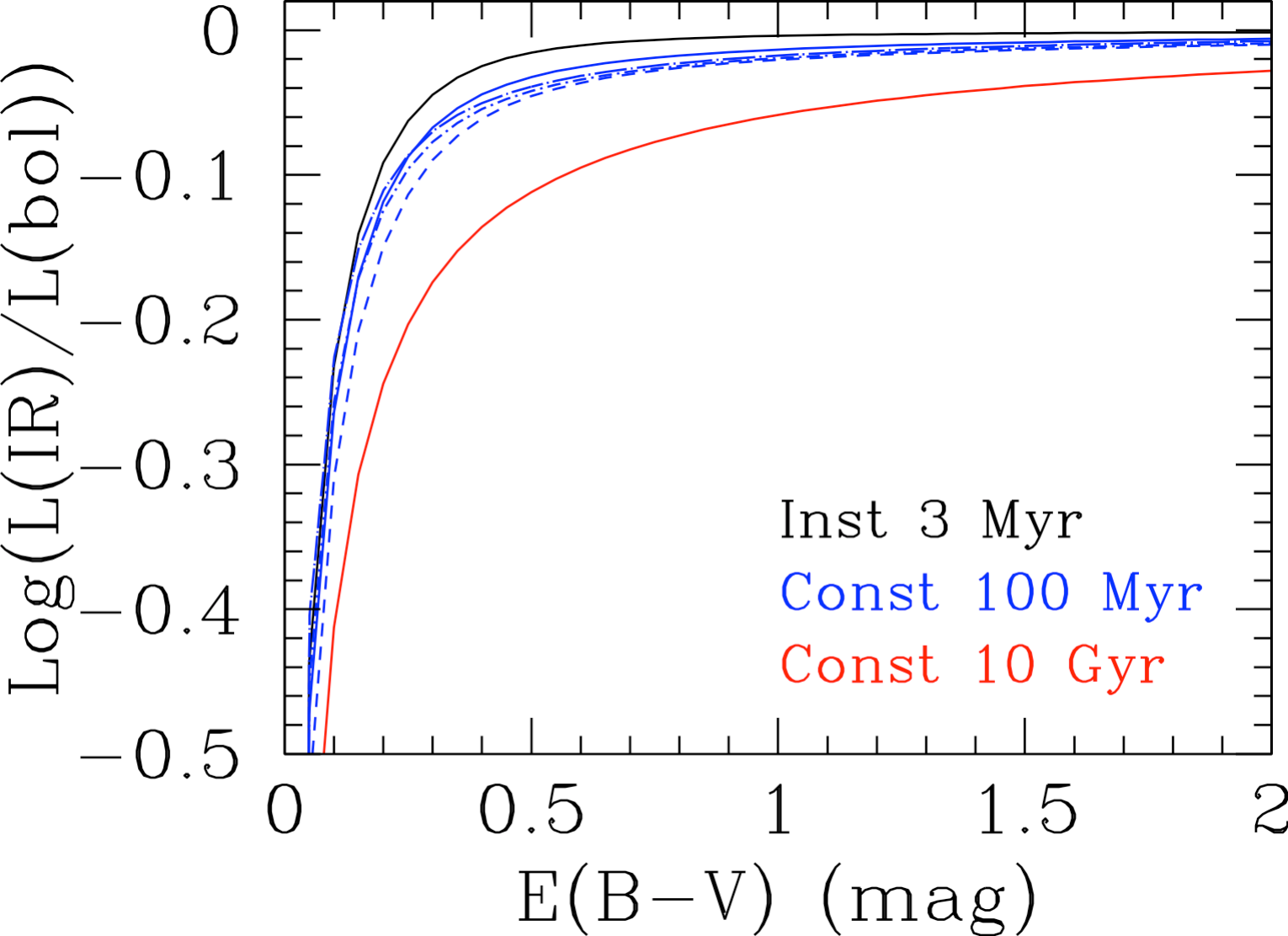}{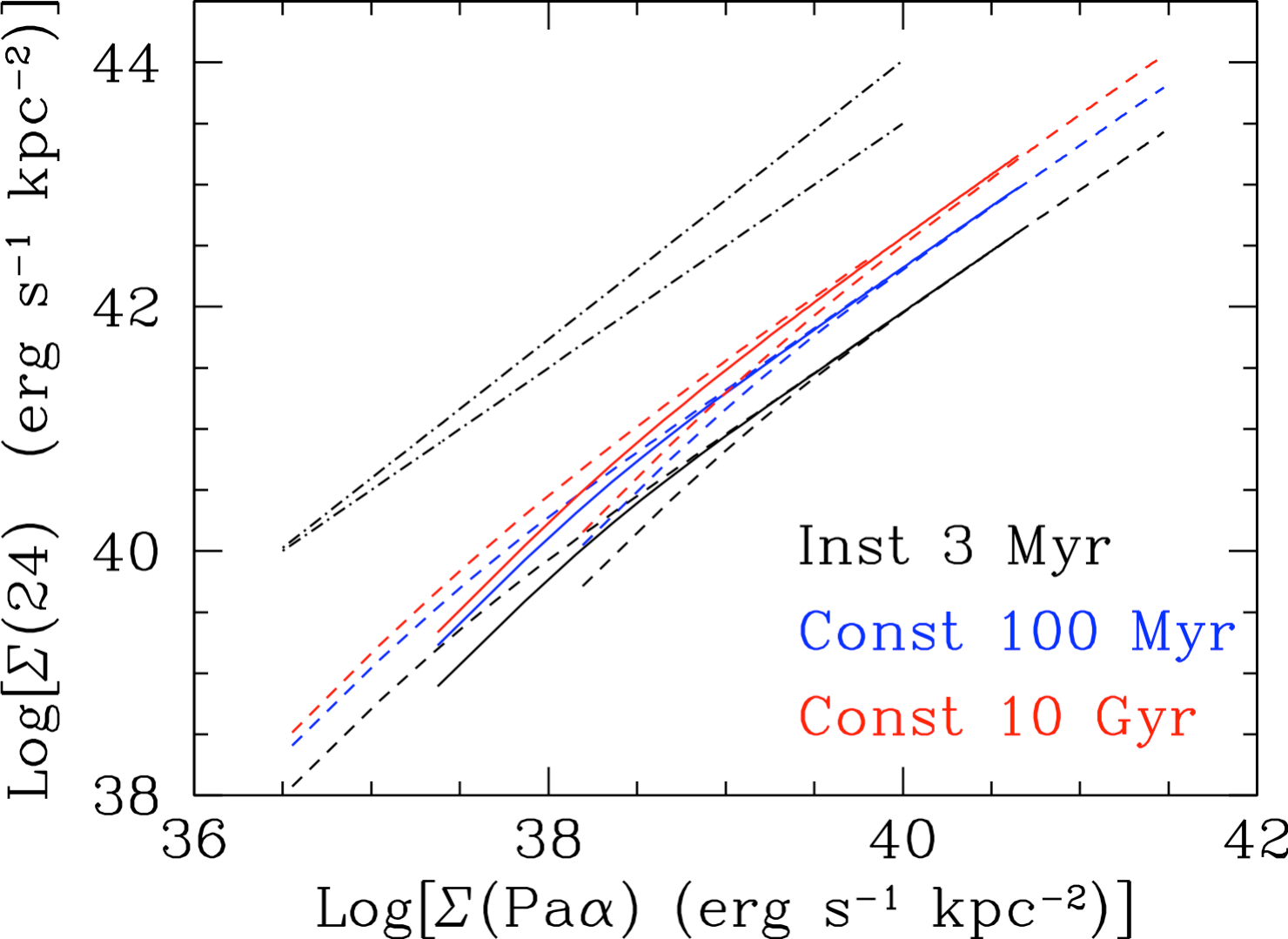}
\caption{{\bf (Left:)} The fraction of bolometric luminosity emerging in the infrared, as a function of color excess E(B--V), for the following stellar population models: 3~Myr old instantaneous burst (black solid line); constant star formation over 100~Myr (blue lines); and constant star formation over 10~Gyr (red solid line). The effect of different extinction/attenuation curve choices are shown for the 100~Myr constant star formation model: starburst attenuation curve (blue solid) and the Milky Way (blue dashed), Large Magellanic Cloud (blue short dot--dashed) and Small Magellanic Cloud (blue long dot--dashed) extinction curves. {\bf (Right:)} The 24~$\mu$m luminosity surface density as a function of the extinction--corrected Pa$\alpha$ luminosity surface density for the three star formation models described in the Left panel. The Pa$\alpha$ surface density is derived via the E(B--V)--$\Sigma$(Pa$\alpha$) relation in equation~\ref{equa:newebv} \citep{Calzetti+2007}. The models for the 3~Myr instantaneous burst (black solid line), $\tau$=100~Myr constant star formation (blue solid line) and $\tau$=10~Gyr constant star formation (red solid line) mark a sequence of increasing $\Sigma$(24) at fixed $\Sigma$(Pa$\alpha$)$_{corr}$. The relation at each age is shown with the 90\% scatter in the E(B--V)--Pa$\alpha$ relation from \citet{Calzetti+2007} as dashed lines.  The black dot--dashed straight lines are examples of linear relations with slopes (bottom to top) of 1 and 1.14, to provide a visual reference.}
 \label{fig:LIR_EBV}
\end{figure}

We use here instantaneous and constant star formation SEDs attenuated by the starburst attenuation curve to evaluate the impact of both mechanisms on the resulting relation between the  luminosity surface densities at 24~$\mu$m and in the Pa$\alpha$ light. Figure~\ref{fig:LIR_EBV}, left, shows the fraction of bolometric light from a stellar population absorbed by dust and re--emitted in the infrared as a function of color excess for three star formation models: a 3~Myr old instantaneous burst and constant star formation with duration 100~Myr and 10~Gyr, respectively. Constant star formation with $\tau$=3~Myr is indistinguishable from the 3~Myr old instantaneous model. The range of color excess shown in the Figure spans 0--2~mag, as the IR fraction of the bolometric light quickly approaches unity for all three SEDs for larger E(B--V) values. The differences in the asymptotic approach of the three SEDs quantify the effect of the increasing contribution of lower mass stars to the bolometric luminosity for increasing $\tau$. These lower mass stars mainly contribute to the optical/near infrared emission which is less sensitive to the effects of dust attenuation than the UV light; thus their presence delays the approach to the asymptotic value of unity of the L(IR)/L(bol) ratio. The trends are only minimally affected by the choice of attenuation/extinction curve. This is shown in Figure~\ref{fig:LIR_EBV}, left, for the $\tau$=100~Myr model, where the L(IR)/L(bol)--vs--E(B--V) trend for our default choice, the starburst attenuation curve, is compared with the analogous trends when using the Milky Way, Large Magellanic Cloud and Small Magellanic Cloud extinction curves \citep{Fitzpatrick1999, Gordon+2003, Fitzpatrick+2007, Fitzpatrick+2019}. For E(B--V)$>$0.15~mag, the differences in the L(IR)/L(bol) ratio are $<$0.05~dex for any choice of the attenuation/extinction curve at a given star formation history.

To convert the trend with color excess to a trend with ionized gas luminosity surface density, we use the empirical relation between E(B--V) and the luminosity surface density $\Sigma$(Pa$\alpha$)$_{corr}$ derived by \citep{Calzetti+2007} with a correction for a typo:
\begin{equation}
E(B-V)=0.21 \Bigl({\Sigma_{ion}\over 1.433 \times 10^{51}}\Bigr)^{0.61},
\end{equation}
which is appropriate for metal rich regions. With the updated relation, we obtain:
\begin{equation}
Log[E(B-V)] = -24.099 +0.61 Log [\Sigma(Pa\alpha_{corr})],
\label{equa:newebv}
\end{equation}
which fits the data in \citet{Calzetti+2007}, and has been used both for the tracks in Figure~\ref{fig:corr_lum} (right) and in \citet{Calzetti+2024}. We combine this relation, and its 90\% range, with the L(IR)/L(bol) trend of Figure~\ref{fig:LIR_EBV}, left, for all three models shown, which result in the plot of Figure~\ref{fig:LIR_EBV}, right. In all models the ratio L(24)/L(IR) is kept constant at the value 0.14. Despite the simplicity of our approach in terms of both star formation history and attenuation recipe (all stellar populations are attenuated by the same color excess), we can already identify and quantify two important trends. The first  is between the two surface densities: in each model, the 24~$\mu$m emission is underluminous at the lowest values of $\Sigma$(Pa$\alpha$)$_{corr}$  creating a non--linear relation, with a slope $>$1 in log--log scale. For reference, Figure~\ref{fig:LIR_EBV}, right, reports two straight lines: one with slope = 1 and one with slope = 1.14, the latter close to what we  find for HII regions and what \citet{Calzetti+2010} find for $\sim$kpc--sized regions. In all cases, the Log[$\Sigma$(24)]--Log[$\Sigma$(Pa$\alpha$)$_{corr}$] relation has an asymptotic behavior towards a slope=1, which is reached sooner for the youngest models (see  Figure~\ref{fig:LIR_EBV}, left). The second trend is with increasing star formation duration $\tau$: the longer $\tau$ the higher the value of $\Sigma(24)$ at fixed $\Sigma$(Pa$\alpha$)$_{corr}$, which mirrors the trends for $b$ with $\tau$ presented in the previous Appendix. Each model is shown for the range E(B--V)=0.05--5~mag; restricting the range in color excess to 0.05--2~mag (0.05--3~mag) would lower the maximum L(Pa$\alpha$) of each model by 0.65~dex (0.37~dex). Finally, the impact of the 90\% scatter in the E(B--V)--vs--$\Sigma$(Pa$\alpha$)$_{corr}$ relation is shown in Figure~\ref{fig:LIR_EBV}, right, as dashed lines: the large scatter between E(B--V) and Pa$\alpha$ has minimal impact on the predicted 24~$\mu$m luminosity surface  density, and it mostly serves to increase the  range for each model. We use these models in Section~\ref{subsec:SFR24model} to provide a comparison with observational data.

\section{Histograms of Slopes and Intercepts}\label{sec:appendixE}

\begin{figure}
\plottwo{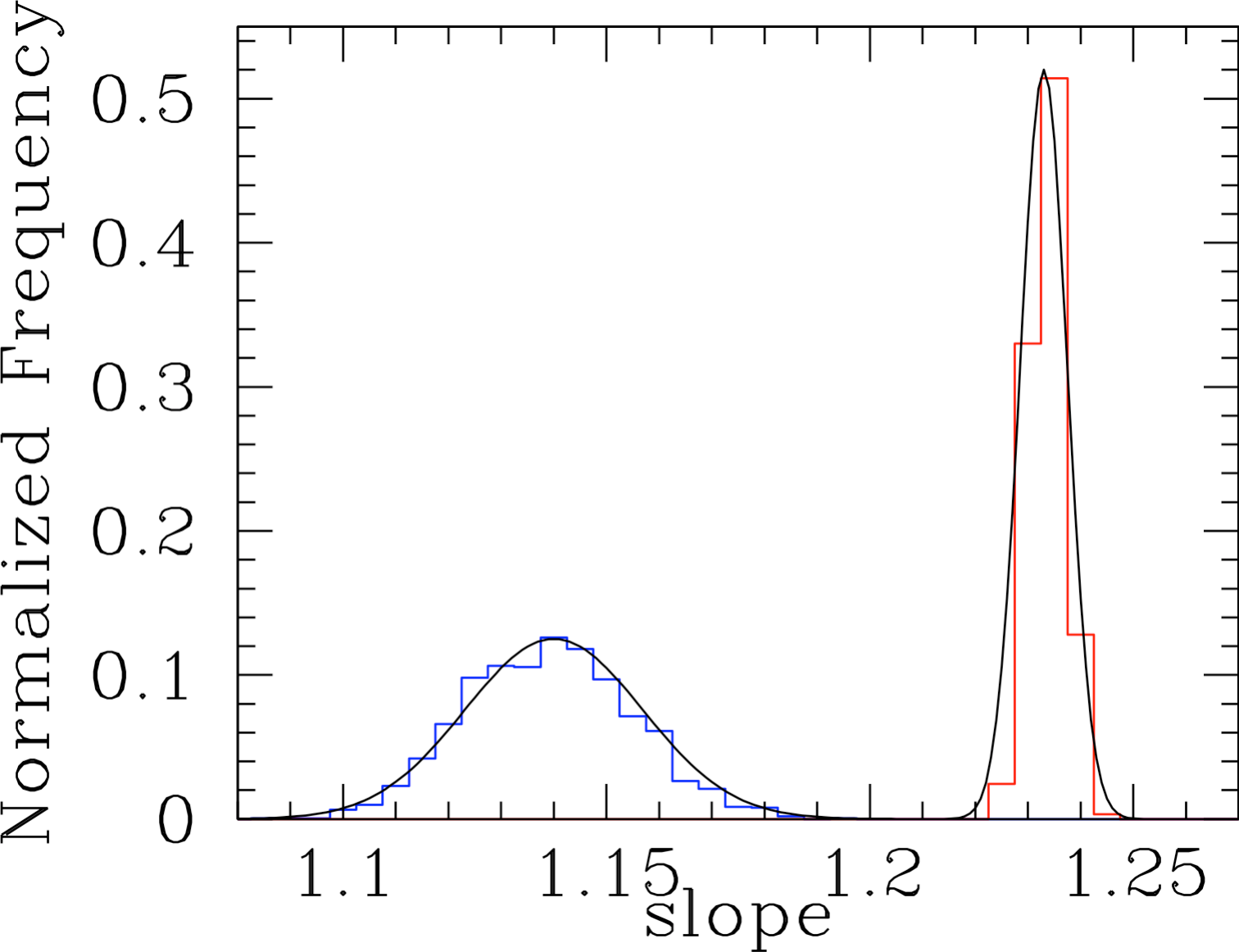}{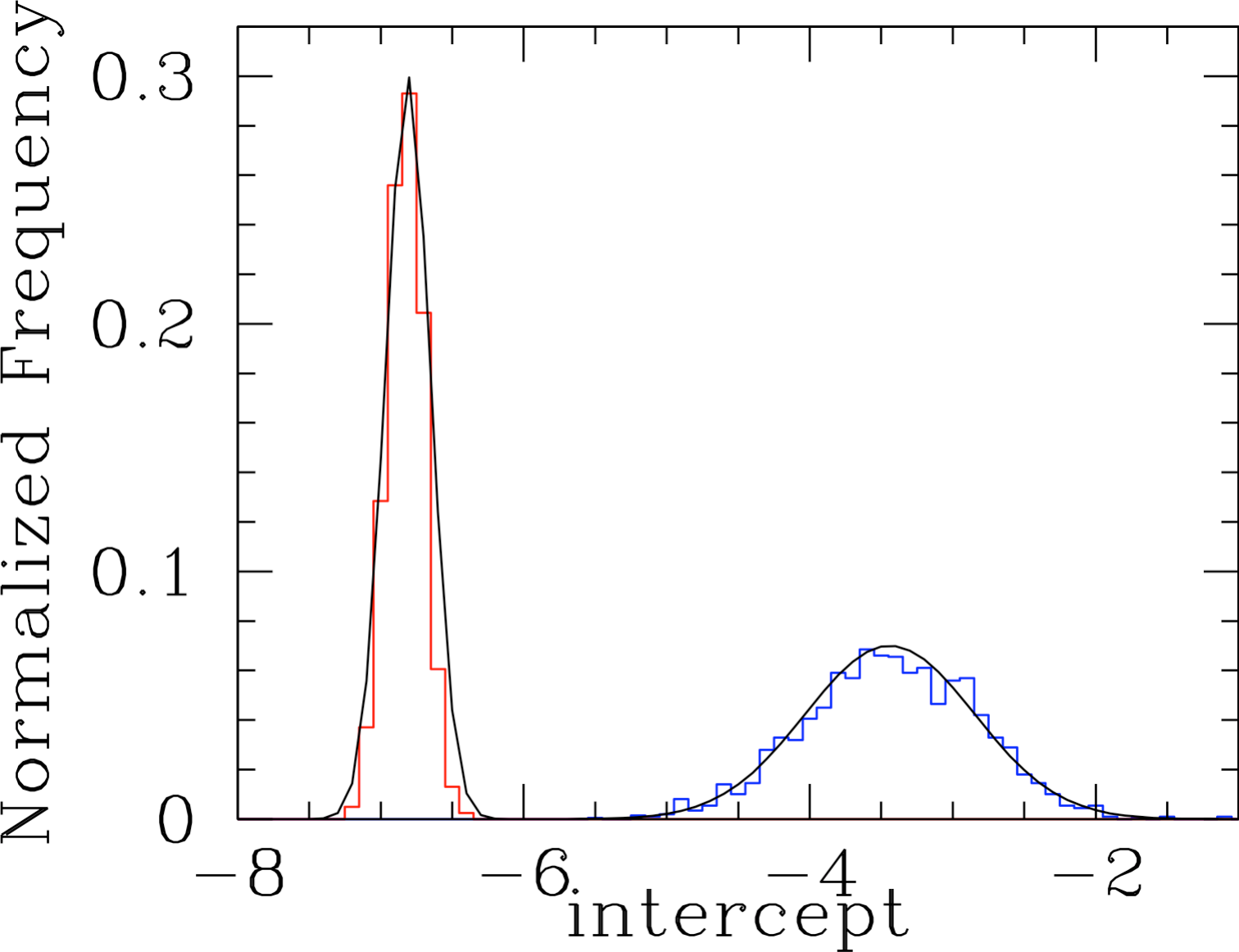}
\caption{Histograms of the slopes (left panels) and intercepts (right panel) data for the HII regions sample (blue) and the full sample of HII regions, kpc--size regions and galaxies (red), with Gaussian distributions over--plotted on the histograms (black).}
 \label{fig:histo_slopes}
\end{figure}

We test how effectively we can discriminate between the slopes 1.14$\pm$0.02 and 1.23$\pm$0.01 found for the Log[L(24)]--Log[L(Pa$\alpha)_{corr}$] relations of HII regions (equation~\ref{equa:l21lpa}) and the full sample of HII regions, kpc--size regions and galaxies (equation~\ref{equa:l24_all}), respectively. We produce 1,000 realizations of each datapoint in each sample by randomly drawing numbers \citep{Matsumoto+1998} within the data's 3~$\sigma$ uncertainties along both the x and y axis. Any assumption on the underlying distribution of the uncertainties is relaxed by adopting uniform priors. We perform a linear fit for each realization, and create histograms of the resulting slopes and intercepts for the two relations (Figure~\ref{fig:histo_slopes}). The data centered around the slope=1.14 (HII regions only) are consistent with a Gaussian distribution with standard deviation $\sigma$=0.017, while the distribution centered on the slope=1.23 (full sample) has standard deviation $\sigma=$0.0045. For the intercepts, the Gaussian consistent with the distribution of the data for the HII regions sample has standard deviation $\sigma$=0.59 and for the full sample the stadanrd deviaiton is $\sigma=$0.16. Thus, even with a non--Gaussian distribution of uncertainties, the two relations can be discriminated to better than 4~$\sigma$.

\section{Scatter in the L(24)--L(Pa$\alpha$) Relation}\label{sec:appendixC}

\begin{figure}
\plottwo{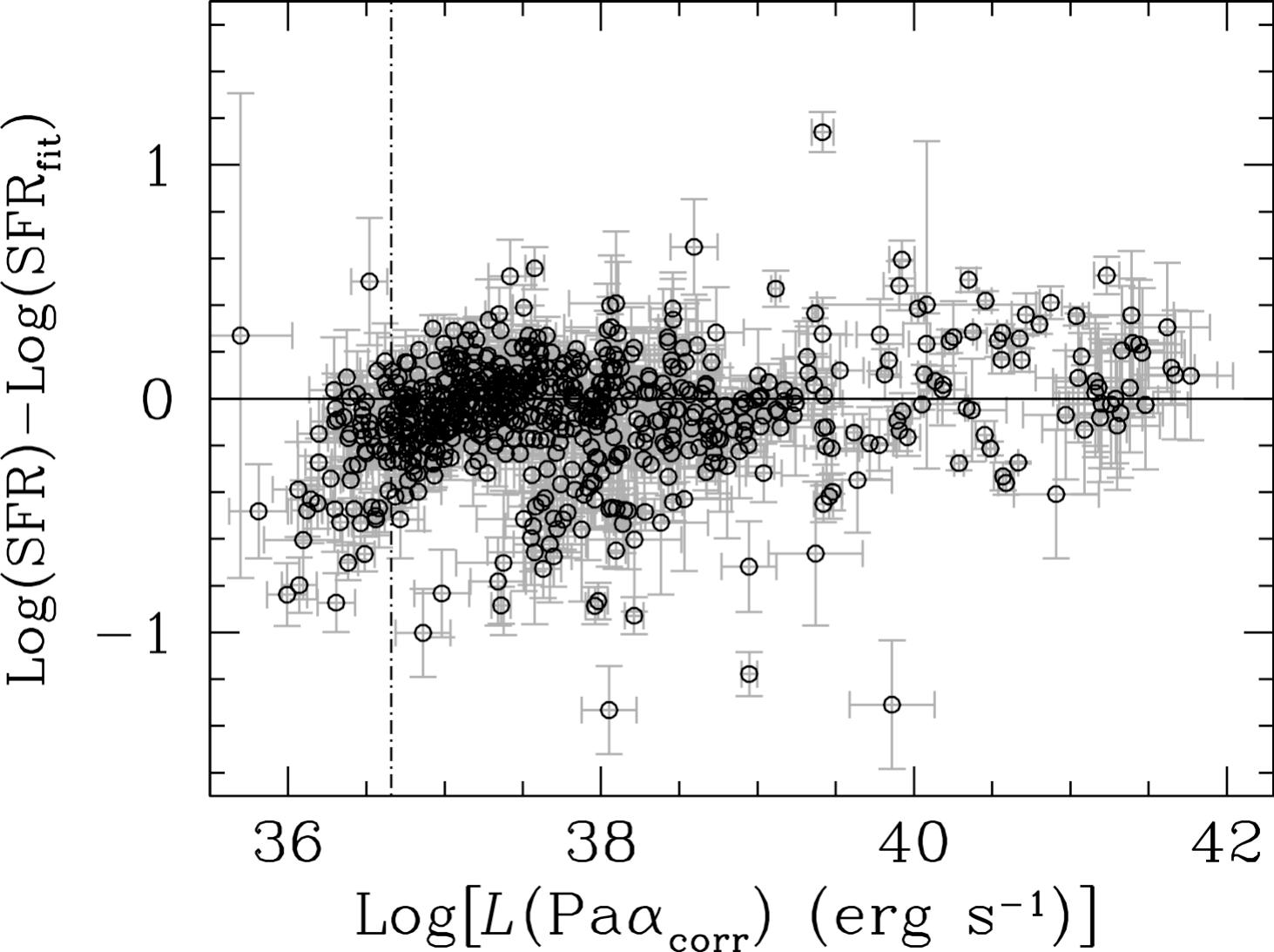}{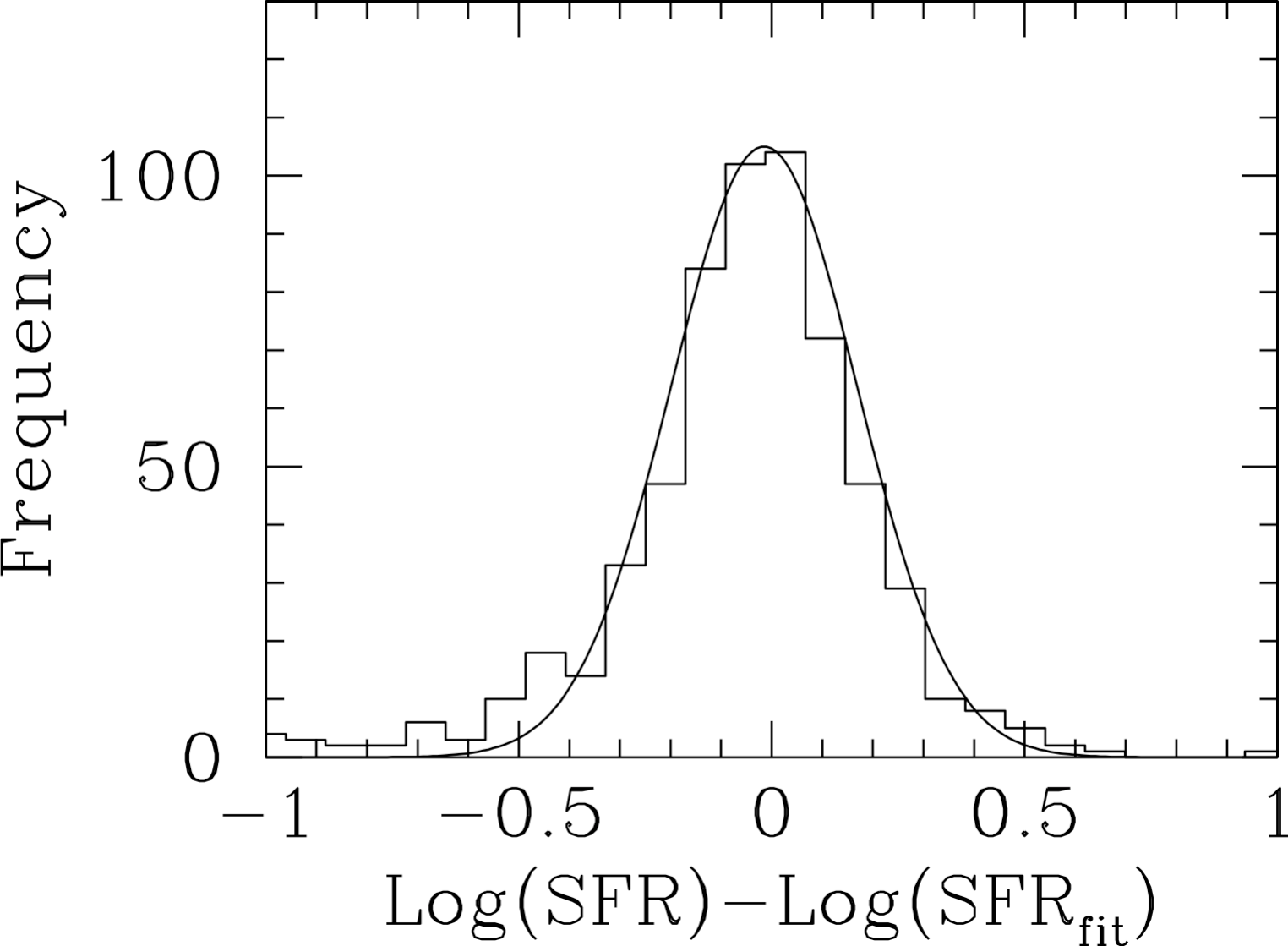}
\caption{{\bf (Left:)} The distribution of the Log(SFR) in our full sample of HII regions, galaxy regions and galaxies, relative to the predicted Log(SFR$_{fit}$) from the best fit line of equation~\ref{equa:SFR24_all}. The SFRs are calculated from the attenuation--corrected Pa$\alpha$ luminosity. The vertical line marks the location of the luminosity below which the stellar IMF is stochastically sampled. Only sources above this limit where used in the derivation of the best fit line. {\bf (Right:)} The histogram of the data to the left with luminosity above the stochastic IMF sampling limit, shown with a best Gaussian fit.}
 \label{fig:scatter}
\end{figure}

Here we show the distribution of the observations about the mean fitted trend of Figure~\ref{fig:SFR24_all}, right, which we express as scatter of the true SFR about the predicted SFR(24) from equation~\ref{equa:SFR24_all} to facilitate evaluation of uncertainties in this parameter. Figure~\ref{fig:scatter} illustrates the distribution of SFR/SFR$_{fit}$ for our entire sample of 658 sources, which include HII regions, galaxy regions, and galaxies across the full luminosity range in the local universe. In this Figure, SFR is the true SFR derived from the attenuation--corrected Pa$\alpha$ luminosity, while SFR$_{fit}$ is the predicted SFR from the 24~$\mu$m luminosity using equation~\ref{equa:SFR24_all}. Only data with Pa$\alpha$ luminosity above the stellar IMF stochastic sampling limit were used in the derivation of this equation (see main text for details). The left panel shows the distribution of individual sources about the best fit line. The only notable deviation from a fairly symmetrical distribution is displayed by the faint end of the galaxy regions from the sample of \citet{Calzetti+2007}. 

The right panel of Figure~\ref{fig:scatter} shows the histogram of the data above the luminosity limit for stochastic IMF sampling, together with the best--fit Gaussian to the distribution. The excess in the tail at negative values (SFR/SFR$_{fit}<$0.4) is due to the same sources from the sample of \citet{Calzetti+2007} discussed above. The standard deviation of the Gaussian is 0.19, which corresponds to a 1$\sigma$ scatter in the SFR/SFR$_{fit}$ ratio of 55\%. We consider this scatter representative of the accuracy of equation~\ref{equa:SFR24_all} in predicting the true SFR from the 24~$\mu$m luminosity from HII regions to luminous infrared galaxies at solar metallicity in the local universe.

\newpage

\startlongtable
\begin{deluxetable}{lrrrrrr}
\tablecolumns{7}
\tabletypesize{\small}
\tablecaption{Source Location, Luminosity and Derived Quantities for the HII Regions in NGC\,5194\label{tab:sources}}
\tablewidth{100pt}
\tablehead{
\colhead{ID} & \colhead{RA(2000),DEC(2000)}  & \colhead{Log[L(H$\alpha$)]} & \colhead{Log[L(Pa$\alpha$)]} & \colhead{Log[L(21)]} & \colhead{Log[EW(Pa$\alpha$)]} & \colhead{E(B$-$V)} 
\\
\colhead{(1)} & \colhead{(2)} & \colhead{(3)} & \colhead{(4)} & \colhead{(5)}  & \colhead{(6)} & \colhead{(7)} 
\\
}
\startdata
    1 &13:29:52.1548,+47:12:44.750 &38.382$\pm$0.040 &37.896$\pm$0.033 &40.097$\pm$0.054 &2.863$\pm$0.052 &0.549$\pm$0.070\\
    2 &13:29:52.0449,+47:12:47.270 &38.481$\pm$0.039 &37.901$\pm$0.033 &39.952$\pm$0.055 &3.254$\pm$0.063 &0.421$\pm$0.069\\
    3 &13:29:52.3785,+47:12:38.550 &37.706$\pm$0.054 &37.682$\pm$0.037 &40.037$\pm$0.054 &3.410$\pm$0.084 &1.174$\pm$0.088\\
    4 &13:29:51.1734,+47:12:45.511 &37.976$\pm$0.047 &37.687$\pm$0.037 &39.911$\pm$0.055 &3.340$\pm$0.079 &0.816$\pm$0.080\\
    5 &13:29:51.0399,+47:12:40.551 &37.478$\pm$0.063 &37.561$\pm$0.039 &39.985$\pm$0.054 &3.073$\pm$0.073 &1.318$\pm$0.100\\
    6 &13:29:55.0593,+47:12:20.546 &37.716$\pm$0.054 &37.250$\pm$0.048 &39.150$\pm$0.068 &2.874$\pm$0.084 &0.577$\pm$0.097\\
    7 &13:29:55.0439,+47:12:32.066 &37.045$\pm$0.088 &36.948$\pm$0.062 &39.250$\pm$0.065 &2.798$\pm$0.105 &1.075$\pm$0.145\\
    8 &13:29:54.6710,+47:12:35.547 &36.828$\pm$0.106 &37.287$\pm$0.047 &39.667$\pm$0.058 &2.632$\pm$0.071 &1.825$\pm$0.156\\
    9 &13:29:54.6904,+47:12:23.787 &37.094$\pm$0.084 &37.057$\pm$0.056 &39.133$\pm$0.068 &2.831$\pm$0.097 &1.156$\pm$0.136\\
   10 &13:29:54.8360,+47:12:40.666 &37.354$\pm$0.068 &37.007$\pm$0.059 &38.924$\pm$0.076 &2.785$\pm$0.099 &0.737$\pm$0.122\\
   11 &13:29:53.2127,+47:12:39.549 &38.360$\pm$0.040 &38.010$\pm$0.032 &40.169$\pm$0.053 &2.969$\pm$0.051 &0.733$\pm$0.069\\
   12 &13:29:50.1056,+47:12:46.911 &37.182$\pm$0.078 &36.907$\pm$0.064 &39.040$\pm$0.071 &2.778$\pm$0.108 &0.835$\pm$0.136\\
   13 &13:29:49.4383,+47:12:37.630 &36.555$\pm$0.137 &37.404$\pm$0.043 &39.677$\pm$0.057 &2.820$\pm$0.071 &2.352$\pm$0.194\\
   14 &13:29:49.4343,+47:12:40.870 &38.453$\pm$0.039 &38.374$\pm$0.029 &40.736$\pm$0.051 &3.326$\pm$0.049 &1.099$\pm$0.066\\
   15 &13:29:49.8976,+47:12:39.950 &36.705$\pm$0.119 &36.987$\pm$0.060 &39.190$\pm$0.067 &2.716$\pm$0.096 &1.585$\pm$0.179\\
   16 &13:29:49.7759,+47:12:35.470 &36.800$\pm$0.109 &36.779$\pm$0.072 &38.752$\pm$0.083 &2.633$\pm$0.112 &1.178$\pm$0.176\\
   17 &13:29:50.4825,+47:12:39.751 &36.604$\pm$0.131 &36.507$\pm$0.095 &38.558$\pm$0.095 &2.704$\pm$0.154 &1.076$\pm$0.218\\
   18 &13:29:48.6807,+47:12:33.349 &36.393$\pm$0.161 &36.397$\pm$0.107 &38.669$\pm$0.088 &2.370$\pm$0.144 &1.212$\pm$0.261\\
   19 &13:29:48.9594,+47:12:31.110 &36.399$\pm$0.160 &36.667$\pm$0.080 &38.662$\pm$0.088 &2.486$\pm$0.115 &1.568$\pm$0.242\\
   20 &13:29:48.1939,+47:12:36.669 &35.962$\pm$0.253 &36.448$\pm$0.101 &38.697$\pm$0.086 &2.628$\pm$0.156 &1.862$\pm$0.367\\
   21 &13:29:49.2105,+47:12:47.550 &36.607$\pm$0.130 &36.141$\pm$0.145 &38.476$\pm$0.101 &2.222$\pm$0.180 &0.577$\pm$0.263\\
   22 &13:29:49.0417,+47:12:48.670 &36.750$\pm$0.114 &36.310$\pm$0.118 &38.201$\pm$0.129 &2.214$\pm$0.148 &0.613$\pm$0.222\\
   23 &13:29:50.1370,+47:12:51.431 &37.292$\pm$0.072 &36.751$\pm$0.074 &38.715$\pm$0.085 &2.811$\pm$0.129 &0.476$\pm$0.139\\
   24 &13:29:50.3333,+47:12:37.071 &37.265$\pm$0.073 &36.824$\pm$0.069 &38.721$\pm$0.085 &2.537$\pm$0.102 &0.609$\pm$0.136\\
   25 &13:29:48.9477,+47:12:23.630 &37.254$\pm$0.074 &36.657$\pm$0.081 &38.732$\pm$0.084 &2.661$\pm$0.128 &0.400$\pm$0.148\\
   26 &13:29:46.4039,+47:12:33.226 &38.127$\pm$0.044 &37.785$\pm$0.035 &40.072$\pm$0.054 &3.300$\pm$0.071 &0.744$\pm$0.075\\
   27 &13:29:47.9744,+47:12:13.909 &36.963$\pm$0.094 &36.753$\pm$0.074 &39.037$\pm$0.071 &2.647$\pm$0.116 &0.923$\pm$0.161\\
   28 &13:29:47.6288,+47:12:23.468 &37.528$\pm$0.060 &36.933$\pm$0.062 &38.968$\pm$0.074 &2.655$\pm$0.098 &0.403$\pm$0.117\\
   29 &13:29:47.3344,+47:12:26.268 &36.822$\pm$0.107 &36.709$\pm$0.077 &38.462$\pm$0.102 &2.459$\pm$0.109 &1.053$\pm$0.177\\
   30 &13:29:47.0204,+47:12:24.227 &37.273$\pm$0.073 &37.103$\pm$0.054 &39.205$\pm$0.066 &2.439$\pm$0.076 &0.977$\pm$0.122\\
   31 &13:29:46.6202,+47:12:16.386 &37.381$\pm$0.067 &37.163$\pm$0.052 &39.138$\pm$0.068 &2.885$\pm$0.091 &0.911$\pm$0.114\\
   32 &13:29:46.5967,+47:12:14.986 &36.454$\pm$0.152 &37.015$\pm$0.058 &39.320$\pm$0.064 &2.862$\pm$0.103 &1.964$\pm$0.219\\
   33 &13:29:46.8282,+47:12:17.867 &37.088$\pm$0.084 &37.058$\pm$0.056 &39.086$\pm$0.070 &2.950$\pm$0.105 &1.165$\pm$0.137\\
   34 &13:29:49.0814,+47:11:59.630 &37.804$\pm$0.051 &37.394$\pm$0.044 &39.670$\pm$0.058 &3.015$\pm$0.081 &0.652$\pm$0.090\\
   35 &13:29:45.5453,+47:11:56.503 &36.922$\pm$0.097 &37.356$\pm$0.045 &39.532$\pm$0.060 &3.254$\pm$0.100 &1.792$\pm$0.145\\
   36 &13:29:45.0940,+47:11:53.222 &36.468$\pm$0.149 &36.820$\pm$0.069 &38.868$\pm$0.078 &2.572$\pm$0.104 &1.681$\pm$0.222\\
   37 &13:29:44.8426,+47:12:01.061 &36.569$\pm$0.135 &36.368$\pm$0.111 &38.571$\pm$0.094 &2.166$\pm$0.137 &0.934$\pm$0.236\\
   38 &13:29:45.6710,+47:11:51.744 &36.962$\pm$0.094 &36.970$\pm$0.060 &39.077$\pm$0.070 &2.926$\pm$0.112 &1.216$\pm$0.151\\
   39 &13:29:45.7294,+47:12:08.144 &37.118$\pm$0.082 &36.776$\pm$0.072 &38.926$\pm$0.076 &2.811$\pm$0.125 &0.743$\pm$0.148\\
   40 &13:29:43.0342,+47:11:37.974 &38.005$\pm$0.046 &37.503$\pm$0.041 &39.602$\pm$0.058 &2.993$\pm$0.073 &0.529$\pm$0.083\\
   41 &13:29:44.9293,+47:11:47.341 &37.505$\pm$0.061 &37.177$\pm$0.051 &39.265$\pm$0.065 &2.663$\pm$0.079 &0.764$\pm$0.108\\
   42 &13:29:42.6063,+47:11:40.172 &37.482$\pm$0.062 &36.909$\pm$0.064 &38.830$\pm$0.080 &2.675$\pm$0.101 &0.432$\pm$0.120\\
   43 &13:29:44.9572,+47:11:33.821 &37.139$\pm$0.081 &36.529$\pm$0.093 &38.636$\pm$0.090 &2.223$\pm$0.118 &0.382$\pm$0.166\\
   44 &13:29:45.0906,+47:11:35.342 &37.018$\pm$0.090 &36.535$\pm$0.092 &38.513$\pm$0.098 &2.411$\pm$0.127 &0.554$\pm$0.174\\
   45 &13:29:43.2977,+47:11:22.655 &37.122$\pm$0.082 &37.029$\pm$0.057 &39.099$\pm$0.069 &2.844$\pm$0.100 &1.081$\pm$0.135\\
   46 &13:29:44.6084,+47:11:20.580 &37.266$\pm$0.073 &36.569$\pm$0.089 &38.800$\pm$0.081 &2.316$\pm$0.117 &0.265$\pm$0.155\\
   47 &13:29:42.9453,+47:11:05.294 &37.730$\pm$0.053 &37.436$\pm$0.042 &39.774$\pm$0.056 &3.045$\pm$0.080 &0.808$\pm$0.092\\
   48 &13:29:42.6317,+47:10:58.813 &37.472$\pm$0.063 &37.398$\pm$0.043 &39.690$\pm$0.057 &2.982$\pm$0.079 &1.106$\pm$0.103\\
   49 &13:29:42.7728,+47:11:00.093 &37.398$\pm$0.066 &37.035$\pm$0.057 &39.201$\pm$0.067 &2.985$\pm$0.110 &0.716$\pm$0.118\\
   50 &13:29:42.6630,+47:11:00.173 &37.439$\pm$0.064 &36.958$\pm$0.061 &38.985$\pm$0.073 &2.916$\pm$0.113 &0.557$\pm$0.120\\
   51 &13:29:41.9484,+47:11:11.130 &37.942$\pm$0.047 &37.372$\pm$0.044 &39.323$\pm$0.064 &3.091$\pm$0.087 &0.436$\pm$0.087\\
   52 &13:29:44.0641,+47:10:47.658 &36.979$\pm$0.093 &36.408$\pm$0.106 &38.336$\pm$0.113 &2.187$\pm$0.132 &0.435$\pm$0.190\\
   53 &13:29:43.0289,+47:10:35.814 &35.870$\pm$0.279 &36.034$\pm$0.166 &38.624$\pm$0.091 &2.168$\pm$0.201 &1.427$\pm$0.438\\
   54 &13:29:43.2212,+47:10:33.775 &37.311$\pm$0.071 &37.339$\pm$0.045 &39.482$\pm$0.060 &3.273$\pm$0.103 &1.244$\pm$0.113\\
   55 &13:29:44.0099,+47:10:29.178 &36.859$\pm$0.103 &36.499$\pm$0.096 &38.578$\pm$0.094 &2.275$\pm$0.124 &0.719$\pm$0.190\\
   56 &13:29:43.3352,+47:10:27.496 &37.046$\pm$0.087 &36.421$\pm$0.104 &37.550$\pm$0.296 &1.894$\pm$0.120 &0.361$\pm$0.184\\
   57 &13:29:43.2176,+47:10:23.975 &38.243$\pm$0.042 &37.786$\pm$0.035 &40.030$\pm$0.054 &3.300$\pm$0.071 &0.589$\pm$0.073\\
   58 &13:29:43.3507,+47:10:34.016 &37.057$\pm$0.087 &37.109$\pm$0.054 &39.289$\pm$0.064 &2.913$\pm$0.097 &1.276$\pm$0.138\\
   59 &13:29:44.0809,+47:10:18.138 &38.169$\pm$0.043 &37.809$\pm$0.034 &39.979$\pm$0.054 &2.959$\pm$0.057 &0.720$\pm$0.074\\
   60 &13:29:44.0650,+47:10:22.698 &38.703$\pm$0.037 &38.365$\pm$0.029 &40.662$\pm$0.051 &3.387$\pm$0.051 &0.749$\pm$0.064\\
   61 &13:29:44.1591,+47:10:23.579 &38.642$\pm$0.038 &38.156$\pm$0.030 &40.331$\pm$0.052 &3.050$\pm$0.049 &0.550$\pm$0.065\\
   62 &13:29:44.1747,+47:10:25.299 &38.124$\pm$0.044 &38.107$\pm$0.031 &40.213$\pm$0.053 &3.532$\pm$0.065 &1.183$\pm$0.072\\
   63 &13:29:43.8454,+47:10:20.898 &37.468$\pm$0.063 &37.238$\pm$0.049 &39.335$\pm$0.063 &3.014$\pm$0.093 &0.895$\pm$0.107\\
   64 &13:29:45.1124,+47:10:22.302 &36.506$\pm$0.144 &36.883$\pm$0.065 &38.740$\pm$0.084 &2.879$\pm$0.118 &1.715$\pm$0.213\\
   65 &13:29:45.1673,+47:10:23.662 &36.810$\pm$0.108 &36.590$\pm$0.087 &38.880$\pm$0.077 &2.446$\pm$0.122 &0.909$\pm$0.187\\
   66 &13:29:44.3788,+47:10:23.900 &38.045$\pm$0.045 &37.472$\pm$0.041 &39.360$\pm$0.063 &2.997$\pm$0.075 &0.433$\pm$0.083\\
   67 &13:29:43.7084,+47:10:12.137 &36.900$\pm$0.099 &36.582$\pm$0.088 &38.753$\pm$0.083 &2.448$\pm$0.123 &0.777$\pm$0.179\\
   68 &13:29:43.5789,+47:10:14.296 &37.039$\pm$0.088 &36.382$\pm$0.109 &38.269$\pm$0.121 &2.495$\pm$0.156 &0.318$\pm$0.189\\
   69 &13:29:46.8975,+47:10:12.667 &37.627$\pm$0.057 &37.373$\pm$0.044 &39.251$\pm$0.065 &3.140$\pm$0.090 &0.862$\pm$0.097\\
   70 &13:29:45.4149,+47:10:05.983 &37.325$\pm$0.070 &36.974$\pm$0.060 &39.013$\pm$0.072 &2.381$\pm$0.082 &0.733$\pm$0.125\\
   71 &13:29:45.3523,+47:10:01.943 &37.944$\pm$0.047 &37.643$\pm$0.037 &39.654$\pm$0.058 &3.217$\pm$0.075 &0.799$\pm$0.082\\
   72 &13:29:45.1602,+47:09:56.902 &37.974$\pm$0.047 &38.270$\pm$0.029 &40.662$\pm$0.051 &3.390$\pm$0.054 &1.606$\pm$0.075\\
   73 &13:29:43.6814,+47:10:00.897 &38.011$\pm$0.046 &37.826$\pm$0.034 &39.859$\pm$0.055 &3.188$\pm$0.064 &0.956$\pm$0.077\\
   74 &13:29:43.6697,+47:09:59.217 &37.341$\pm$0.069 &37.160$\pm$0.052 &39.342$\pm$0.063 &3.257$\pm$0.121 &0.962$\pm$0.116\\
   75 &13:29:43.5287,+47:09:55.456 &37.298$\pm$0.071 &37.037$\pm$0.057 &38.977$\pm$0.074 &2.992$\pm$0.110 &0.854$\pm$0.123\\
   76 &13:29:43.9168,+47:10:00.618 &37.381$\pm$0.067 &36.811$\pm$0.070 &38.600$\pm$0.092 &2.495$\pm$0.101 &0.436$\pm$0.131\\
   77 &13:29:43.7007,+47:10:09.217 &36.598$\pm$0.132 &37.044$\pm$0.057 &39.291$\pm$0.064 &2.773$\pm$0.094 &1.807$\pm$0.193\\
   78 &13:29:46.0538,+47:10:36.388 &37.245$\pm$0.074 &36.791$\pm$0.071 &38.973$\pm$0.074 &2.709$\pm$0.116 &0.594$\pm$0.139\\
   79 &13:29:47.9134,+47:10:28.232 &37.518$\pm$0.061 &37.144$\pm$0.052 &39.140$\pm$0.068 &2.829$\pm$0.089 &0.701$\pm$0.108\\
   80 &13:29:48.0389,+47:10:30.793 &37.650$\pm$0.056 &37.173$\pm$0.051 &39.383$\pm$0.062 &3.149$\pm$0.109 &0.561$\pm$0.102\\
   81 &13:29:46.3601,+47:10:24.549 &36.267$\pm$0.183 &36.375$\pm$0.110 &38.474$\pm$0.101 &3.002$\pm$0.224 &1.353$\pm$0.288\\
   82 &13:29:46.4382,+47:10:37.509 &36.844$\pm$0.104 &36.353$\pm$0.112 &38.430$\pm$0.105 &2.371$\pm$0.151 &0.543$\pm$0.207\\
   83 &13:29:44.7434,+47:10:38.145 &37.029$\pm$0.089 &36.311$\pm$0.118 &37.876$\pm$0.186 &2.159$\pm$0.145 &0.237$\pm$0.199\\
   84 &13:29:47.7957,+47:10:31.352 &36.815$\pm$0.107 &36.359$\pm$0.112 &38.288$\pm$0.118 &2.369$\pm$0.150 &0.590$\pm$0.209\\
   85 &13:29:48.9644,+47:10:37.670 &36.951$\pm$0.095 &36.567$\pm$0.089 &38.504$\pm$0.099 &2.564$\pm$0.133 &0.687$\pm$0.176\\
   86 &13:29:50.6592,+47:10:42.071 &37.008$\pm$0.090 &36.769$\pm$0.073 &38.935$\pm$0.075 &2.559$\pm$0.108 &0.883$\pm$0.156\\
   87 &13:29:50.4552,+47:10:49.191 &37.339$\pm$0.069 &36.766$\pm$0.073 &38.767$\pm$0.083 &2.273$\pm$0.095 &0.433$\pm$0.136\\
   88 &13:29:50.2355,+47:10:51.191 &37.533$\pm$0.060 &37.098$\pm$0.054 &38.872$\pm$0.078 &2.878$\pm$0.096 &0.620$\pm$0.109\\
   89 &13:29:45.3941,+47:10:51.423 &37.434$\pm$0.064 &36.866$\pm$0.066 &38.790$\pm$0.082 &2.660$\pm$0.104 &0.439$\pm$0.125\\
   90 &13:29:50.7063,+47:11:10.551 &37.221$\pm$0.076 &37.268$\pm$0.048 &39.440$\pm$0.061 &3.122$\pm$0.098 &1.270$\pm$0.121\\
   91 &13:29:50.8161,+47:10:43.031 &36.837$\pm$0.105 &36.479$\pm$0.098 &38.508$\pm$0.099 &2.473$\pm$0.139 &0.722$\pm$0.194\\
   92 &13:29:52.0058,+47:10:54.303 &38.136$\pm$0.044 &37.668$\pm$0.037 &39.926$\pm$0.055 &3.345$\pm$0.081 &0.573$\pm$0.077\\
   93 &13:29:51.0171,+47:11:25.743 &37.205$\pm$0.077 &37.202$\pm$0.050 &39.615$\pm$0.058 &2.987$\pm$0.094 &1.203$\pm$0.124\\
   94 &13:29:51.4096,+47:11:27.583 &36.833$\pm$0.106 &37.082$\pm$0.055 &39.096$\pm$0.070 &2.931$\pm$0.101 &1.542$\pm$0.161\\
   95 &13:29:51.0249,+47:11:31.423 &36.753$\pm$0.114 &36.905$\pm$0.064 &39.242$\pm$0.065 &2.211$\pm$0.082 &1.411$\pm$0.176\\
   96 &13:29:50.9543,+47:11:33.343 &37.773$\pm$0.052 &37.568$\pm$0.039 &39.554$\pm$0.059 &2.945$\pm$0.067 &0.929$\pm$0.088\\
   97 &13:29:50.4049,+47:11:24.383 &37.443$\pm$0.064 &37.040$\pm$0.057 &38.990$\pm$0.073 &2.546$\pm$0.084 &0.662$\pm$0.116\\
   98 &13:29:51.5194,+47:11:20.303 &37.589$\pm$0.058 &37.226$\pm$0.049 &39.200$\pm$0.067 &2.958$\pm$0.090 &0.716$\pm$0.103\\
   99 &13:29:51.7391,+47:11:23.503 &37.214$\pm$0.076 &37.186$\pm$0.051 &39.287$\pm$0.064 &2.477$\pm$0.072 &1.168$\pm$0.123\\
  100 &13:29:52.1786,+47:11:27.183 &36.885$\pm$0.101 &36.930$\pm$0.063 &39.484$\pm$0.060 &2.869$\pm$0.112 &1.267$\pm$0.160\\
  101 &13:29:53.5552,+47:11:22.669 &38.045$\pm$0.045 &37.799$\pm$0.035 &39.947$\pm$0.055 &3.152$\pm$0.064 &0.873$\pm$0.077\\
  102 &13:29:54.0887,+47:11:19.228 &37.276$\pm$0.073 &37.329$\pm$0.046 &39.527$\pm$0.060 &3.086$\pm$0.090 &1.278$\pm$0.116\\
  103 &13:29:50.0705,+47:11:36.911 &37.892$\pm$0.049 &37.713$\pm$0.036 &39.787$\pm$0.056 &3.585$\pm$0.094 &0.963$\pm$0.082\\
  104 &13:29:49.6231,+47:11:40.670 &37.418$\pm$0.065 &37.109$\pm$0.054 &39.261$\pm$0.065 &2.397$\pm$0.074 &0.788$\pm$0.114\\
  105 &13:29:51.2086,+47:11:38.751 &36.399$\pm$0.160 &36.430$\pm$0.103 &38.811$\pm$0.081 &2.773$\pm$0.176 &1.248$\pm$0.257\\
  106 &13:29:51.2949,+47:11:40.911 &37.258$\pm$0.074 &37.271$\pm$0.047 &39.590$\pm$0.059 &3.486$\pm$0.133 &1.224$\pm$0.118\\
  107 &13:29:53.8928,+47:11:34.668 &37.232$\pm$0.075 &37.060$\pm$0.056 &39.198$\pm$0.067 &2.165$\pm$0.071 &0.974$\pm$0.126\\
  108 &13:29:53.7672,+47:11:35.388 &37.076$\pm$0.085 &36.966$\pm$0.061 &39.182$\pm$0.067 &2.614$\pm$0.093 &1.058$\pm$0.141\\
  109 &13:29:54.1438,+47:11:29.148 &37.104$\pm$0.083 &36.704$\pm$0.077 &38.859$\pm$0.078 &2.623$\pm$0.119 &0.667$\pm$0.154\\
  110 &13:29:57.1960,+47:11:35.014 &37.656$\pm$0.056 &37.275$\pm$0.047 &39.191$\pm$0.067 &2.864$\pm$0.081 &0.691$\pm$0.099\\
  111 &13:29:54.0258,+47:12:12.862 &38.246$\pm$0.042 &37.783$\pm$0.035 &39.965$\pm$0.054 &3.046$\pm$0.061 &0.581$\pm$0.073\\
  112 &13:29:54.3555,+47:12:13.181 &37.848$\pm$0.050 &37.591$\pm$0.039 &39.766$\pm$0.056 &3.228$\pm$0.079 &0.859$\pm$0.085\\
  113 &13:29:53.2564,+47:12:09.183 &37.631$\pm$0.056 &37.575$\pm$0.039 &39.930$\pm$0.055 &3.208$\pm$0.079 &1.130$\pm$0.093\\
  114 &13:29:53.1937,+47:12:11.743 &37.553$\pm$0.059 &37.166$\pm$0.051 &39.295$\pm$0.064 &2.731$\pm$0.083 &0.684$\pm$0.106\\
  115 &13:29:52.1576,+47:12:21.744 &37.258$\pm$0.074 &36.859$\pm$0.067 &39.116$\pm$0.069 &2.510$\pm$0.097 &0.666$\pm$0.134\\
  116 &13:29:56.2082,+47:12:15.817 &37.236$\pm$0.075 &36.834$\pm$0.068 &38.779$\pm$0.082 &2.458$\pm$0.097 &0.663$\pm$0.137\\
  117 &13:29:55.6740,+47:12:02.618 &37.637$\pm$0.056 &37.144$\pm$0.052 &39.155$\pm$0.068 &2.836$\pm$0.090 &0.541$\pm$0.104\\
  118 &13:29:51.1605,+47:12:10.225 &36.269$\pm$0.183 &36.704$\pm$0.078 &38.799$\pm$0.081 &2.628$\pm$0.120 &1.792$\pm$0.268\\
  119 &13:29:53.2407,+47:12:05.423 &36.945$\pm$0.095 &36.965$\pm$0.061 &38.976$\pm$0.074 &3.060$\pm$0.125 &1.233$\pm$0.153\\
  120 &13:29:55.1714,+47:11:50.060 &37.209$\pm$0.076 &37.026$\pm$0.058 &39.186$\pm$0.067 &2.503$\pm$0.083 &0.959$\pm$0.129\\
  121 &13:29:54.5672,+47:12:00.941 &37.625$\pm$0.057 &37.011$\pm$0.058 &38.885$\pm$0.077 &2.969$\pm$0.111 &0.377$\pm$0.110\\
  122 &13:29:52.7148,+47:12:12.384 &36.148$\pm$0.207 &36.666$\pm$0.080 &38.829$\pm$0.080 &2.347$\pm$0.108 &1.905$\pm$0.300\\
  123 &13:29:51.2311,+47:12:04.225 &37.129$\pm$0.082 &36.745$\pm$0.074 &38.955$\pm$0.074 &2.255$\pm$0.097 &0.688$\pm$0.149\\
  124 &13:29:50.0930,+47:11:42.305 &37.501$\pm$0.062 &37.585$\pm$0.039 &39.904$\pm$0.055 &3.103$\pm$0.073 &1.319$\pm$0.098\\
  125 &13:29:51.4901,+47:11:51.745 &37.601$\pm$0.058 &37.279$\pm$0.047 &39.472$\pm$0.061 &2.572$\pm$0.070 &0.771$\pm$0.101\\
  126 &13:29:51.4901,+47:11:53.665 &37.016$\pm$0.090 &37.044$\pm$0.057 &39.121$\pm$0.069 &2.513$\pm$0.082 &1.243$\pm$0.143\\
  127 &13:29:52.2751,+47:12:00.864 &35.780$\pm$0.308 &36.740$\pm$0.075 &38.628$\pm$0.090 &2.201$\pm$0.095 &2.501$\pm$0.428\\
  128 &13:29:54.3550,+47:11:47.741 &37.430$\pm$0.065 &37.033$\pm$0.057 &39.349$\pm$0.063 &2.510$\pm$0.083 &0.670$\pm$0.117\\
  129 &13:29:54.2294,+47:11:46.942 &36.783$\pm$0.110 &36.982$\pm$0.060 &39.180$\pm$0.067 &2.439$\pm$0.084 &1.474$\pm$0.170\\
  130 &13:29:50.1950,+47:11:51.345 &37.481$\pm$0.062 &37.555$\pm$0.039 &39.976$\pm$0.054 &3.229$\pm$0.082 &1.306$\pm$0.100\\
  131 &13:29:50.2422,+47:11:19.265 &38.157$\pm$0.043 &37.785$\pm$0.035 &40.074$\pm$0.054 &3.209$\pm$0.067 &0.704$\pm$0.075\\
  132 &13:29:49.9361,+47:11:30.945 &36.863$\pm$0.103 &37.768$\pm$0.035 &40.188$\pm$0.053 &3.371$\pm$0.076 &2.427$\pm$0.147\\
  133 &13:29:50.8700,+47:11:56.865 &37.934$\pm$0.048 &37.550$\pm$0.039 &39.893$\pm$0.055 &2.584$\pm$0.058 &0.688$\pm$0.083\\
  134 &13:29:51.5843,+47:11:57.505 &37.677$\pm$0.055 &37.549$\pm$0.039 &39.720$\pm$0.057 &2.713$\pm$0.061 &1.033$\pm$0.091\\
  135 &13:29:51.5101,+47:12:00.516 &37.876$\pm$0.049 &37.703$\pm$0.036 &40.254$\pm$0.053 &2.932$\pm$0.061 &0.972$\pm$0.082\\
  136 &13:29:51.7378,+47:12:02.036 &37.583$\pm$0.058 &37.720$\pm$0.036 &40.021$\pm$0.054 &3.787$\pm$0.112 &1.391$\pm$0.092\\
  137 &13:29:52.5149,+47:12:03.075 &37.931$\pm$0.048 &37.777$\pm$0.035 &40.185$\pm$0.053 &3.117$\pm$0.064 &0.998$\pm$0.080\\
  138 &13:29:52.0518,+47:12:04.516 &37.257$\pm$0.074 &37.563$\pm$0.039 &39.904$\pm$0.055 &2.683$\pm$0.059 &1.620$\pm$0.113\\
  139 &13:29:51.0627,+47:11:58.116 &37.372$\pm$0.067 &37.184$\pm$0.051 &39.511$\pm$0.060 &2.451$\pm$0.071 &0.952$\pm$0.114\\
  140 &13:29:50.4818,+47:12:08.116 &37.231$\pm$0.075 &36.718$\pm$0.076 &38.781$\pm$0.082 &2.554$\pm$0.113 &0.514$\pm$0.145\\
  141 &13:29:48.6843,+47:11:56.755 &36.697$\pm$0.120 &36.782$\pm$0.072 &39.012$\pm$0.072 &2.649$\pm$0.112 &1.320$\pm$0.188\\
  142 &13:29:55.4501,+47:11:40.190 &37.812$\pm$0.051 &37.804$\pm$0.035 &39.983$\pm$0.054 &2.869$\pm$0.055 &1.195$\pm$0.083\\
  143 &13:29:55.3715,+47:11:36.430 &37.369$\pm$0.068 &37.586$\pm$0.039 &39.900$\pm$0.055 &2.929$\pm$0.066 &1.498$\pm$0.105\\
  144 &13:29:56.3524,+47:11:31.388 &37.418$\pm$0.065 &36.767$\pm$0.073 &38.777$\pm$0.082 &2.491$\pm$0.105 &0.327$\pm$0.132\\
  145 &13:29:55.9137,+47:11:59.549 &37.030$\pm$0.089 &36.785$\pm$0.072 &38.956$\pm$0.074 &2.287$\pm$0.094 &0.875$\pm$0.154\\
  146 &13:29:53.1896,+47:11:29.554 &36.789$\pm$0.110 &37.075$\pm$0.055 &38.901$\pm$0.077 &3.467$\pm$0.160 &1.592$\pm$0.166\\
  147 &13:29:55.8168,+47:11:44.867 &38.442$\pm$0.039 &38.369$\pm$0.029 &40.745$\pm$0.051 &3.187$\pm$0.047 &1.107$\pm$0.066\\
  148 &13:29:52.8104,+47:11:23.993 &37.148$\pm$0.080 &36.867$\pm$0.066 &38.872$\pm$0.078 &2.050$\pm$0.081 &0.826$\pm$0.141\\
  149 &13:29:56.6104,+47:12:12.225 &37.344$\pm$0.069 &37.257$\pm$0.048 &39.243$\pm$0.065 &3.067$\pm$0.095 &1.088$\pm$0.113\\
  150 &13:30:01.1402,+47:12:18.685 &37.127$\pm$0.082 &36.619$\pm$0.084 &38.581$\pm$0.093 &2.484$\pm$0.121 &0.520$\pm$0.159\\
  151 &13:30:01.8163,+47:12:34.841 &37.645$\pm$0.056 &37.318$\pm$0.046 &39.319$\pm$0.064 &3.031$\pm$0.087 &0.765$\pm$0.098\\
  152 &13:30:01.3128,+47:12:16.684 &37.138$\pm$0.081 &36.619$\pm$0.084 &38.450$\pm$0.103 &2.330$\pm$0.112 &0.506$\pm$0.158\\
  153 &13:30:02.1539,+47:12:36.919 &37.054$\pm$0.087 &36.702$\pm$0.078 &38.827$\pm$0.080 &2.250$\pm$0.100 &0.731$\pm$0.157\\
  154 &13:30:01.4004,+47:12:39.163 &37.375$\pm$0.067 &37.154$\pm$0.052 &39.321$\pm$0.064 &2.756$\pm$0.085 &0.907$\pm$0.115\\
  155 &13:30:01.1963,+47:12:38.765 &36.024$\pm$0.236 &36.292$\pm$0.121 &39.043$\pm$0.071 &2.066$\pm$0.144 &1.567$\pm$0.358\\
  156 &13:30:01.4316,+47:12:34.763 &36.928$\pm$0.097 &36.994$\pm$0.059 &38.999$\pm$0.073 &2.897$\pm$0.107 &1.295$\pm$0.153\\
  157 &13:30:01.6046,+47:12:39.562 &36.911$\pm$0.098 &36.709$\pm$0.077 &38.818$\pm$0.080 &2.236$\pm$0.099 &0.933$\pm$0.169\\
  158 &13:30:03.6998,+47:12:36.186 &37.472$\pm$0.063 &37.284$\pm$0.047 &39.333$\pm$0.063 &3.158$\pm$0.099 &0.951$\pm$0.106\\
  159 &13:30:04.0067,+47:12:46.343 &36.823$\pm$0.106 &36.660$\pm$0.081 &38.718$\pm$0.085 &2.563$\pm$0.121 &0.986$\pm$0.181\\
  160 &13:30:03.4026,+47:12:53.708 &37.681$\pm$0.055 &37.141$\pm$0.052 &39.059$\pm$0.071 &3.024$\pm$0.102 &0.477$\pm$0.102\\
  161 &13:30:04.2742,+47:12:54.262 &37.433$\pm$0.065 &37.026$\pm$0.058 &38.696$\pm$0.086 &2.545$\pm$0.085 &0.655$\pm$0.117\\
  162 &13:30:03.2465,+47:13:08.349 &37.062$\pm$0.086 &36.906$\pm$0.064 &39.221$\pm$0.066 &2.938$\pm$0.121 &0.995$\pm$0.145\\
  163 &13:29:59.3672,+47:13:01.731 &37.281$\pm$0.072 &36.877$\pm$0.066 &38.981$\pm$0.073 &2.630$\pm$0.101 &0.661$\pm$0.132\\
  164 &13:29:59.3354,+47:12:54.291 &36.797$\pm$0.109 &36.659$\pm$0.081 &38.476$\pm$0.101 &2.797$\pm$0.140 &1.020$\pm$0.183\\
  165 &13:30:02.5160,+47:13:04.034 &37.685$\pm$0.055 &37.187$\pm$0.051 &39.178$\pm$0.067 &2.628$\pm$0.077 &0.533$\pm$0.101\\
  166 &13:30:01.3933,+47:13:07.640 &37.820$\pm$0.051 &37.320$\pm$0.046 &39.461$\pm$0.061 &2.948$\pm$0.082 &0.531$\pm$0.092\\
  167 &13:30:01.2128,+47:13:10.201 &37.290$\pm$0.072 &36.894$\pm$0.065 &38.926$\pm$0.076 &2.379$\pm$0.088 &0.672$\pm$0.131\\
  168 &13:30:04.7615,+47:13:01.458 &37.789$\pm$0.051 &37.473$\pm$0.041 &39.356$\pm$0.063 &2.972$\pm$0.074 &0.780$\pm$0.089\\
  169 &13:30:01.0389,+47:12:48.602 &38.201$\pm$0.042 &37.690$\pm$0.036 &39.761$\pm$0.056 &3.342$\pm$0.079 &0.515$\pm$0.076\\
  170 &13:30:01.4709,+47:12:51.560 &38.082$\pm$0.045 &38.456$\pm$0.028 &41.240$\pm$0.051 &3.396$\pm$0.049 &1.710$\pm$0.071\\
  171 &13:30:03.6525,+47:12:34.346 &37.487$\pm$0.062 &36.838$\pm$0.068 &38.632$\pm$0.090 &2.786$\pm$0.116 &0.331$\pm$0.124\\
  172 &13:30:03.6838,+47:12:32.746 &37.284$\pm$0.072 &36.601$\pm$0.086 &38.424$\pm$0.105 &2.406$\pm$0.118 &0.284$\pm$0.151\\
  173 &13:30:05.5833,+47:13:17.171 &37.092$\pm$0.084 &36.513$\pm$0.094 &38.411$\pm$0.106 &2.571$\pm$0.141 &0.423$\pm$0.171\\
  174 &13:30:05.2092,+47:13:02.974 &37.289$\pm$0.072 &36.811$\pm$0.070 &38.844$\pm$0.079 &2.649$\pm$0.109 &0.560$\pm$0.135\\
  175 &13:30:05.3789,+47:13:14.773 &37.192$\pm$0.078 &36.643$\pm$0.082 &38.492$\pm$0.100 &2.643$\pm$0.129 &0.465$\pm$0.153\\
  176 &13:30:03.6405,+47:13:28.466 &37.132$\pm$0.081 &36.672$\pm$0.080 &38.767$\pm$0.083 &2.906$\pm$0.150 &0.584$\pm$0.154\\
  177 &13:30:02.9526,+47:13:17.871 &37.419$\pm$0.065 &36.945$\pm$0.062 &38.881$\pm$0.077 &2.780$\pm$0.104 &0.565$\pm$0.121\\
  178 &13:30:01.1105,+47:13:06.482 &37.573$\pm$0.059 &37.308$\pm$0.046 &39.386$\pm$0.062 &2.561$\pm$0.068 &0.848$\pm$0.101\\
  179 &13:29:59.7294,+47:13:25.609 &37.338$\pm$0.069 &37.141$\pm$0.052 &39.292$\pm$0.064 &2.706$\pm$0.083 &0.941$\pm$0.117\\
  180 &13:29:59.8078,+47:13:22.169 &37.193$\pm$0.077 &37.291$\pm$0.047 &39.229$\pm$0.066 &3.182$\pm$0.100 &1.338$\pm$0.122\\
  181 &13:29:58.8178,+47:13:10.813 &37.502$\pm$0.061 &37.060$\pm$0.056 &38.981$\pm$0.073 &2.927$\pm$0.103 &0.609$\pm$0.112\\
  182 &13:29:58.6692,+47:13:23.774 &37.241$\pm$0.075 &36.849$\pm$0.067 &38.866$\pm$0.078 &2.755$\pm$0.112 &0.676$\pm$0.136\\
  183 &13:29:59.1761,+47:13:33.012 &37.406$\pm$0.066 &37.117$\pm$0.053 &39.081$\pm$0.070 &2.711$\pm$0.085 &0.815$\pm$0.115\\
  184 &13:29:59.3101,+47:13:44.331 &37.514$\pm$0.061 &37.438$\pm$0.042 &39.509$\pm$0.060 &3.077$\pm$0.081 &1.103$\pm$0.100\\
  185 &13:29:59.7816,+47:13:49.449 &37.190$\pm$0.078 &37.204$\pm$0.050 &39.172$\pm$0.067 &3.053$\pm$0.098 &1.225$\pm$0.125\\
  186 &13:30:00.5195,+47:13:42.005 &37.329$\pm$0.070 &36.605$\pm$0.086 &38.460$\pm$0.102 &2.589$\pm$0.130 &0.229$\pm$0.149\\
  187 &13:29:57.9632,+47:13:46.137 &37.360$\pm$0.068 &37.018$\pm$0.058 &39.218$\pm$0.066 &2.929$\pm$0.107 &0.745$\pm$0.121\\
  188 &13:30:02.3154,+47:14:04.195 &38.114$\pm$0.044 &37.495$\pm$0.041 &39.169$\pm$0.067 &2.950$\pm$0.071 &0.371$\pm$0.081\\
  189 &13:30:00.9449,+47:14:05.563 &37.461$\pm$0.063 &36.996$\pm$0.059 &38.840$\pm$0.079 &2.883$\pm$0.106 &0.579$\pm$0.117\\
  190 &13:30:00.8075,+47:14:07.084 &37.470$\pm$0.063 &36.954$\pm$0.061 &38.701$\pm$0.086 &2.974$\pm$0.118 &0.509$\pm$0.119\\
  191 &13:29:58.5769,+47:14:11.214 &38.118$\pm$0.044 &37.505$\pm$0.041 &39.319$\pm$0.064 &3.224$\pm$0.085 &0.378$\pm$0.081\\
  192 &13:29:59.1463,+47:14:09.772 &37.220$\pm$0.076 &36.896$\pm$0.065 &38.941$\pm$0.075 &3.024$\pm$0.130 &0.768$\pm$0.134\\
  193 &13:29:58.9810,+47:14:02.453 &37.016$\pm$0.090 &36.613$\pm$0.085 &38.611$\pm$0.091 &2.678$\pm$0.136 &0.661$\pm$0.167\\
  194 &13:29:58.5018,+47:14:00.175 &37.561$\pm$0.059 &37.194$\pm$0.050 &39.411$\pm$0.062 &2.940$\pm$0.092 &0.711$\pm$0.105\\
  195 &13:29:58.0740,+47:14:07.216 &37.932$\pm$0.048 &37.491$\pm$0.041 &39.405$\pm$0.062 &3.094$\pm$0.078 &0.611$\pm$0.085\\
  196 &13:29:58.4666,+47:14:03.695 &37.626$\pm$0.057 &37.368$\pm$0.044 &39.446$\pm$0.061 &2.786$\pm$0.072 &0.857$\pm$0.097\\
  197 &13:29:57.4652,+47:14:05.179 &37.488$\pm$0.062 &37.062$\pm$0.056 &39.097$\pm$0.070 &3.021$\pm$0.110 &0.631$\pm$0.113\\
  198 &13:29:57.5512,+47:13:54.858 &37.716$\pm$0.054 &37.563$\pm$0.039 &39.888$\pm$0.055 &2.838$\pm$0.064 &0.999$\pm$0.090\\
  199 &13:29:57.4217,+47:13:56.699 &36.819$\pm$0.107 &36.696$\pm$0.078 &38.900$\pm$0.077 &2.527$\pm$0.114 &1.039$\pm$0.179\\
  200 &13:29:58.3522,+47:13:51.095 &37.782$\pm$0.052 &37.230$\pm$0.049 &39.275$\pm$0.065 &3.006$\pm$0.093 &0.460$\pm$0.096\\
  201 &13:29:57.0604,+47:13:37.431 &37.415$\pm$0.065 &37.138$\pm$0.053 &39.031$\pm$0.072 &2.940$\pm$0.097 &0.832$\pm$0.113\\
  202 &13:29:57.6814,+47:13:53.109 &37.453$\pm$0.064 &37.196$\pm$0.050 &39.178$\pm$0.067 &2.742$\pm$0.081 &0.858$\pm$0.109\\
  203 &13:29:58.1917,+47:13:48.307 &38.035$\pm$0.045 &37.643$\pm$0.037 &39.626$\pm$0.058 &3.159$\pm$0.072 &0.676$\pm$0.079\\
  204 &13:29:56.9338,+47:13:06.711 &37.160$\pm$0.080 &36.639$\pm$0.083 &38.579$\pm$0.094 &2.616$\pm$0.127 &0.502$\pm$0.155\\
  205 &13:29:57.0984,+47:12:58.471 &37.048$\pm$0.087 &36.630$\pm$0.083 &38.904$\pm$0.076 &2.742$\pm$0.139 &0.641$\pm$0.163\\
  206 &13:29:55.9517,+47:12:48.074 &37.279$\pm$0.072 &36.741$\pm$0.075 &38.791$\pm$0.081 &2.637$\pm$0.116 &0.479$\pm$0.140\\
  207 &13:29:52.6894,+47:12:44.081 &36.984$\pm$0.092 &36.832$\pm$0.069 &39.002$\pm$0.073 &2.295$\pm$0.090 &1.001$\pm$0.155\\
  208 &13:29:37.4764,+47:10:36.633 &37.148$\pm$0.080 &37.222$\pm$0.049 &38.779$\pm$0.082 &3.199$\pm$0.108 &1.306$\pm$0.127\\
  209 &13:29:37.3818,+47:10:42.553 &35.903$\pm$0.269 &36.380$\pm$0.109 &38.610$\pm$0.092 &2.512$\pm$0.157 &1.850$\pm$0.392\\
  210 &13:29:52.2745,+47:11:07.321 &37.370$\pm$0.068 &37.084$\pm$0.055 &39.112$\pm$0.069 &2.691$\pm$0.087 &0.820$\pm$0.117\\
  211 &13:29:54.2599,+47:11:06.518 &37.282$\pm$0.072 &36.677$\pm$0.080 &38.545$\pm$0.096 &2.462$\pm$0.113 &0.390$\pm$0.145\\
  212 &13:30:07.3768,+47:13:22.206 &38.130$\pm$0.044 &38.178$\pm$0.030 &40.364$\pm$0.052 &3.536$\pm$0.062 &1.271$\pm$0.072\\
  213 &13:30:06.2005,+47:13:43.257 &37.253$\pm$0.074 &36.779$\pm$0.072 &38.759$\pm$0.083 &2.892$\pm$0.132 &0.566$\pm$0.139\\
  214 &13:29:49.9980,+47:11:25.031 &37.917$\pm$0.048 &37.553$\pm$0.039 &39.798$\pm$0.056 &3.104$\pm$0.075 &0.714$\pm$0.084\\
  215 &13:30:00.3372,+47:13:09.646 &37.777$\pm$0.052 &37.532$\pm$0.040 &39.626$\pm$0.058 &3.150$\pm$0.079 &0.875$\pm$0.088\\
  216 &13:30:00.3848,+47:13:19.086 &38.227$\pm$0.042 &37.792$\pm$0.035 &39.987$\pm$0.054 &2.943$\pm$0.057 &0.618$\pm$0.074\\
  217 &13:30:00.0909,+47:13:30.447 &38.040$\pm$0.045 &37.729$\pm$0.036 &39.917$\pm$0.055 &3.336$\pm$0.076 &0.786$\pm$0.078\\
  218 &13:29:58.9578,+47:14:09.333 &37.183$\pm$0.078 &37.553$\pm$0.039 &40.024$\pm$0.054 &3.307$\pm$0.087 &1.706$\pm$0.118\\
  219 &13:29:58.8714,+47:14:08.493 &38.139$\pm$0.043 &37.771$\pm$0.035 &39.917$\pm$0.055 &2.978$\pm$0.059 &0.709$\pm$0.075\\
  220 &13:29:53.3150,+47:11:56.270 &37.137$\pm$0.081 &36.757$\pm$0.074 &38.902$\pm$0.077 &2.214$\pm$0.094 &0.694$\pm$0.148\\
  221 &13:29:52.5152,+47:12:01.039 &36.868$\pm$0.102 &36.674$\pm$0.080 &39.013$\pm$0.072 &2.245$\pm$0.103 &0.943$\pm$0.175\\
  222 &13:29:52.1126,+47:11:58.600 &37.046$\pm$0.087 &36.986$\pm$0.060 &39.167$\pm$0.067 &2.573$\pm$0.089 &1.125$\pm$0.143\\
  223 &13:29:50.3986,+47:12:10.024 &36.851$\pm$0.104 &36.312$\pm$0.118 &38.046$\pm$0.151 &2.086$\pm$0.142 &0.479$\pm$0.212\\
  224 &13:29:50.2517,+47:12:10.301 &37.067$\pm$0.086 &36.153$\pm$0.143 &38.236$\pm$0.124 &1.946$\pm$0.164 &-.028$\pm$0.225\\
  225 &13:29:49.3267,+47:12:01.872 &37.168$\pm$0.079 &36.610$\pm$0.085 &38.451$\pm$0.103 &2.727$\pm$0.140 &0.453$\pm$0.157\\
  226 &13:29:49.5442,+47:12:25.662 &36.689$\pm$0.121 &36.165$\pm$0.141 &38.016$\pm$0.157 &2.102$\pm$0.168 &0.499$\pm$0.250\\
  227 &13:29:49.3973,+47:12:10.080 &36.293$\pm$0.178 &35.616$\pm$0.305 &36.821$\pm$1.178 &1.475$\pm$0.317 &0.293$\pm$0.477\\
  228 &13:29:47.4162,+47:12:35.254 &37.649$\pm$0.056 &37.012$\pm$0.058 &38.805$\pm$0.081 &2.763$\pm$0.097 &0.345$\pm$0.109\\
  229 &13:29:47.2099,+47:12:10.077 &37.097$\pm$0.084 &36.472$\pm$0.099 &38.082$\pm$0.145 &2.363$\pm$0.132 &0.362$\pm$0.175\\
  230 &13:29:47.4711,+47:12:05.197 &36.962$\pm$0.094 &36.279$\pm$0.122 &38.063$\pm$0.149 &2.077$\pm$0.147 &0.285$\pm$0.208\\
  231 &13:29:52.4012,+47:12:34.480 &37.171$\pm$0.079 &36.722$\pm$0.076 &38.575$\pm$0.094 &2.646$\pm$0.119 &0.600$\pm$0.148\\
  232 &13:29:59.2853,+47:12:26.530 &36.801$\pm$0.109 &36.264$\pm$0.125 &38.197$\pm$0.129 &2.256$\pm$0.159 &0.481$\pm$0.223\\
  233 &13:29:58.0663,+47:12:25.592 &36.695$\pm$0.120 &36.161$\pm$0.141 &37.841$\pm$0.194 &2.176$\pm$0.173 &0.485$\pm$0.250\\
  234 &13:29:54.0220,+47:12:48.978 &37.249$\pm$0.074 &36.559$\pm$0.090 &38.385$\pm$0.109 &2.432$\pm$0.125 &0.274$\pm$0.157\\
  235 &13:29:56.4017,+47:12:49.554 &36.945$\pm$0.095 &36.313$\pm$0.118 &38.382$\pm$0.109 &2.342$\pm$0.156 &0.352$\pm$0.205\\
  236 &13:29:56.7341,+47:12:55.012 &36.999$\pm$0.091 &36.375$\pm$0.110 &38.166$\pm$0.133 &2.439$\pm$0.152 &0.363$\pm$0.192\\
  237 &13:30:02.5483,+47:13:26.905 &37.271$\pm$0.073 &36.651$\pm$0.082 &38.257$\pm$0.122 &2.319$\pm$0.108 &0.368$\pm$0.148\\
  238 &13:30:02.1890,+47:13:25.465 &37.222$\pm$0.076 &36.671$\pm$0.080 &38.494$\pm$0.100 &2.268$\pm$0.104 &0.462$\pm$0.149\\
  239 &13:29:58.0644,+47:13:57.286 &37.886$\pm$0.049 &37.204$\pm$0.050 &38.982$\pm$0.073 &2.943$\pm$0.091 &0.285$\pm$0.094\\
  240 &13:29:58.6415,+47:13:55.038 &37.486$\pm$0.062 &36.896$\pm$0.065 &38.677$\pm$0.087 &2.605$\pm$0.098 &0.409$\pm$0.121\\
  241 &13:29:58.3897,+47:13:21.489 &36.988$\pm$0.092 &36.479$\pm$0.098 &38.092$\pm$0.144 &2.459$\pm$0.138 &0.519$\pm$0.181\\
  242 &13:29:51.8192,+47:13:05.090 &37.369$\pm$0.068 &36.797$\pm$0.071 &38.726$\pm$0.085 &2.715$\pm$0.115 &0.433$\pm$0.132\\
  243 &13:29:51.7266,+47:13:07.280 &37.159$\pm$0.080 &36.897$\pm$0.065 &38.849$\pm$0.079 &2.800$\pm$0.110 &0.852$\pm$0.138\\
  244 &13:29:56.2821,+47:12:51.662 &37.046$\pm$0.087 &36.512$\pm$0.094 &38.456$\pm$0.103 &2.111$\pm$0.116 &0.484$\pm$0.174\\
  245 &13:29:55.5680,+47:12:09.907 &37.305$\pm$0.071 &36.452$\pm$0.101 &38.229$\pm$0.125 &2.028$\pm$0.120 &0.053$\pm$0.166\\
  246 &13:29:44.7023,+47:11:37.352 &36.985$\pm$0.092 &36.307$\pm$0.119 &38.209$\pm$0.128 &2.168$\pm$0.146 &0.289$\pm$0.203\\
  247 &13:29:43.6467,+47:11:42.450 &36.872$\pm$0.102 &36.497$\pm$0.096 &38.622$\pm$0.091 &2.161$\pm$0.119 &0.699$\pm$0.189\\
  248 &13:29:50.3823,+47:11:15.734 &37.462$\pm$0.063 &36.969$\pm$0.061 &38.690$\pm$0.087 &2.313$\pm$0.081 &0.540$\pm$0.118\\
  249 &13:29:50.5796,+47:10:52.140 &37.135$\pm$0.081 &36.112$\pm$0.150 &38.084$\pm$0.145 &2.013$\pm$0.175 &-.176$\pm$0.231\\
  250 &13:29:46.7164,+47:10:43.899 &37.293$\pm$0.072 &36.593$\pm$0.087 &38.198$\pm$0.129 &2.634$\pm$0.135 &0.261$\pm$0.152\\
  251 &13:29:46.4601,+47:11:15.120 &37.228$\pm$0.075 &36.888$\pm$0.065 &38.634$\pm$0.090 &2.568$\pm$0.097 &0.746$\pm$0.134\\
  252 &13:29:44.5954,+47:10:38.848 &37.365$\pm$0.068 &36.630$\pm$0.083 &38.288$\pm$0.118 &2.514$\pm$0.121 &0.214$\pm$0.145\\
  253 &13:30:01.6927,+47:12:15.232 &37.530$\pm$0.060 &37.140$\pm$0.052 &39.202$\pm$0.067 &2.851$\pm$0.091 &0.680$\pm$0.108\\
  254 &13:30:01.2654,+47:12:12.656 &37.601$\pm$0.058 &37.195$\pm$0.050 &39.113$\pm$0.069 &2.945$\pm$0.092 &0.658$\pm$0.103\\
\hline
\enddata
\tablenotetext{}{(1) The identification number of the source.}
\tablenotetext{}{(2) Right Ascension  and Declination in J2000 coordinates.}
\tablenotetext{}{(3)--(5) Logarithm of the luminosity  of each source at the indicated wavelength, in units of erg~s$^{-1}$. The photometry is measured in circular apertures with 0$^{\prime\prime}$.7 radius on the plane of the sky. The H$\alpha$ and Pa$\alpha$ luminosities are corrected for the Milky Way foreground extinction. See text for more details.}
\tablenotetext{}{(6) The logarithm of the equivalent width (EW) of Pa$\alpha$, in \AA, calculated from the ratio of the emission line flux  to the stellar continuum flux density. }
\tablenotetext{}{(7) The color excess, E(B$-$V), in mag, derived from the H$\alpha$/Pa$\alpha$ luminosity ratio.}
\end{deluxetable}

\bibliographystyle{aasjournal}
\bibliography{bibliography_ngc5194}{}

\begin{thebibliography}{}
\expandafter\ifx\csname natexlab\endcsname\relax\def\natexlab#1{#1}\fi

\bibitem[{{Adamo} {et~al.}(2017){Adamo}, {Ryon}, {Messa}, {Kim}, {Grasha},
  {Cook}, {Calzetti}, {Lee}, {Whitmore}, {Elmegreen}, {Ubeda}, {Smith},
  {Bright}, {Runnholm}, {Andrews}, {Fumagalli}, {Gouliermis}, {Kahre}, {Nair},
  {Thilker}, {Walterbos}, {Wofford}, {Aloisi}, {Ashworth}, {Brown}, {Chandar},
  {Christian}, {Cignoni}, {Clayton}, {Dale}, {de Mink}, {Dobbs}, {Elmegreen},
  {Evans}, {Gallagher}, {Grebel}, {Herrero}, {Hunter}, {Johnson}, {Kennicutt},
  {Krumholz}, {Lennon}, {Levay}, {Martin}, {Nota}, {{\"O}stlin}, {Pellerin},
  {Prieto}, {Regan}, {Sabbi}, {Sacchi}, {Schaerer}, {Schiminovich}, {Shabani},
  {Tosi}, {Van Dyk}, \& {Zackrisson}}]{Adamo+2017}
{Adamo}, A., {Ryon}, J.~E., {Messa}, M., {et~al.} 2017, \apj, 841, 131

\bibitem[{{Adamo} {et~al.}(2020){Adamo}, {Zeidler}, {Kruijssen}, {Chevance},
  {Gieles}, {Calzetti}, {Charbonnel}, {Zinnecker}, \& {Krause}}]{Adamo+2020}
{Adamo}, A., {Zeidler}, P., {Kruijssen}, J.~M.~D., {et~al.} 2020, \ssr, 216, 69

\bibitem[{{Alonso-Herrero} {et~al.}(2006){Alonso-Herrero}, {Rieke}, {Rieke},
  {Colina}, {P{\'e}rez-Gonz{\'a}lez}, \& {Ryder}}]{Alonso+2006}
{Alonso-Herrero}, A., {Rieke}, G.~H., {Rieke}, M.~J., {et~al.} 2006, \apj, 650,
  835

\bibitem[{{Asplund} {et~al.}(2009){Asplund}, {Grevesse}, {Sauval}, \&
  {Scott}}]{Asplund+2009}
{Asplund}, M., {Grevesse}, N., {Sauval}, A.~J., \& {Scott}, P. 2009, \araa, 47,
  481

\bibitem[{{Belfiore} {et~al.}(2023){Belfiore}, {Leroy}, {Williams}, {Barnes},
  {Bigiel}, {Boquien}, {Cao}, {Chastenet}, {Congiu}, {Dale}, {Egorov},
  {Eibensteiner}, {Emsellem}, {Glover}, {Groves}, {Hassani}, {Klessen},
  {Kreckel}, {Neumann}, {Neumann}, {Querejeta}, {Rosolowsky},
  {Sanchez-Blazquez}, {Sandstrom}, {Schinnerer}, {Sun}, {Sutter}, \&
  {Watkins}}]{Belfiore+2023}
{Belfiore}, F., {Leroy}, A.~K., {Williams}, T.~G., {et~al.} 2023, \aap, 678,
  A129

\bibitem[{{Bendo} {et~al.}(2012){Bendo}, {Boselli}, {Dariush}, {Pohlen},
  {Roussel}, {Sauvage}, {Smith}, {Wilson}, {Baes}, {Cooray}, {Clements},
  {Cortese}, {Foyle}, {Galametz}, {Gomez}, {Lebouteiller}, {Lu}, {Madden},
  {Mentuch}, {O'Halloran}, {Page}, {Remy}, {Schulz}, \&
  {Spinoglio}}]{Bendo+2012}
{Bendo}, G.~J., {Boselli}, A., {Dariush}, A., {et~al.} 2012, \mnras, 419, 1833

\bibitem[{{Berg} {et~al.}(2020){Berg}, {Pogge}, {Skillman}, {Croxall},
  {Moustakas}, {Rogers}, \& {Sun}}]{Berg+2020}
{Berg}, D.~A., {Pogge}, R.~W., {Skillman}, E.~D., {et~al.} 2020, \apj, 893, 96

\bibitem[{{Boquien} {et~al.}(2016){Boquien}, {Kennicutt}, {Calzetti}, {Dale},
  {Galametz}, {Sauvage}, {Croxall}, {Draine}, {Kirkpatrick}, {Kumari}, {Hunt},
  {De Looze}, {Pellegrini}, {Rela{\~n}o}, {Smith}, \&
  {Tabatabaei}}]{Boquien+2016}
{Boquien}, M., {Kennicutt}, R., {Calzetti}, D., {et~al.} 2016, \aap, 591, A6

\bibitem[{{Boselli} {et~al.}(2004){Boselli}, {Lequeux}, \&
  {Gavazzi}}]{Boselli+2004}
{Boselli}, A., {Lequeux}, J., \& {Gavazzi}, G. 2004, \aap, 428, 409

\bibitem[{{Bouwens} {et~al.}(2020){Bouwens}, {Gonzalez-Lopez}, {Aravena},
  {Decarli}, {Novak}, {Stefanon}, {Walter}, {Boogaard}, {Carilli},
  {Dudzeviciute}, {Smail}, {Daddi}, {da Cunha}, {Ivison}, {Nanayakkara},
  {Cortes}, {Cox}, {Inami}, {Oesch}, {Popping}, {Riechers}, {van der Werf},
  {Weiss}, {Fudamoto}, \& {Wagg}}]{Bouwens+2020}
{Bouwens}, R., {Gonzalez-Lopez}, J., {Aravena}, M., {et~al.} 2020, arXiv
  e-prints, arXiv:2009.10727

\bibitem[{{Brown} \& {Gnedin}(2021)}]{Brown+2021}
{Brown}, G., \& {Gnedin}, O.~Y. 2021, \mnras, 508, 5935

\bibitem[{{Buat} \& {Deharveng}(1988)}]{Buat+1988}
{Buat}, V., \& {Deharveng}, J.~M. 1988, \aap, 195, 60

\bibitem[{{Buat} {et~al.}(1999){Buat}, {Donas}, {Milliard}, \&
  {Xu}}]{Buat+1999}
{Buat}, V., {Donas}, J., {Milliard}, B., \& {Xu}, C. 1999, \aap, 352, 371

\bibitem[{{Buat} \& {Xu}(1996)}]{Buat+1996}
{Buat}, V., \& {Xu}, C. 1996, \aap, 306, 61

\bibitem[{{Bushouse} {et~al.}(2022){Bushouse}, {Eisenhamer}, {Dencheva},
  {Davies}, {Greenfield}, {Morrison}, {Hodge}, {Simon}, {Grumm}, {Droettboom},
  {Slavich}, {Sosey}, {Pauly}, {Miller}, {Jedrzejewski}, {Hack}, {Davis},
  {Crawford}, {Law}, {Gordon}, {Regan}, {Cara}, {MacDonald}, {Bradley},
  {Shanahan}, {Jamieson}, {Teodoro}, \& {Williams}}]{Bushouse+2022}
{Bushouse}, H., {Eisenhamer}, J., {Dencheva}, N., {et~al.} 2022,
  doi:10.5281/zenodo.7487203

\bibitem[{{Calzetti}(2001)}]{Calzetti2001}
{Calzetti}, D. 2001, \pasp, 113, 1449

\bibitem[{{Calzetti}(2013)}]{Calzetti2013}
---. 2013, {Star Formation Rate Indicators}, ed. J.~{Falc{\'o}n-Barroso} \&
  J.~H. {Knapen}, 419

\bibitem[{{Calzetti} {et~al.}(2000){Calzetti}, {Armus}, {Bohlin}, {Kinney},
  {Koornneef}, \& {Storchi-Bergmann}}]{Calzetti+2000}
{Calzetti}, D., {Armus}, L., {Bohlin}, R.~C., {et~al.} 2000, \apj, 533, 682

\bibitem[{{Calzetti} {et~al.}(1994){Calzetti}, {Kinney}, \&
  {Storchi-Bergmann}}]{Calzetti+1994}
{Calzetti}, D., {Kinney}, A.~L., \& {Storchi-Bergmann}, T. 1994, \apj, 429, 582

\bibitem[{{Calzetti} {et~al.}(2005){Calzetti}, {Kennicutt}, {Bianchi},
  {Thilker}, {Dale}, {Engelbracht}, {Leitherer}, {Meyer}, {Sosey}, {Mutchler},
  {Regan}, {Thornley}, {Armus}, {Bendo}, {Boissier}, {Boselli}, {Draine},
  {Gordon}, {Helou}, {Hollenbach}, {Kewley}, {Madore}, {Martin}, {Murphy},
  {Rieke}, {Rieke}, {Roussel}, {Sheth}, {Smith}, {Walter}, {White}, {Yi},
  {Scoville}, {Polletta}, \& {Lindler}}]{Calzetti+2005}
{Calzetti}, D., {Kennicutt}, R.~C., J., {Bianchi}, L., {et~al.} 2005, \apj,
  633, 871

\bibitem[{{Calzetti} {et~al.}(2007){Calzetti}, {Kennicutt}, {Engelbracht},
  {Leitherer}, {Draine}, {Kewley}, {Moustakas}, {Sosey}, {Dale}, {Gordon},
  {Helou}, {Hollenbach}, {Armus}, {Bendo}, {Bot}, {Buckalew}, {Jarrett}, {Li},
  {Meyer}, {Murphy}, {Prescott}, {Regan}, {Rieke}, {Roussel}, {Sheth}, {Smith},
  {Thornley}, \& {Walter}}]{Calzetti+2007}
{Calzetti}, D., {Kennicutt}, R.~C., {Engelbracht}, C.~W., {et~al.} 2007, \apj,
  666, 870

\bibitem[{{Calzetti} {et~al.}(2010){Calzetti}, {Wu}, {Hong}, {Kennicutt},
  {Lee}, {Dale}, {Engelbracht}, {van Zee}, {Draine}, {Hao}, {Gordon},
  {Moustakas}, {Murphy}, {Regan}, {Begum}, {Block}, {Dalcanton}, {Funes}, {Gil
  de Paz}, {Johnson}, {Sakai}, {Skillman}, {Walter}, {Weisz}, {Williams}, \&
  {Wu}}]{Calzetti+2010}
{Calzetti}, D., {Wu}, S.~Y., {Hong}, S., {et~al.} 2010, \apj, 714, 1256

\bibitem[{{Calzetti} {et~al.}(2015){Calzetti}, {Lee}, {Sabbi}, {Adamo},
  {Smith}, {Andrews}, {Ubeda}, {Bright}, {Thilker}, {Aloisi}, {Brown},
  {Chandar}, {Christian}, {Cignoni}, {Clayton}, {da Silva}, {de Mink}, {Dobbs},
  {Elmegreen}, {Elmegreen}, {Evans}, {Fumagalli}, {Gallagher}, {Gouliermis},
  {Grebel}, {Herrero}, {Hunter}, {Johnson}, {Kennicutt}, {Kim}, {Krumholz},
  {Lennon}, {Levay}, {Martin}, {Nair}, {Nota}, {{\"O}stlin}, {Pellerin},
  {Prieto}, {Regan}, {Ryon}, {Schaerer}, {Schiminovich}, {Tosi}, {Van Dyk},
  {Walterbos}, {Whitmore}, \& {Wofford}}]{Calzetti+2015a}
{Calzetti}, D., {Lee}, J.~C., {Sabbi}, E., {et~al.} 2015, \aj, 149, 51

\bibitem[{{Calzetti} {et~al.}(2024){Calzetti}, {Adamo}, {Linden}, {Gregg},
  {Krumholz}, {Bajaj}, {Bik}, {Cignoni}, {Correnti}, {Elmegreen}, {Faustino
  Vieira}, {Gallagher}, {Grasha}, {Gutermuth}, {Johnson}, {Messa}, {Melinder},
  {{\"O}stlin}, {Pedrini}, {Sabbi}, {Smith}, \& {Tosi}}]{Calzetti+2024}
{Calzetti}, D., {Adamo}, A., {Linden}, S.~T., {et~al.} 2024, \apj, 971, 118

\bibitem[{{Casey} {et~al.}(2018){Casey}, {Zavala}, {Spilker}, {da Cunha},
  {Hodge}, {Hung}, {Staguhn}, {Finkelstein}, \& {Drew}}]{Casey+2018}
{Casey}, C.~M., {Zavala}, J.~A., {Spilker}, J., {et~al.} 2018, \apj, 862, 77

\bibitem[{{Cervi{\~n}o} {et~al.}(2002){Cervi{\~n}o}, {Valls-Gabaud},
  {Luridiana}, \& {Mas-Hesse}}]{Cervino+2002}
{Cervi{\~n}o}, M., {Valls-Gabaud}, D., {Luridiana}, V., \& {Mas-Hesse}, J.~M.
  2002, \aap, 381, 51

\bibitem[{{Chisholm} {et~al.}(2019){Chisholm}, {Rigby}, {Bayliss}, {Berg},
  {Dahle}, {Gladders}, \& {Sharon}}]{Chisholm+2019}
{Chisholm}, J., {Rigby}, J.~R., {Bayliss}, M., {et~al.} 2019, \apj, 882, 182

\bibitem[{{Chomiuk} \& {Povich}(2011)}]{Chomiuk+2011}
{Chomiuk}, L., \& {Povich}, M.~S. 2011, \aj, 142, 197

\bibitem[{{Colombo} {et~al.}(2014){Colombo}, {Meidt}, {Schinnerer},
  {Garc{\'\i}a-Burillo}, {Hughes}, {Pety}, {Leroy}, {Dobbs}, {Dumas},
  {Thompson}, {Schuster}, \& {Kramer}}]{Colombo+2014}
{Colombo}, D., {Meidt}, S.~E., {Schinnerer}, E., {et~al.} 2014, \apj, 784, 4

\bibitem[{{Cook} {et~al.}(2014){Cook}, {Dale}, {Johnson}, {Van Zee}, {Lee},
  {Kennicutt}, {Calzetti}, {Staudaher}, \& {Engelbracht}}]{Cook+2014}
{Cook}, D.~O., {Dale}, D.~A., {Johnson}, B.~D., {et~al.} 2014, \mnras, 445, 899

\bibitem[{{Cortijo-Ferrero} {et~al.}(2017){Cortijo-Ferrero}, {Gonz{\'a}lez
  Delgado}, {P{\'e}rez}, {Cid Fernandes}, {Garc{\'\i}a-Benito}, {Di Matteo},
  {S{\'a}nchez}, {de Amorim}, {Lacerda}, {L{\'o}pez Fern{\'a}ndez}, \&
  {Tadhunter}}]{Cortijo+2017}
{Cortijo-Ferrero}, C., {Gonz{\'a}lez Delgado}, R.~M., {P{\'e}rez}, E., {et~al.}
  2017, \aap, 607, A70

\bibitem[{{Croxall} {et~al.}(2015){Croxall}, {Pogge}, {Berg}, {Skillman}, \&
  {Moustakas}}]{Croxall+2015}
{Croxall}, K.~V., {Pogge}, R.~W., {Berg}, D.~A., {Skillman}, E.~D., \&
  {Moustakas}, J. 2015, \apj, 808, 42

\bibitem[{{Cs{\"o}rnyei} {et~al.}(2023){Cs{\"o}rnyei}, {Anderson}, {Vogl},
  {Taubenberger}, {Blondin}, {Leibundgut}, \& {Hillebrandt}}]{Csornyei+2023}
{Cs{\"o}rnyei}, G., {Anderson}, R.~I., {Vogl}, C., {et~al.} 2023, \aap, 678,
  A44

\bibitem[{{Dale} {et~al.}(2023){Dale}, {Boquien}, {Turner}, {Calzetti},
  {Kennicutt}, \& {Lee}}]{Dale+2023}
{Dale}, D.~A., {Boquien}, M., {Turner}, J.~A., {et~al.} 2023, \aj, 165, 260

\bibitem[{{Dale} \& {Helou}(2002)}]{DaleHelou2002}
{Dale}, D.~A., \& {Helou}, G. 2002, \apj, 576, 159

\bibitem[{{Dale} {et~al.}(2009){Dale}, {Cohen}, {Johnson}, {Schuster},
  {Calzetti}, {Engelbracht}, {Gil de Paz}, {Kennicutt}, {Lee}, {Begum},
  {Block}, {Dalcanton}, {Funes}, {Gordon}, {Johnson}, {Marble}, {Sakai},
  {Skillman}, {van Zee}, {Walter}, {Weisz}, {Williams}, {Wu}, \&
  {Wu}}]{Dale+2009}
{Dale}, D.~A., {Cohen}, S.~A., {Johnson}, L.~C., {et~al.} 2009, \apj, 703, 517

\bibitem[{{Dale} {et~al.}(2012){Dale}, {Aniano}, {Engelbracht}, {Hinz},
  {Krause}, {Montiel}, {Roussel}, {Appleton}, {Armus}, {Beir{\~a}o}, {Bolatto},
  {Brandl}, {Calzetti}, {Crocker}, {Croxall}, {Draine}, {Galametz}, {Gordon},
  {Groves}, {Hao}, {Helou}, {Hunt}, {Johnson}, {Kennicutt}, {Koda}, {Leroy},
  {Li}, {Meidt}, {Miller}, {Murphy}, {Rahman}, {Rix}, {Sandstrom}, {Sauvage},
  {Schinnerer}, {Skibba}, {Smith}, {Tabatabaei}, {Walter}, {Wilson}, {Wolfire},
  \& {Zibetti}}]{Dale+2012}
{Dale}, D.~A., {Aniano}, G., {Engelbracht}, C.~W., {et~al.} 2012, \apj, 745, 95

\bibitem[{{Dayal} {et~al.}(2022){Dayal}, {Ferrara}, {Sommovigo}, {Bouwens},
  {Oesch}, {Smit}, {Gonzalez}, {Schouws}, {Stefanon}, {Kobayashi}, {Bremer},
  {Algera}, {Aravena}, {Bowler}, {da Cunha}, {Fudamoto}, {Graziani}, {Hodge},
  {Inami}, {De Looze}, {Pallottini}, {Riechers}, {Schneider}, {Stark}, \&
  {Endsley}}]{Dayal+2022}
{Dayal}, P., {Ferrara}, A., {Sommovigo}, L., {et~al.} 2022, \mnras, 512, 989

\bibitem[{{de la Fuente Marcos} \& {de la Fuente
  Marcos}(2009)}]{Delafuente+2009}
{de la Fuente Marcos}, R., \& {de la Fuente Marcos}, C. 2009, \apj, 700, 436

\bibitem[{{de Vaucouleurs} {et~al.}(1991){de Vaucouleurs}, {de Vaucouleurs},
  {Corwin}, {Buta}, {Paturel}, \& {Fouque}}]{deVaucouleurs+1991}
{de Vaucouleurs}, G., {de Vaucouleurs}, A., {Corwin}, Herold~G., J., {et~al.}
  1991, {Third Reference Catalogue of Bright Galaxies}

\bibitem[{{Dicken} {et~al.}(2024){Dicken}, {Mar{\'\i}n}, {Shivaei}, {Guillard},
  {Libralato}, {Glasse}, {Gordon}, {Cossou}, {Kavanagh}, {Temim}, {Flagey},
  {Klaassen}, {Rieke}, {Wright}, {Alberts}, {Azzollini},
  {{\'A}lvarez-M{\'a}rquez}, {Bouchet}, {Bright}, {Cracraft}, {Coulais},
  {Detre}, {Engesser}, {Fox}, {Gaspar}, {Gastaud}, {Glauser}, {Hines},
  {Kendrew}, {Labiano}, {Lagage}, {Lee}, {Law}, {Morrison}, {Noriega-Crespo},
  {Jones}, {Patapis}, {Scheithauer}, {Sloan}, \& {Tamas}}]{Dicken+2024}
{Dicken}, D., {Mar{\'\i}n}, M.~G., {Shivaei}, I., {et~al.} 2024, \aap, 689, A5

\bibitem[{{Dong} \& {Draine}(2011)}]{Dong+2011}
{Dong}, R., \& {Draine}, B.~T. 2011, \apj, 727, 35

\bibitem[{{Dopita} {et~al.}(2003){Dopita}, {Groves}, {Sutherland}, \&
  {Kewley}}]{Dopita+2003}
{Dopita}, M.~A., {Groves}, B.~A., {Sutherland}, R.~S., \& {Kewley}, L.~J. 2003,
  \apj, 583, 727

\bibitem[{{Draine}(2011)}]{Draine+2011}
{Draine}, B.~T. 2011, \apj, 732, 100

\bibitem[{{Draine} {et~al.}(2007){Draine}, {Dale}, {Bendo}, {Gordon}, {Smith},
  {Armus}, {Engelbracht}, {Helou}, {Kennicutt}, {Li}, {Roussel}, {Walter},
  {Calzetti}, {Moustakas}, {Murphy}, {Rieke}, {Bot}, {Hollenbach}, {Sheth}, \&
  {Teplitz}}]{Draine+2007}
{Draine}, B.~T., {Dale}, D.~A., {Bendo}, G., {et~al.} 2007, \apj, 663, 866

\bibitem[{{Efremov} \& {Elmegreen}(1998)}]{Efremov+1998}
{Efremov}, Y.~N., \& {Elmegreen}, B.~G. 1998, \mnras, 299, 588

\bibitem[{{Egorov} {et~al.}(2023){Egorov}, {Kreckel}, {Sandstrom}, {Leroy},
  {Glover}, {Groves}, {Kruijssen}, {Barnes}, {Belfiore}, {Bigiel}, {Blanc},
  {Boquien}, {Cao}, {Chastenet}, {Chevance}, {Congiu}, {Dale}, {Emsellem},
  {Grasha}, {Klessen}, {Larson}, {Liu}, {Murphy}, {Pan}, {Pessa}, {Pety},
  {Rosolowsky}, {Scheuermann}, {Schinnerer}, {Sutter}, {Thilker}, {Watkins}, \&
  {Williams}}]{Egorov+2023}
{Egorov}, O.~V., {Kreckel}, K., {Sandstrom}, K.~M., {et~al.} 2023, \apjl, 944,
  L16

\bibitem[{{Elia} {et~al.}(2025){Elia}, {Evans}, {Soler}, {Strafella},
  {Schisano}, {Molinari}, {Giannetti}, \& {Patra}}]{Elia+2025}
{Elia}, D., {Evans}, N.~J., {Soler}, J.~D., {et~al.} 2025, \apj, 980, 216

\bibitem[{{Elia} {et~al.}(2022){Elia}, {Molinari}, {Schisano}, {Soler},
  {Merello}, {Russeil}, {Veneziani}, {Zavagno}, {Noriega-Crespo}, {Olmi},
  {Benedettini}, {Hennebelle}, {Klessen}, {Leurini}, {Paladini}, {Pezzuto},
  {Traficante}, {Eden}, {Martin}, {Sormani}, {Coletta}, {Colman}, {Plume},
  {Maruccia}, {Mininni}, \& {Liu}}]{Elia+2022}
{Elia}, D., {Molinari}, S., {Schisano}, E., {et~al.} 2022, \apj, 941, 162

\bibitem[{{Elmegreen}(1989)}]{Elmegreen1989}
{Elmegreen}, B.~G. 1989, \apj, 338, 178

\bibitem[{{Elmegreen}(2011)}]{Elmegreen+2011}
{Elmegreen}, B.~G. 2011, in EAS Publications Series, Vol.~51, EAS Publications
  Series, ed. C.~{Charbonnel} \& T.~{Montmerle}, 31--44

\bibitem[{{Elmegreen}(2018)}]{Elmegreen+2018}
---. 2018, \apj, 853, 88

\bibitem[{{Engelbracht} {et~al.}(2008){Engelbracht}, {Rieke}, {Gordon},
  {Smith}, {Werner}, {Moustakas}, {Willmer}, \& {Vanzi}}]{Engelbracht+2008}
{Engelbracht}, C.~W., {Rieke}, G.~H., {Gordon}, K.~D., {et~al.} 2008, \apj,
  678, 804

\bibitem[{{Fahrion} \& {De Marchi}(2023)}]{Fahrion+2023}
{Fahrion}, K., \& {De Marchi}, G. 2023, \aap, 671, L14

\bibitem[{{Ferguson} {et~al.}(1996){Ferguson}, {Wyse}, {Gallagher}, \&
  {Hunter}}]{Ferguson+1996}
{Ferguson}, A. M.~N., {Wyse}, R. F.~G., {Gallagher}, J.~S., I., \& {Hunter},
  D.~A. 1996, \aj, 111, 2265

\bibitem[{{Fitzpatrick}(1999)}]{Fitzpatrick1999}
{Fitzpatrick}, E.~L. 1999, \pasp, 111, 63

\bibitem[{{Fitzpatrick} \& {Massa}(2007)}]{Fitzpatrick+2007}
{Fitzpatrick}, E.~L., \& {Massa}, D. 2007, \apj, 663, 320

\bibitem[{{Fitzpatrick} {et~al.}(2019){Fitzpatrick}, {Massa}, {Gordon},
  {Bohlin}, \& {Clayton}}]{Fitzpatrick+2019}
{Fitzpatrick}, E.~L., {Massa}, D., {Gordon}, K.~D., {Bohlin}, R., \& {Clayton},
  G.~C. 2019, \apj, 886, 108

\bibitem[{{Fumagalli} {et~al.}(2011){Fumagalli}, {da Silva}, \&
  {Krumholz}}]{Fumagalli+2011}
{Fumagalli}, M., {da Silva}, R.~L., \& {Krumholz}, M.~R. 2011, \apjl, 741, L26

\bibitem[{{Gallagher} \& {Hunter}(1983)}]{Gallagher+1983}
{Gallagher}, J.~S., \& {Hunter}, D.~A. 1983, \apj, 274, 141

\bibitem[{{Gardner} {et~al.}(2023){Gardner}, {Mather}, {Abbott}, {Abell},
  {Abernathy}, {Abney}, {Abraham}, {Abraham}, {Abul-Huda}, {Acton}, {Adams},
  {Adams}, {Adler}, {Adriaensen}, {Aguilar}, {Ahmed}, {Ahmed}, {Ahmed},
  {Albat}, {Albert}, {Alberts}, {Aldridge}, {Allen}, {Allen}, {Altenburg},
  {Altunc}, {Alvarez}, {{\'A}lvarez-M{\'a}rquez}, {Alves de Oliveira},
  {Ambrose}, {Anandakrishnan}, {Andersen}, {Anderson}, {Anderson}, {Anderson},
  {Anderson}, {Aprea}, {Archer}, {Arenberg}, {Argyriou}, {Arribas}, {Artigau},
  {Arvai}, {Atcheson}, {Atkinson}, {Averbukh}, {Aymergen}, {Bacinski},
  {Baggett}, {Bagnasco}, {Baker}, {Balzano}, {Banks}, {Baran}, {Barker},
  {Barrett}, {Barringer}, {Barto}, {Bast}, {Baudoz}, {Baum}, {Beatty},
  {Beaulieu}, {Bechtold}, {Beck}, {Beddard}, {Beichman}, {Bellagama}, {Bely},
  {Berger}, {Bergeron}, {Bernier}, {Bertch}, {Beskow}, {Betz}, {Biagetti},
  {Birkmann}, {Bjorklund}, {Blackwood}, {Blazek}, {Blossfeld}, {Bluth},
  {Boccaletti}, {Boegner}, {Bohlin}, {Boia}, {B{\"o}ker}, {Bonaventura},
  {Bond}, {Bosley}, {Boucarut}, {Bouchet}, {Bouwman}, {Bower}, {Bowers},
  {Bowers}, {Boyce}, {Boyer}, {Boyer}, {Boyer}, {Boyer}, {Bradley}, {Brady},
  {Brandl}, {Brannen}, {Breda}, {Bremmer}, {Brennan}, {Bresnahan}, {Bright},
  {Broiles}, {Bromenschenkel}, {Brooks}, {Brooks}, {Brown}, {Brown}, {Brown},
  {Bruce}, {Bryson}, {Bujanda}, {Bullock}, {Bunker}, {Bureo}, {Burt}, {Bush},
  {Bushouse}, {Bussman}, {Cabaud}, {Cale}, {Calhoon}, {Calvani}, {Canipe},
  {Caputo}, {Cara}, {Carey}, {Case}, {Cesari}, {Cetorelli}, {Chance},
  {Chandler}, {Chaney}, {Chapman}, {Charlot}, {Chayer}, {Cheezum}, {Chen},
  {Chen}, {Cherinka}, {Chichester}, {Chilton}, {Chittiraibalan}, {Clampin},
  {Clark}, {Clark}, {Clark}, {Claybrooks}, {Cleveland}, {Cohen}, {Cohen},
  {Col{\'o}n}, {Coleman}, {Colina}, {Comber}, {Comeau}, {Comer}, {Conde Reis},
  {Connolly}, {Conroy}, {Contos}, {Contreras}, {Cook}, {Cooper}, {Cooper},
  {Correia}, {Correnti}, {Cossou}, {Costanza}, {Coulais}, {Cox}, {Coyle},
  {Cracraft}, {Crew}, {Curtis}, {Cusveller}, {Da Costa Maciel}, {Dailey},
  {Daugeron}, {Davidson}, {Davies}, {Davis}, {Davis}, {Day}, {de Chambure}, {de
  Jong}, {De Marchi}, {Dean}, {Decker}, {Delisa}, {Dell}, \&
  {Dellagatta}}]{Gardner+2023}
{Gardner}, J.~P., {Mather}, J.~C., {Abbott}, R., {et~al.} 2023, \pasp, 135,
  068001

\bibitem[{{Garn} \& {Best}(2010)}]{Garn+2010}
{Garn}, T., \& {Best}, P.~N. 2010, \mnras, 409, 421

\bibitem[{{Girardi} {et~al.}(2000){Girardi}, {Bressan}, {Bertelli}, \&
  {Chiosi}}]{Girardi+2000}
{Girardi}, L., {Bressan}, A., {Bertelli}, G., \& {Chiosi}, C. 2000, \aaps, 141,
  371

\bibitem[{{Gonzaga} {et~al.}(2012){Gonzaga}, {Hack}, {Fruchter}, \&
  {Mack}}]{Gonzaga+2012}
{Gonzaga}, S., {Hack}, W., {Fruchter}, A., \& {Mack}, J. 2012, {The DrizzlePac
  Handbook}

\bibitem[{{Gordon} {et~al.}(2003){Gordon}, {Clayton}, {Misselt}, {Land olt}, \&
  {Wolff}}]{Gordon+2003}
{Gordon}, K.~D., {Clayton}, G.~C., {Misselt}, K.~A., {Land olt}, A.~U., \&
  {Wolff}, M.~J. 2003, \apj, 594, 279

\bibitem[{{Gouliermis}(2018)}]{Gouliermis+2018}
{Gouliermis}, D.~A. 2018, \pasp, 130, 072001

\bibitem[{{Gouliermis} {et~al.}(2015){Gouliermis}, {Thilker}, {Elmegreen},
  {Elmegreen}, {Calzetti}, {Lee}, {Adamo}, {Aloisi}, {Cignoni}, {Cook}, {Dale},
  {Gallagher}, {Grasha}, {Grebel}, {Dav{\'o}}, {Hunter}, {Johnson}, {Kim},
  {Nair}, {Nota}, {Pellerin}, {Ryon}, {Sabbi}, {Sacchi}, {Smith}, {Tosi},
  {Ubeda}, \& {Whitmore}}]{Gouliermis+2015}
{Gouliermis}, D.~A., {Thilker}, D., {Elmegreen}, B.~G., {et~al.} 2015, \mnras,
  452, 3508

\bibitem[{{Grasha} {et~al.}(2015){Grasha}, {Calzetti}, {Adamo}, {Kim},
  {Elmegreen}, {Gouliermis}, {Aloisi}, {Bright}, {Christian}, {Cignoni},
  {Dale}, {Dobbs}, {Elmegreen}, {Fumagalli}, {Gallagher}, {Grebel}, {Johnson},
  {Lee}, {Messa}, {Smith}, {Ryon}, {Thilker}, {Ubeda}, \&
  {Wofford}}]{Grasha+2015}
{Grasha}, K., {Calzetti}, D., {Adamo}, A., {et~al.} 2015, \apj, 815, 93

\bibitem[{{Grasha} {et~al.}(2017{\natexlab{a}}){Grasha}, {Elmegreen},
  {Calzetti}, {Adamo}, {Aloisi}, {Bright}, {Cook}, {Dale}, {Fumagalli},
  {Gallagher}, {Gouliermis}, {Grebel}, {Kahre}, {Kim}, {Krumholz}, {Lee},
  {Messa}, {Ryon}, \& {Ubeda}}]{Grasha+2017}
{Grasha}, K., {Elmegreen}, B.~G., {Calzetti}, D., {et~al.} 2017{\natexlab{a}},
  \apj, 842, 25

\bibitem[{{Grasha} {et~al.}(2017{\natexlab{b}}){Grasha}, {Calzetti}, {Adamo},
  {Kim}, {Elmegreen}, {Gouliermis}, {Dale}, {Fumagalli}, {Grebel}, {Johnson},
  {Kahre}, {Kennicutt}, {Messa}, {Pellerin}, {Ryon}, {Smith}, {Shabani},
  {Thilker}, \& {Ubeda}}]{Grasha+2017a}
{Grasha}, K., {Calzetti}, D., {Adamo}, A., {et~al.} 2017{\natexlab{b}}, \apj,
  840, 113

\bibitem[{{Greenfield} \& {Miller}(2016)}]{Greenfield+2016}
{Greenfield}, P., \& {Miller}, T. 2016, Astronomy and Computing, 16, 41

\bibitem[{{Gregg} {et~al.}(2022){Gregg}, {Calzetti}, \& {Heyer}}]{Gregg+2022}
{Gregg}, B., {Calzetti}, D., \& {Heyer}, M. 2022, \apj, 928, 120

\bibitem[{{Hannon} {et~al.}(2022){Hannon}, {Lee}, {Whitmore}, {Mobasher},
  {Thilker}, {Chandar}, {Adamo}, {Wofford}, {Orozco-Duarte}, {Calzetti}, {Della
  Bruna}, {Kreckel}, {Groves}, {Barnes}, {Boquien}, {Belfiore}, \&
  {Linden}}]{Hannon+2022}
{Hannon}, S., {Lee}, J.~C., {Whitmore}, B.~C., {et~al.} 2022, \mnras, 512, 1294

\bibitem[{{Hao} {et~al.}(2011){Hao}, {Kennicutt}, {Johnson}, {Calzetti},
  {Dale}, \& {Moustakas}}]{Hao+2011}
{Hao}, C.-N., {Kennicutt}, R.~C., {Johnson}, B.~D., {et~al.} 2011, \apj, 741,
  124

\bibitem[{{Helou}(1986)}]{Helou1986}
{Helou}, G. 1986, \apjl, 311, L33

\bibitem[{{Heyer} {et~al.}(2009){Heyer}, {Krawczyk}, {Duval}, \&
  {Jackson}}]{Heyer+2009}
{Heyer}, M., {Krawczyk}, C., {Duval}, J., \& {Jackson}, J.~M. 2009, \apj, 699,
  1092

\bibitem[{{Heyer} \& {Brunt}(2004)}]{Heyer+2004}
{Heyer}, M.~H., \& {Brunt}, C.~M. 2004, \apjl, 615, L45

\bibitem[{{Hirashita} {et~al.}(2003){Hirashita}, {Buat}, \&
  {Inoue}}]{Hirashita+2003}
{Hirashita}, H., {Buat}, V., \& {Inoue}, A.~K. 2003, \aap, 410, 83

\bibitem[{{Hoopes} \& {Walterbos}(2003)}]{Hoopes+2003}
{Hoopes}, C.~G., \& {Walterbos}, R. A.~M. 2003, \apj, 586, 902

\bibitem[{{Hoopes} {et~al.}(1996){Hoopes}, {Walterbos}, \&
  {Greenwalt}}]{Hoopes+1996}
{Hoopes}, C.~G., {Walterbos}, R. A.~M., \& {Greenwalt}, B.~E. 1996, \aj, 112,
  1429

\bibitem[{{Howell} {et~al.}(2010){Howell}, {Armus}, {Mazzarella}, {Evans},
  {Surace}, {Sanders}, {Petric}, {Appleton}, {Bothun}, {Bridge}, {Chan},
  {Charmandaris}, {Frayer}, {Haan}, {Inami}, {Kim}, {Lord}, {Madore},
  {Melbourne}, {Schulz}, {U}, {Vavilkin}, {Veilleux}, \& {Xu}}]{Howell+2010}
{Howell}, J.~H., {Armus}, L., {Mazzarella}, J.~M., {et~al.} 2010, \apj, 715,
  572

\bibitem[{{Hunter} {et~al.}(1989){Hunter}, {Gallagher}, {Rice}, \&
  {Gillett}}]{Hunter+1989}
{Hunter}, D.~A., {Gallagher}, III, J.~S., {Rice}, W.~L., \& {Gillett}, F.~C.
  1989, \apj, 336, 152

\bibitem[{{Inoue} {et~al.}(2001){Inoue}, {Hirashita}, \& {Kamaya}}]{Inoue+2001}
{Inoue}, A.~K., {Hirashita}, H., \& {Kamaya}, H. 2001, \apj, 555, 613

\bibitem[{{Isobe} {et~al.}(1990){Isobe}, {Feigelson}, {Akritas}, \&
  {Babu}}]{Isobe+1990}
{Isobe}, T., {Feigelson}, E.~D., {Akritas}, M.~G., \& {Babu}, G.~J. 1990, \apj,
  364, 104

\bibitem[{{Joye} \& {Mandel}(2003)}]{Joye+2003}
{Joye}, W.~A., \& {Mandel}, E. 2003, in Astronomical Society of the Pacific
  Conference Series, Vol. 295, Astronomical Data Analysis Software and Systems
  XII, ed. H.~E. {Payne}, R.~I. {Jedrzejewski}, \& R.~N. {Hook}, 489

\bibitem[{{Kelly}(2007)}]{Kelly2007}
{Kelly}, B.~C. 2007, \apj, 665, 1489

\bibitem[{{Kennicutt} {et~al.}(2008){Kennicutt}, {Lee}, {Funes}, {J.}, {Sakai},
  \& {Akiyama}}]{Kennicutt+2008}
{Kennicutt}, Robert~C., J., {Lee}, J.~C., {Funes}, J.~G., {et~al.} 2008, \apjs,
  178, 247

\bibitem[{{Kennicutt} {et~al.}(2007){Kennicutt}, {Calzetti}, {Walter}, {Helou},
  {Hollenbach}, {Armus}, {Bendo}, {Dale}, {Draine}, {Engelbracht}, {Gordon},
  {Prescott}, {Regan}, {Thornley}, {Bot}, {Brinks}, {de Blok}, {de Mello},
  {Meyer}, {Moustakas}, {Murphy}, {Sheth}, \& {Smith}}]{Kennicutt+2007}
{Kennicutt}, Robert~C., J., {Calzetti}, D., {Walter}, F., {et~al.} 2007, \apj,
  671, 333

\bibitem[{{Kennicutt} {et~al.}(2009){Kennicutt}, {Hao}, {Calzetti},
  {Moustakas}, {Dale}, {Bendo}, {Engelbracht}, {Johnson}, \&
  {Lee}}]{Kennicutt+2009}
{Kennicutt}, Robert~C., J., {Hao}, C.-N., {Calzetti}, D., {et~al.} 2009, \apj,
  703, 1672

\bibitem[{{Kennicutt} \& {Evans}(2012)}]{KennicuttEvans2012}
{Kennicutt}, R.~C., \& {Evans}, N.~J. 2012, \araa, 50, 531

\bibitem[{{Kroupa}(2001)}]{Kroupa2001}
{Kroupa}, P. 2001, \mnras, 322, 231

\bibitem[{{Krumholz} \& {Matzner}(2009)}]{Krumholz+2009}
{Krumholz}, M.~R., \& {Matzner}, C.~D. 2009, \apj, 703, 1352

\bibitem[{{Leitherer} {et~al.}(1999){Leitherer}, {Schaerer}, {Goldader},
  {Delgado}, {Robert}, {Kune}, {de Mello}, {Devost}, \&
  {Heckman}}]{Leitherer+1999}
{Leitherer}, C., {Schaerer}, D., {Goldader}, J.~D., {et~al.} 1999, \apjs, 123,
  3

\bibitem[{{Leroy} {et~al.}(2023){Leroy}, {Sandstrom}, {Rosolowsky}, {Belfiore},
  {Bolatto}, {Cao}, {Koch}, {Schinnerer}, {Barnes}, {Be{\v{s}}li{\'c}},
  {Bigiel}, {Blanc}, {Chastenet}, {Chen}, {Chevance}, {Chown}, {Congiu},
  {Dale}, {Egorov}, {Emsellem}, {Eibensteiner}, {Faesi}, {Glover}, {Grasha},
  {Groves}, {Hassani}, {Henshaw}, {Hughes}, {Jim{\'e}nez-Donaire}, {Kim},
  {Klessen}, {Kreckel}, {Kruijssen}, {Larson}, {Lee}, {Levy}, {Liu}, {Lopez},
  {Meidt}, {Murphy}, {Neumann}, {Pessa}, {Pety}, {Saito}, {Sardone}, {Sun},
  {Thilker}, {Usero}, {Watkins}, {Whitcomb}, \& {Williams}}]{Leroy+2023}
{Leroy}, A.~K., {Sandstrom}, K., {Rosolowsky}, E., {et~al.} 2023, \apjl, 944,
  L9

\bibitem[{{Li} {et~al.}(2013){Li}, {Crocker}, {Calzetti}, {Wilson},
  {Kennicutt}, {Murphy}, {Brandl}, {Draine}, {Galametz}, {Johnson}, {Armus},
  {Gordon}, {Croxall}, {Dale}, {Engelbracht}, {Groves}, {Hao}, {Helou}, {Hinz},
  {Hunt}, {Krause}, {Roussel}, {Sauvage}, \& {Smith}}]{Li+2013}
{Li}, Y., {Crocker}, A.~F., {Calzetti}, D., {et~al.} 2013, \apj, 768, 180

\bibitem[{{Libralato} {et~al.}(2024){Libralato}, {Argyriou}, {Dicken},
  {Garc{\'\i}a Mar{\'\i}n}, {Guillard}, {Hines}, {Kavanagh}, {Kendrew}, {Law},
  {Noriega-Crespo}, \& {{\'A}lvarez-M{\'a}rquez}}]{Libralato+2024}
{Libralato}, M., {Argyriou}, I., {Dicken}, D., {et~al.} 2024, \pasp, 136,
  034502

\bibitem[{{Liu} {et~al.}(2025){Liu}, {Dunlop}, {McLure}, {McLeod}, {Barrufet},
  {Carnall}, {Begley}, {P{\'e}rez-Gonz{\'a}lez}, {Donnan}, {Ellis}, {Grogin},
  {Magee}, {Illingworth}, {Cullen}, {Stevenson}, {Koekemoer}, {Fontana}, \&
  {Bowler}}]{Liu+2025}
{Liu}, F.-Y., {Dunlop}, J.~S., {McLure}, R.~J., {et~al.} 2025, arXiv e-prints,
  arXiv:2503.07774

\bibitem[{{Liu} {et~al.}(2011){Liu}, {Koda}, {Calzetti}, {Fukuhara}, \&
  {Momose}}]{Liu+2011}
{Liu}, G., {Koda}, J., {Calzetti}, D., {Fukuhara}, M., \& {Momose}, R. 2011,
  \apj, 735, 63

\bibitem[{{Lonsdale Persson} \& {Helou}(1987)}]{LonsdaleHelou1987}
{Lonsdale Persson}, C.~J., \& {Helou}, G. 1987, \apj, 314, 513

\bibitem[{{Madau} \& {Dickinson}(2014)}]{MadauDickinson2014}
{Madau}, P., \& {Dickinson}, M. 2014, \araa, 52, 415

\bibitem[{{Magnelli} {et~al.}(2014){Magnelli}, {Lutz}, {Saintonge}, {Berta},
  {Santini}, {Symeonidis}, {Altieri}, {Andreani}, {Aussel}, {B{\'e}thermin},
  {Bock}, {Bongiovanni}, {Cepa}, {Cimatti}, {Conley}, {Daddi}, {Elbaz},
  {F{\"o}rster Schreiber}, {Genzel}, {Ivison}, {Le Floc'h}, {Magdis},
  {Maiolino}, {Nordon}, {Oliver}, {Page}, {P{\'e}rez Garc{\'\i}a}, {Poglitsch},
  {Popesso}, {Pozzi}, {Riguccini}, {Rodighiero}, {Rosario}, {Roseboom},
  {Sanchez-Portal}, {Scott}, {Sturm}, {Tacconi}, {Valtchanov}, {Wang}, \&
  {Wuyts}}]{Magnelli+2014}
{Magnelli}, B., {Lutz}, D., {Saintonge}, A., {et~al.} 2014, \aap, 561, A86

\bibitem[{{Marcillac} {et~al.}(2006){Marcillac}, {Elbaz}, {Chary}, {Dickinson},
  {Galliano}, \& {Morrison}}]{Marcillac+2006}
{Marcillac}, D., {Elbaz}, D., {Chary}, R.~R., {et~al.} 2006, \aap, 451, 57

\bibitem[{{Mart{\'\i}nez-Garc{\'\i}a}
  {et~al.}(2018){Mart{\'\i}nez-Garc{\'\i}a}, {Bruzual}, {Magris C.}, \&
  {Gonz{\'a}lez-L{\'o}pezlira}}]{Martinez+2018}
{Mart{\'\i}nez-Garc{\'\i}a}, E.~E., {Bruzual}, G., {Magris C.}, G., \&
  {Gonz{\'a}lez-L{\'o}pezlira}, R.~A. 2018, \mnras, 474, 1862

\bibitem[{Matsumoto \& Nishimura(1998)}]{Matsumoto+1998}
Matsumoto, M., \& Nishimura, T. 1998, {ACM} Trans. Model. Comput. Simul., 8, 3

\bibitem[{{McCallum} {et~al.}(2025){McCallum}, {Wood}, {Benjamin},
  {Krishnarao}, {Zucker}, {Edenhofer}, \& {Haffner}}]{McCallum+2025}
{McCallum}, L., {Wood}, K., {Benjamin}, R.~A., {et~al.} 2025, \mnras, 540, L21

\bibitem[{{McQuinn} {et~al.}(2016){McQuinn}, {Skillman}, {Dolphin}, {Berg}, \&
  {Kennicutt}}]{McQuinn+2016}
{McQuinn}, K. B.~W., {Skillman}, E.~D., {Dolphin}, A.~E., {Berg}, D., \&
  {Kennicutt}, R. 2016, \apj, 826, 21

\bibitem[{{Meidt} {et~al.}(2012){Meidt}, {Schinnerer}, {Knapen}, {Bosma},
  {Athanassoula}, {Sheth}, {Buta}, {Zaritsky}, {Laurikainen}, {Elmegreen},
  {Elmegreen}, {Gadotti}, {Salo}, {Regan}, {Ho}, {Madore}, {Hinz}, {Skibba},
  {Gil de Paz}, {Mu{\~n}oz-Mateos}, {Men{\'e}ndez-Delmestre}, {Seibert}, {Kim},
  {Mizusawa}, {Laine}, \& {Comer{\'o}n}}]{Meidt+2012}
{Meidt}, S.~E., {Schinnerer}, E., {Knapen}, J.~H., {et~al.} 2012, \apj, 744, 17

\bibitem[{{Mendigut{\'\i}a} {et~al.}(2018){Mendigut{\'\i}a}, {Lada}, \&
  {Oudmaijer}}]{Mendigutia+2018}
{Mendigut{\'\i}a}, I., {Lada}, C.~J., \& {Oudmaijer}, R.~D. 2018, \aap, 618,
  A119

\bibitem[{{Messa} {et~al.}(2021){Messa}, {Calzetti}, {Adamo}, {Grasha},
  {Johnson}, {Sabbi}, {Smith}, {Bajaj}, {Finn}, \& {Lin}}]{Messa+2021}
{Messa}, M., {Calzetti}, D., {Adamo}, A., {et~al.} 2021, \apj, 909, 121

\bibitem[{{Meurer} {et~al.}(1999){Meurer}, {Heckman}, \&
  {Calzetti}}]{Meurer+1999}
{Meurer}, G.~R., {Heckman}, T.~M., \& {Calzetti}, D. 1999, \apj, 521, 64

\bibitem[{{Moustakas} \& {Kennicutt}(2006)}]{Moustakas+2006}
{Moustakas}, J., \& {Kennicutt}, Jr., R.~C. 2006, \apjs, 164, 81

\bibitem[{{Oey} {et~al.}(2007){Oey}, {Meurer}, {Yelda}, {Furst},
  {Caballero-Nieves}, {Hanish}, {Levesque}, {Thilker}, {Walth},
  {Bland-Hawthorn}, {Dopita}, {Ferguson}, {Heckman}, {Doyle}, {Drinkwater},
  {Freeman}, {Kennicutt}, {Kilborn}, {Knezek}, {Koribalski}, {Meyer}, {Putman},
  {Ryan-Weber}, {Smith}, {Staveley-Smith}, {Webster}, {Werk}, \&
  {Zwaan}}]{Oey+2007}
{Oey}, M.~S., {Meurer}, G.~R., {Yelda}, S., {et~al.} 2007, \apj, 661, 801

\bibitem[{{Osterbrock} \& {Ferland}(2006)}]{Osterbrock+2006}
{Osterbrock}, D.~E., \& {Ferland}, G.~J. 2006, {Astrophysics of gaseous nebulae
  and active galactic nuclei}

\bibitem[{{Paspaliaris} {et~al.}(2021){Paspaliaris}, {Xilouris}, {Nersesian},
  {Masoura}, {Plionis}, {Georgantopoulos}, {Bianchi}, {Katsioli}, \&
  {Mountrichas}}]{Paspaliaris+2021}
{Paspaliaris}, E.~D., {Xilouris}, E.~M., {Nersesian}, A., {et~al.} 2021, \aap,
  649, A137

\bibitem[{{Pedrini} {et~al.}(2024){Pedrini}, {Adamo}, {Calzetti}, {Bik},
  {Gregg}, {Linden}, {Bajaj}, {Ryon}, {Ali}, {Bortolini}, {Correnti},
  {Elmegreen}, {Elmegreen}, {Gallagher}, {Grasha}, {Gutermuth}, {Johnson},
  {Melinder}, {Messa}, {{\"O}stlin}, {Sabbi}, {Smith}, {Tosi}, \&
  {Vieira}}]{Pedrini+2024}
{Pedrini}, A., {Adamo}, A., {Calzetti}, D., {et~al.} 2024, \apj, 971, 32

\bibitem[{{Pellegrini} {et~al.}(2012){Pellegrini}, {Oey}, {Winkler}, {Points},
  {Smith}, {Jaskot}, \& {Zastrow}}]{Pellegrini+2012}
{Pellegrini}, E.~W., {Oey}, M.~S., {Winkler}, P.~F., {et~al.} 2012, \apj, 755,
  40

\bibitem[{{Pereira-Santaella} {et~al.}(2015){Pereira-Santaella},
  {Alonso-Herrero}, {Colina}, {Miralles-Caballero}, {P{\'e}rez-Gonz{\'a}lez},
  {Arribas}, {Bellocchi}, {Cazzoli}, {D{\'\i}az-Santos}, \& {Piqueras
  L{\'o}pez}}]{Pereira+2015}
{Pereira-Santaella}, M., {Alonso-Herrero}, A., {Colina}, L., {et~al.} 2015,
  \aap, 577, A78

\bibitem[{{Press} {et~al.}(1992){Press}, {Teukolsky}, {Vetterling}, \&
  {Flannery}}]{Press+1992}
{Press}, W.~H., {Teukolsky}, S.~A., {Vetterling}, W.~T., \& {Flannery}, B.~P.
  1992, {Numerical recipes in FORTRAN. The art of scientific computing}

\bibitem[{{Querejeta} {et~al.}(2023){Querejeta}, {Pety}, {Schruba}, {Leroy},
  {Herrera}, {Chiang}, {Meidt}, {Rosolowsky}, {Schinnerer}, {Schuster}, {Sun},
  {Herrmann}, {Barnes}, {Be{\v{s}}li{\'c}}, {Bigiel}, {Cao}, {Chevance},
  {Eibensteiner}, {Emsellem}, {Faesi}, {Hughes}, {Kim}, {Klessen}, {Kreckel},
  {Kruijssen}, {Liu}, {Neumayer}, {Pan}, {Saito}, {Sandstrom}, {Teng}, {Usero},
  {Williams}, \& {Zakardjian}}]{Querejeta+2023}
{Querejeta}, M., {Pety}, J., {Schruba}, A., {et~al.} 2023, \aap, 680, A4

\bibitem[{{Rela{\~n}o} {et~al.}(2007){Rela{\~n}o}, {Lisenfeld},
  {P{\'e}rez-Gonz{\'a}lez}, {V{\'\i}lchez}, \& {Battaner}}]{Relano+2007}
{Rela{\~n}o}, M., {Lisenfeld}, U., {P{\'e}rez-Gonz{\'a}lez}, P.~G.,
  {V{\'\i}lchez}, J.~M., \& {Battaner}, E. 2007, \apjl, 667, L141

\bibitem[{{Rela{\~n}o} {et~al.}(2013){Rela{\~n}o}, {Verley}, {P{\'e}rez},
  {Kramer}, {Calzetti}, {Xilouris}, {Boquien}, {Abreu-Vicente}, {Combes},
  {Israel}, {Tabatabaei}, {Braine}, {Buchbender}, {Gonz{\'a}lez}, {Gratier},
  {Lord}, {Mookerjea}, {Quintana-Lacaci}, \& {van der Werf}}]{Relano+2013}
{Rela{\~n}o}, M., {Verley}, S., {P{\'e}rez}, I., {et~al.} 2013, \aap, 552, A140

\bibitem[{{Rela{\~n}o} {et~al.}(2018){Rela{\~n}o}, {De Looze}, {Kennicutt},
  {Lisenfeld}, {Dariush}, {Verley}, {Braine}, {Tabatabaei}, {Kramer},
  {Boquien}, {Xilouris}, \& {Gratier}}]{Relano+2018}
{Rela{\~n}o}, M., {De Looze}, I., {Kennicutt}, R.~C., {et~al.} 2018, \aap, 613,
  A43

\bibitem[{{R{\'e}my-Ruyer} {et~al.}(2015){R{\'e}my-Ruyer}, {Madden},
  {Galliano}, {Lebouteiller}, {Baes}, {Bendo}, {Boselli}, {Ciesla}, {Cormier},
  {Cooray}, {Cortese}, {De Looze}, {Doublier-Pritchard}, {Galametz}, {Jones},
  {Karczewski}, {Lu}, \& {Spinoglio}}]{RemyRuyer+2015}
{R{\'e}my-Ruyer}, A., {Madden}, S.~C., {Galliano}, F., {et~al.} 2015, \aap,
  582, A121

\bibitem[{{Renzini} \& {Peng}(2015)}]{Renzini+2015}
{Renzini}, A., \& {Peng}, Y.-j. 2015, \apjl, 801, L29

\bibitem[{{Reynolds}(1984)}]{Reynolds+1984}
{Reynolds}, R.~J. 1984, \apj, 282, 191

\bibitem[{{Reynolds}(1990)}]{Reynolds+1990}
---. 1990, \apjl, 349, L17

\bibitem[{{Rieke} {et~al.}(2009){Rieke}, {Alonso-Herrero}, {Weiner},
  {P{\'e}rez-Gonz{\'a}lez}, {Blaylock}, {Donley}, \& {Marcillac}}]{Rieke+2009}
{Rieke}, G.~H., {Alonso-Herrero}, A., {Weiner}, B.~J., {et~al.} 2009, \apj,
  692, 556

\bibitem[{{Rieke} {et~al.}(2004){Rieke}, {Young}, {Engelbracht}, {Kelly},
  {Low}, {Haller}, {Beeman}, {Gordon}, {Stansberry}, {Misselt}, {Cadien},
  {Morrison}, {Rivlis}, {Latter}, {Noriega-Crespo}, {Padgett}, {Stapelfeldt},
  {Hines}, {Egami}, {Muzerolle}, {Alonso-Herrero}, {Blaylock}, {Dole}, {Hinz},
  {Le Floc'h}, {Papovich}, {P{\'e}rez-Gonz{\'a}lez}, {Smith}, {Su}, {Bennett},
  {Frayer}, {Henderson}, {Lu}, {Masci}, {Pesenson}, {Rebull}, {Rho}, {Keene},
  {Stolovy}, {Wachter}, {Wheaton}, {Werner}, \& {Richards}}]{Rieke+2004}
{Rieke}, G.~H., {Young}, E.~T., {Engelbracht}, C.~W., {et~al.} 2004, \apjs,
  154, 25

\bibitem[{{Rieke} {et~al.}(2015){Rieke}, {Wright}, {B{\"o}ker}, {Bouwman},
  {Colina}, {Glasse}, {Gordon}, {Greene}, {G{\"u}del}, {Henning}, {Justtanont},
  {Lagage}, {Meixner}, {N{\o}rgaard-Nielsen}, {Ray}, {Ressler}, {van Dishoeck},
  \& {Waelkens}}]{Rieke+2015}
{Rieke}, G.~H., {Wright}, G.~S., {B{\"o}ker}, T., {et~al.} 2015, \pasp, 127,
  584

\bibitem[{{Rieke} {et~al.}(2005){Rieke}, {Kelly}, \& {Horner}}]{Rieke+2005}
{Rieke}, M.~J., {Kelly}, D., \& {Horner}, S. 2005, in Society of Photo-Optical
  Instrumentation Engineers (SPIE) Conference Series, Vol. 5904, Cryogenic
  Optical Systems and Instruments XI, ed. J.~B. {Heaney} \& L.~G. {Burriesci},
  1--8

\bibitem[{{Rieke} {et~al.}(2023){Rieke}, {Kelly}, {Misselt}, {Stansberry},
  {Boyer}, {Beatty}, {Egami}, {Florian}, {Greene}, {Hainline}, {Leisenring},
  {Roellig}, {Schlawin}, {Sun}, {Tinnin}, {Williams}, {Willmer}, {Wilson},
  {Clark}, {Rohrbach}, {Brooks}, {Canipe}, {Correnti}, {DiFelice}, {Gennaro},
  {Girard}, {Hartig}, {Hilbert}, {Koekemoer}, {Nikolov}, {Pirzkal}, {Rest},
  {Robberto}, {Sunnquist}, {Telfer}, {Wu}, {Ferry}, {Lewis}, {Baum},
  {Beichman}, {Doyon}, {Dressler}, {Eisenstein}, {Ferrarese}, {Hodapp},
  {Horner}, {Jaffe}, {Johnstone}, {Krist}, {Martin}, {McCarthy}, {Meyer},
  {Rieke}, {Trauger}, \& {Young}}]{Rieke+2023}
{Rieke}, M.~J., {Kelly}, D.~M., {Misselt}, K., {et~al.} 2023, \pasp, 135,
  028001

\bibitem[{{Rigby} {et~al.}(2023){Rigby}, {Perrin}, {McElwain}, {Kimble},
  {Friedman}, {Lallo}, {Doyon}, {Feinberg}, {Ferruit}, {Glasse}, {Rieke},
  {Rieke}, {Wright}, {Willott}, {Colon}, {Milam}, {Neff}, {Stark}, {Valenti},
  {Abell}, {Abney}, {Abul-Huda}, {Acton}, {Adams}, {Adler}, {Aguilar}, {Ahmed},
  {Albert}, {Alberts}, {Aldridge}, {Allen}, {Altenburg},
  {{\'A}lvarez-M{\'a}rquez}, {Alves de Oliveira}, {Andersen}, {Anderson},
  {Anderson}, {Argyriou}, {Armstrong}, {Arribas}, {Artigau}, {Arvai},
  {Atkinson}, {Bacon}, {Bair}, {Banks}, {Barrientes}, {Barringer}, {Bartosik},
  {Bast}, {Baudoz}, {Beatty}, {Bechtold}, {Beck}, {Bergeron}, {Bergkoetter},
  {Bhatawdekar}, {Birkmann}, {Blazek}, {Blome}, {Boccaletti}, {B{\"o}ker},
  {Boia}, {Bonaventura}, {Bond}, {Bosley}, {Boucarut}, {Bourque}, {Bouwman},
  {Bower}, {Bowers}, {Boyer}, {Bradley}, {Brady}, {Braun}, {Breda},
  {Bresnahan}, {Bright}, {Britt}, {Bromenschenkel}, {Brooks}, {Brooks},
  {Brown}, {Brown}, {Brown}, {Bunker}, {Burger}, {Bushouse}, {Cale}, {Cameron},
  {Cameron}, {Canipe}, {Caplinger}, {Caputo}, {Cara}, {Carey}, {Carniani},
  {Carrasquilla}, {Carruthers}, {Case}, {Catherine}, {Chance}, {Chapman},
  {Charlot}, {Charlow}, {Chayer}, {Chen}, {Cherinka}, {Chichester}, {Chilton},
  {Chonis}, {Clampin}, {Clark}, {Clark}, {Coe}, {Coleman}, {Comber}, {Comeau},
  {Connolly}, {Cooper}, {Cooper}, {Coppock}, {Correnti}, {Cossou}, {Coulais},
  {Coyle}, {Cracraft}, {Curti}, {Cuturic}, {Davis}, {Davis}, {Dean}, {DeLisa},
  {deMeester}, {Dencheva}, {Dencheva}, {DePasquale}, {Deschenes}, {Hunor
  Detre}, {Diaz}, {Dicken}, {DiFelice}, {Dillman}, {Dixon}, {Doggett},
  {Donaldson}, {Douglas}, {DuPrie}, {Dupuis}, {Durning}, {Easmin}, {Eck},
  {Edeani}, {Egami}, {Ehrenwinkler}, {Eisenhamer}, {Eisenhower}, {Elie},
  {Elliott}, {Elliott}, {Ellis}, {Engesser}, {Espinoza}, {Etienne}, {Etxaluze},
  {Falini}, {Feeney}, {Ferry}, {Filippazzo}, {Fincham}, {Fix}, {Flagey},
  {Florian}, {Flynn}, {Fontanella}, {Ford}, {Forshay}, {Fox}, {Franz}, {Fu},
  {Fullerton}, {Galkin}, {Galyer}, {Garc{\'\i}a Mar{\'\i}n}, {Gardner},
  {Gardner}, {Garland}, {Garrett}, {Gasman}, {Gaspar}, {Gaudreau}, {Gauthier},
  {Geers}, {Geithner}, {Gennaro}, {Giardino}, {Girard}, {Giuliano},
  {Glassmire}, \& {Glauser}}]{Rigby+2023}
{Rigby}, J., {Perrin}, M., {McElwain}, M., {et~al.} 2023, \pasp, 135, 048001

\bibitem[{{Rowan-Robinson} \& {Crawford}(1989)}]{RowanRobinson+1989}
{Rowan-Robinson}, M., \& {Crawford}, J. 1989, \mnras, 238, 523

\bibitem[{{Ryon} {et~al.}(2015){Ryon}, {Bastian}, {Adamo}, {Konstantopoulos},
  {Gallagher}, {Larsen}, {Hollyhead}, {Silva-Villa}, \& {Smith}}]{Ryon+2015}
{Ryon}, J.~E., {Bastian}, N., {Adamo}, A., {et~al.} 2015, \mnras, 452, 525

\bibitem[{{Ryon} {et~al.}(2017){Ryon}, {Gallagher}, {Smith}, {Adamo},
  {Calzetti}, {Bright}, {Cignoni}, {Cook}, {Dale}, {Elmegreen}, {Fumagalli},
  {Gouliermis}, {Grasha}, {Grebel}, {Kim}, {Messa}, {Thilker}, \&
  {Ubeda}}]{Ryon+2017}
{Ryon}, J.~E., {Gallagher}, J.~S., {Smith}, L.~J., {et~al.} 2017, \apj, 841, 92

\bibitem[{{Sabbi} {et~al.}(2018){Sabbi}, {Calzetti}, {Ubeda}, {Adamo},
  {Cignoni}, {Thilker}, {Aloisi}, {Elmegreen}, {Elmegreen}, {Gouliermis},
  {Grebel}, {Messa}, {Smith}, {Tosi}, {Dolphin}, {Andrews}, {Ashworth},
  {Bright}, {Brown}, {Chandar}, {Christian}, {Clayton}, {Cook}, {Dale}, {de
  Mink}, {Dobbs}, {Evans}, {Fumagalli}, {Gallagher}, {Grasha}, {Herrero},
  {Hunter}, {Johnson}, {Kahre}, {Kennicutt}, {Kim}, {Krumholz}, {Lee},
  {Lennon}, {Martin}, {Nair}, {Nota}, {{\"O}stlin}, {Pellerin}, {Prieto},
  {Regan}, {Ryon}, {Sacchi}, {Schaerer}, {Schiminovich}, {Shabani}, {Van Dyk},
  {Walterbos}, {Whitmore}, \& {Wofford}}]{Sabbi+2018}
{Sabbi}, E., {Calzetti}, D., {Ubeda}, L., {et~al.} 2018, \apjs, 235, 23

\bibitem[{{Sauvage} \& {Thuan}(1992)}]{Sauvage+1992}
{Sauvage}, M., \& {Thuan}, T.~X. 1992, \apjl, 396, L69

\bibitem[{{Schlafly} \& {Finkbeiner}(2011)}]{Schlafly+2011}
{Schlafly}, E.~F., \& {Finkbeiner}, D.~P. 2011, \apj, 737, 103

\bibitem[{{Shashank} {et~al.}(2025){Shashank}, {Subramanian}, {Muraleedharan},
  {Menon}, {Mondal}, {Krishna}, {Das}, \& {Subramaniam}}]{Shashank+2025}
{Shashank}, G., {Subramanian}, S., {Muraleedharan}, S., {et~al.} 2025, \aap,
  693, A188

\bibitem[{{Sirianni} {et~al.}(2005){Sirianni}, {Jee}, {Ben{\'\i}tez},
  {Blakeslee}, {Martel}, {Meurer}, {Clampin}, {De Marchi}, {Ford}, {Gilliland},
  {Hartig}, {Illingworth}, {Mack}, \& {McCann}}]{Sirianni+2005}
{Sirianni}, M., {Jee}, M.~J., {Ben{\'\i}tez}, N., {et~al.} 2005, \pasp, 117,
  1049

\bibitem[{{Smith} {et~al.}(2012){Smith}, {Dunne}, {da Cunha}, {Rowlands},
  {Maddox}, {Gomez}, {Bonfield}, {Charlot}, {Driver}, {Popescu}, {Tuffs},
  {Dunlop}, {Jarvis}, {Seymour}, {Symeonidis}, {Baes}, {Bourne}, {Clements},
  {Cooray}, {De Zotti}, {Dye}, {Eales}, {Scott}, {Verma}, {van der Werf},
  {Andrae}, {Auld}, {Buttiglione}, {Cava}, {Dariush}, {Fritz}, {Hopwood},
  {Ibar}, {Ivison}, {Kelvin}, {Madore}, {Pohlen}, {Rigby}, {Robotham},
  {Seibert}, \& {Temi}}]{SmithDunne+2012}
{Smith}, D.~J.~B., {Dunne}, L., {da Cunha}, E., {et~al.} 2012, \mnras, 427, 703

\bibitem[{{Smith} \& {Weedman}(1970)}]{Smith+1970}
{Smith}, M.~G., \& {Weedman}, D.~W. 1970, \apj, 161, 33

\bibitem[{{Soler} {et~al.}(2023){Soler}, {Zari}, {Elia}, {Molinari}, {Mininni},
  {Schisano}, {Traficante}, {Klessen}, {Glover}, {Hennebelle}, {Colman},
  {Frankel}, \& {Wenger}}]{Soler+2023}
{Soler}, J.~D., {Zari}, E., {Elia}, D., {et~al.} 2023, \aap, 678, A95

\bibitem[{{Terlevich} \& {Melnick}(1981)}]{Terlevich+1981}
{Terlevich}, R., \& {Melnick}, J. 1981, \mnras, 195, 839

\bibitem[{{Tody}(1986)}]{Tody1986}
{Tody}, D. 1986, in Society of Photo-Optical Instrumentation Engineers (SPIE)
  Conference Series, Vol. 627, Instrumentation in astronomy VI, ed. D.~L.
  {Crawford}, 733

\bibitem[{{Tody}(1993)}]{Tody1993}
{Tody}, D. 1993, in Astronomical Society of the Pacific Conference Series,
  Vol.~52, Astronomical Data Analysis Software and Systems II, ed. R.~J.
  {Hanisch}, R.~J.~V. {Brissenden}, \& J.~{Barnes}, 173

\bibitem[{{Tomi{\v{c}}i{\'c}} {et~al.}(2019){Tomi{\v{c}}i{\'c}}, {Ho},
  {Kreckel}, {Schinnerer}, {Leroy}, {Groves}, {Sandstrom}, {Blanc}, {Jarrett},
  {Thilker}, {Kapala}, \& {McElroy}}]{Tomicic+2019}
{Tomi{\v{c}}i{\'c}}, N., {Ho}, I.~T., {Kreckel}, K., {et~al.} 2019, \apj, 873,
  3

\bibitem[{{Treyer} {et~al.}(2010){Treyer}, {Schiminovich}, {Johnson}, {O'Dowd},
  {Martin}, {Wyder}, {Charlot}, {Heckman}, {Martins}, {Seibert}, \& {van der
  Hulst}}]{Treyer+2010}
{Treyer}, M., {Schiminovich}, D., {Johnson}, B.~D., {et~al.} 2010, \apj, 719,
  1191

\bibitem[{{van der Hulst} {et~al.}(1988){van der Hulst}, {Kennicutt}, {Crane},
  \& {Rots}}]{vanderHulst+1988}
{van der Hulst}, J.~M., {Kennicutt}, R.~C., {Crane}, P.~C., \& {Rots}, A.~H.
  1988, \aap, 195, 38

\bibitem[{{V{\'a}zquez} \& {Leitherer}(2005)}]{Vazquez+2005}
{V{\'a}zquez}, G.~A., \& {Leitherer}, C. 2005, \apj, 621, 695

\bibitem[{{Voges} \& {Walterbos}(2006)}]{Voges+2006}
{Voges}, E.~S., \& {Walterbos}, R.~A.~M. 2006, \apjl, 644, L29

\bibitem[{{Walterbos} \& {Greenawalt}(1996)}]{Walterbos+1996}
{Walterbos}, R. A.~M., \& {Greenawalt}, B. 1996, \apj, 460, 696

\bibitem[{{Wang} \& {Heckman}(1996)}]{Wang+1996}
{Wang}, B., \& {Heckman}, T.~M. 1996, \apj, 457, 645

\bibitem[{{Weingartner} \& {Draine}(2001)}]{Weingartner+2001}
{Weingartner}, J.~C., \& {Draine}, B.~T. 2001, \apj, 548, 296

\bibitem[{{Whitmore} {et~al.}(2005){Whitmore}, {Gilmore}, {Leitherer}, {Fall},
  {Chandar}, {Blair}, {Schweizer}, {Zhang}, \& {Miller}}]{Whitmore+2005}
{Whitmore}, B.~C., {Gilmore}, D., {Leitherer}, C., {et~al.} 2005, \aj, 130,
  2104

\bibitem[{{Whitmore} {et~al.}(2011){Whitmore}, {Chandar}, {Kim}, {Kaleida},
  {Mutchler}, {Stankiewicz}, {Calzetti}, {Saha}, {O'Connell}, {Balick}, {Bond},
  {Carollo}, {Disney}, {Dopita}, {Frogel}, {Hall}, {Holtzman}, {Kimble},
  {McCarthy}, {Paresce}, {Silk}, {Trauger}, {Walker}, {Windhorst}, \&
  {Young}}]{Whitmore+2011}
{Whitmore}, B.~C., {Chandar}, R., {Kim}, H., {et~al.} 2011, \apj, 729, 78

\bibitem[{{Wofford} {et~al.}(2016){Wofford}, {Charlot}, {Bruzual}, {Eldridge},
  {Calzetti}, {Adamo}, {Cignoni}, {de Mink}, {Gouliermis}, {Grasha}, {Grebel},
  {Lee}, {{\"O}stlin}, {Smith}, {Ubeda}, \& {Zackrisson}}]{Wofford+2016}
{Wofford}, A., {Charlot}, S., {Bruzual}, G., {et~al.} 2016, \mnras, 457, 4296

\bibitem[{{Zavala} {et~al.}(2021){Zavala}, {Casey}, {Manning}, {Aravena},
  {Bethermin}, {Caputi}, {Clements}, {Cunha}, {Drew}, {Finkelstein},
  {Fujimoto}, {Hayward}, {Hodge}, {Kartaltepe}, {Knudsen}, {Koekemoer}, {Long},
  {Magdis}, {Man}, {Popping}, {Sanders}, {Scoville}, {Sheth}, {Staguhn},
  {Toft}, {Treister}, {Vieira}, \& {Yun}}]{Zavala+2021}
{Zavala}, J.~A., {Casey}, C.~M., {Manning}, S.~M., {et~al.} 2021, \apj, 909,
  165

\bibitem[{{Zhang} {et~al.}(2017){Zhang}, {Yan}, {Bundy}, {Bershady}, {Haffner},
  {Walterbos}, {Maiolino}, {Tremonti}, {Thomas}, {Drory}, {Jones}, {Belfiore},
  {S{\'a}nchez}, {Diamond-Stanic}, {Bizyaev}, {Nitschelm}, {Andrews},
  {Brinkmann}, {Brownstein}, {Cheung}, {Li}, {Law}, {Roman Lopes}, {Oravetz},
  {Pan}, {Storchi Bergmann}, \& {Simmons}}]{Zhang+2017}
{Zhang}, K., {Yan}, R., {Bundy}, K., {et~al.} 2017, \mnras, 466, 3217

\bibitem[{{Zhu} {et~al.}(2008){Zhu}, {Wu}, {Cao}, \& {Li}}]{Zhu+2008}
{Zhu}, Y.-N., {Wu}, H., {Cao}, C., \& {Li}, H.-N. 2008, \apj, 686, 155

\end{thebibliography}

\end{document}